\documentclass{article}

\usepackage{arxiv}

\usepackage[english]{babel}
\usepackage{amssymb}
\usepackage{amsmath}
\newcommand*{\R}{\mathbb{R}}
\usepackage{url}
\usepackage{booktabs, nicematrix}
\usepackage{pifont}
\usepackage{xcolor}
\usepackage{algorithm}
\usepackage{algpseudocode}
\usepackage{graphicx}
\usepackage[graphicx]{realboxes}
\usepackage{hyperref}
\hypersetup{colorlinks = true, citecolor = blue}
\usepackage{caption}
\usepackage{subcaption}
\usepackage[authoryear,round,semicolon,nobibstyle]{natbib}

\usepackage{afterpage}
\usepackage{multirow}
\usepackage{multicol}
\usepackage{array}
\usepackage{tabularx}  
\usepackage{makecell}  
\usepackage{longtable} 
\usepackage[figuresright]{rotating} 

\title{GARG-AML Against Smurfing: A Scalable and Interpretable Graph-Based Framework for Anti-Money Laundering}

\author{
	Bruno Deprez\thanks{Corresponding author: \href{mailto:bruno.deprez@kuleuven.be}{\texttt{bruno.deprez@kuleuven.be}}} \\
	KU Leuven\\
	University of Antwerp - imec
	\And
	Bart Baesens\\
	KU Leuven \\
	University of Southampton 
	\And
	Tim Verdonck\\
	University of Antwerp - imec\\
	KU Leuven 
	\And
	Wouter Verbeke\\
	KU Leuven
}

\begin{document}
\maketitle

\begin{abstract}
	Money laundering, a global problem estimated to account for $2\%$-$5\%$ of the global GDP, required regulators to enforce stringent rules on financial institutions to combat it. One prominent laundering method, called smurfing, relies on breaking up transactions into smaller amounts to evade detection. Given the complexity of laundering schemes, involving numerous transactions across multiple parties, network analytics is an important tool for anti-money laundering. However, recent advances in network analytics have leaned heavily on black-box network embedding methods, limiting their utility in businesses.
	To address the specific nature of smurfing and banks' operational requirements,
	this paper introduces GARG-AML, a novel graph-based method that quantifies smurfing risk through a single interpretable metric derived from the structure of the second-order transaction network.
	Unlike traditional methods, GARG-AML provides a balance between computational efficiency and transparency, enabling integration into existing AML workflows. We demonstrate the scalability of GARG-AML on large random networks, and its performance on two open-source money laundering datasets. Our results highlight the potential of GARG-AML to enhance smurfing detection and support investigative decision-making.
\end{abstract}


	\section{Introduction}
	Criminals launder an estimated USD~2~trillion, up to $5\%$ of global GDP, a year to disguise their illicit profits and fund future crimes~\citep{UNODC}. 
	To mitigate this, regulators have mandated that financial institutions serve as the primary front in anti-money laundering (AML) efforts~\citep{FATF, levi2006money}. 
	
	Financial institutions have put in place AML detection and reporting processes.
	This often starts with alert generation.
	Alert generation is an automatic process for spotting unusual behaviour, and results in many false positives. 
	Manual investigation filters out legitimate transactions, based on both internal (e.g., transactions and client records) and external data (e.g., public domain information searches on company structure). 
	The remaining alerts are further investigated to build cases, which contain multiple alerts. 
	Expert investigators decide if a suspicious activity report~(SAR) is filed to the regulator~\citep{10.1108/JMLC-09-2024-0152, 10.1108/JMLC-07-2019-0055,FATF, levi2006money}.  
	
	A large portion of generated alerts comes from business rules flagging large transaction amounts in accordance with regulation~\citep{welling1989smurfs,10.1214/ss/1042727940,senator1995financial,oztas2024transaction,10.1108/JMLC-07-2019-0055}. 
	These rules also consider aggregation attempts, where criminals make many transactions each staying under the reporting threshold, but where the aggregated amount passes a certain threshold.
	However, the efficacy of these rules is often compromised by the sheer volume and complexity of global transaction data. 
	
	A popular method to circumvent these business rules is \textit{smurfing}~\citep{madinger2011money, welling1989smurfs, starnini2021smurf, zhdanova2014no}. 
	Criminals split transactions into smaller amounts and transfer them through money mules, called smurfs, to dodge the rules. 
	Smurfing makes AML harder since (1) the chain between the criminal activity and the criminals' bank accounts is interrupted, and (2) smaller transactions through distinct accounts seem less suspicious.
	
	Banks monitor millions of transactions and require multiple departments to report suspicious activity. To be useful, any new tool must meet at least these two criteria: 
	\begin{itemize}
		\item The method must process large volumes of transactions in a memory and time-efficient manner; 
		\item The predictions must be interpretable for smooth integration in SAR filing. 
	\end{itemize}
	Despite these business requirements, the current network-based AML literature focuses on black box graph neural networks~(GNNs)~\citep{deprez2024networkevaluation}. 
	Consequently, many institutions rely on business rules and manual feature engineering based on expert knowledge.  
	
	To address the specific nature of smurfing and the operational requirements, we introduce \textbf{G}raph-\textbf{A}ided \textbf{R}isk \textbf{G}rading for \textbf{A}nti-\textbf{M}oney \textbf{L}aundering (GARG-AML), a novel method that grades clients based on their second-order neighbourhood in the transaction network. 
	GARG-AML quantifies how closely an account's network structure resembles known smurfing patterns. 
	
	Moreover, we present an extension using tree-based learners to increase smurf detection power. 
	This shows the versatility and power of our method, without compromising on the business need of speed and simplicity. 
	
	The performance is evaluated against state-of-the-art smurfing detection methods using both synthetic data and two open-source network datasets. 
	These datasets, which include millions of clients and transactions, closely resemble real-world financial networks. 
	As a result, our experiments assess both detection power and computational efficiency.
	
	The main contributions of this work are as follows:
	\begin{itemize}
		\item We introduce a novel smurfing detection method that is scalable to large transaction networks, interpretable and easily integrated into existing models;
		\item We evaluate our method on multiple datasets that are uniquely suited for assessing the effectiveness of our approach; 
		\item We compare the current state-of-the-art in smurfing detection methods, and we analyse the scalability of these methods to very large transaction datasets, where the original papers only tested their methods on substantially smaller datasets.
	\end{itemize}
	
	The remainder of this work is organised as follows. 
	Related work on anti-money laundering and smurfing is discussed in Section~\ref{sec: related work}. 
	The methodology behind GARG-AML is described in Section~\ref{sec: methodology}. 
	Section~\ref{sec: experimental setup} outlines the experimental evaluation, with results presented in Section~\ref{sec: results and discussion}. 
	We discuss the main data and methodological limitations in Section~\ref{sec: limitations}. 
	Section~\ref{sec: conclusion} concludes this work. 
	
	The full implementation of the experiments is made available on GitHub\footnote{\url{https://github.com/B-Deprez/GARG-AML}}. 
	The GARG-AML implementation is also published as the open-source Python package \texttt{garg-aml-smurfing}, available via PyPI\footnote{\url{https://pypi.org/project/garg-aml-smurfing/}}, fostering the adoption of our method by researchers and practitioners.
	
	\section{Related Work}
	\label{sec: related work}
	According to Routine Activity Theory (RAT), criminal acts occur when three elements converge: a motivated offender, a suitable target, and the absence of a capable guardian. In the context of money laundering, all three are present~\citep{94159f36-482b-366e-8fe0-8267f3acfe6f}:
	\begin{itemize}
		\item Motivated Offenders: Criminal organisations seeking to integrate ``dirty'' money into the legitimate economy without triggering regulatory alarms.
		\item Suitable Targets: Financial institutions, whose systemic gaps or high transaction volumes can be exploited to hide illicit flows.
		\item Capable Guardians: The AML systems and human investigators tasked with identifying and reporting suspicious activity. 
	\end{itemize}
	
	Current technological ``guardians'' largely rely on IF-THEN business rules that monitor individual transactions or aggregate amounts over time~\citep{10.1214/ss/1042727940, LEONARD1995350, SANCHEZ20093630, welling1989smurfs, cardoso2022laundrograph, deprez2024networkevaluation}. 
	While transparent, these rules are often ineffective as they are easily circumvented once the underlying logic is exposed, reducing their utility as capable guardians~\citep{zhu2013rule,VANBELLE2023113866}. 
	Additionally, the volume and complexity of monitored transactions often compromise the effectiveness of these guardians~\citep{10.1108/JMLC-06-2025-0106,10.1108/JMLC-07-2024-0114}.
	As a result, machine learning has emerged as a preferred approach, offering scalability and the ability to detect complex laundering patterns~\citep{chen2018machine,10.1108/JMLC-07-2019-0055,10.1108/JMLC-04-2024-0083}. 
	
	Network analytics has emerged as a critical tool for AML, shifting the focus from individual transactions to the structural interactions between entities~\citep{deprez2024networkevaluation}.
	Current literature can be categorised into three primary research streams: manual feature engineering, deep learning architectures, and pattern-specific detection frameworks.
	
	Early research emphasised the use of centrality measures for profile definition.
	\citet{DREZEWSKI201518} used centrality measures to define nine roles in criminal networks. 
	\citet{FRONZETTICOLLADON201749} used similar metrics across industries and analysed the correlation with the target label to demonstrate the added value of network analytics for AML. 
	\citet{wu2021detecting} defined temporal motives tailored to money laundering and mixing detection in the Bitcoin network. 
	
	Deep learning has been studied extensively through GNNs~\citep{lo2023inspection, alarab2020competence, weber2019anti,deprez2024networkevaluation}. 
	These methods capture the complex geometry of suspicious money laundering activities in financial networks by learning network topology while also incorporating node features. 
	A drawback is that GNNs are not interpretable~\citep{kute2021deep}, which limits their use by investigators. 
	
	Smurfing, also called microstructuring, has long been a popular laundering method~\citep{welling1989smurfs, senator1995financial}. 
	\citet{welling1989smurfs} sees it as a consequence of the Bank Secrecy Act in the USA, which required reporting of transactions above $\$$10,000. 
	Similar monetary-based systems are still in place that require immediate investigation of transactions that surpass a given amount~\citep{senator1995financial,oztas2024transaction, 10.1214/ss/1042727940}. 
	
\begin{figure}
		\centering
		\includegraphics[width=0.6\linewidth]{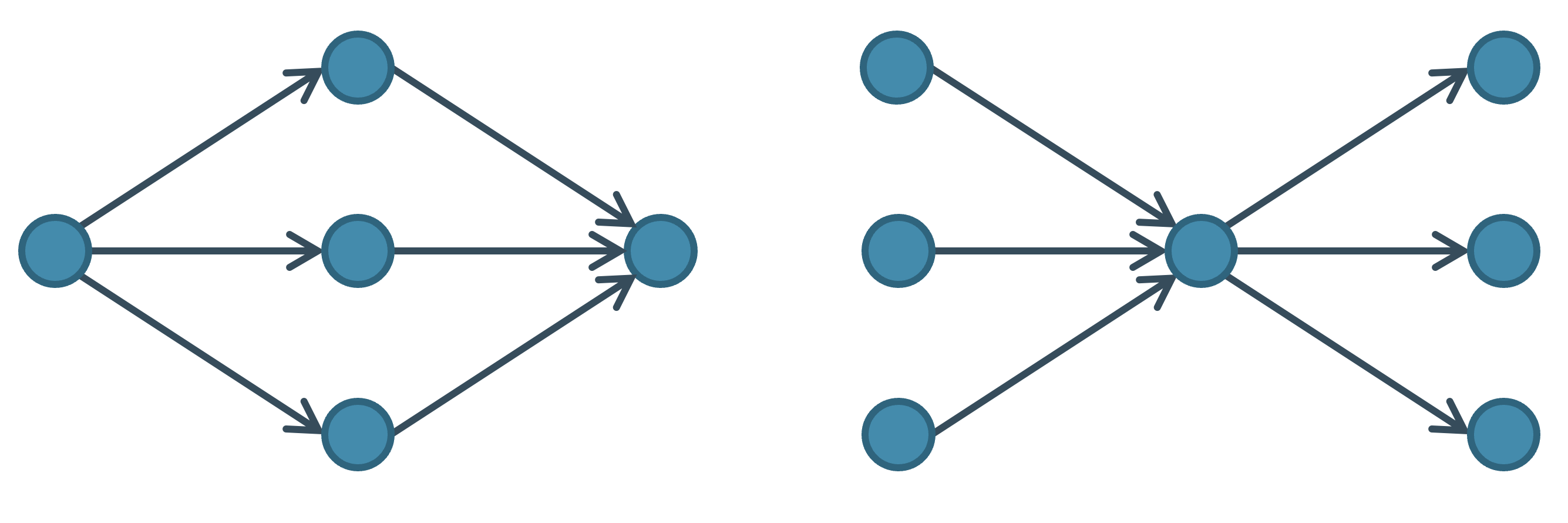}
		\caption{The two smurfing patterns; as scatter-gather (left) and gather-scatter (right).}
		\label{fig:patterns}
	\end{figure}

	Although many money laundering methods and patterns exist~\citep{AMLSim}, two types of patterns are typical for smurfing~\citep{starnini2021smurf,altman2024realistic}: (1) scatter-gather and (2) gather-scatter. 
	Both patterns are visualised in Figure~\ref{fig:patterns}. 
	With scatter-gather, money is sent to its destination account via money mules. 
	Gather-scatter, on the other hand, consists of collecting money from multiple accounts into one (or a few) account(s) and sending it to many destination accounts. 
	Gather-scatter is a popular money laundering method for cryptocurrencies~\citep{wu2021towards}. 
	
	Smurfing makes use of smurfs, who function as money mules. 
	Smurfs are recruited from outside the criminal organisation for a short period of time. 
	Criminals lure the young or financially vulnerable with promises of quick cash, using them to open accounts in the victim's name to funnel illicit funds through~\citep{10.1108/JMLC-05-2020-0053,leukfeldt2015cyber,doi:10.1177/14773708251323349}.
	
	The evolution of laundering techniques like smurfing is best explained by Rational Choice Theory~\citep{McCarthy2014}. 
	Offenders seek to maximise reward while minimising the risk of detection. 
	By splitting large, reportable sums into smaller transactions funnelled through smurfs, criminals exploit the `blind spots' of rule-based systems. 
	This tactic interrupts the visible chain of funds, making illicit flows appear indistinguishable from legitimate retail banking activity.
	
	There is a rich collection of research specifically on smurfing detection.
	PSA@R~\citep{zhdanova2014no} detects smurfing in mobile money networks. 
	It uses different underlying workflow models. 
	Transaction streams that differ from the process model are flagged as anomalous. 
	FlowScope~\citep{li2020flowscope} identifies anomalous sub-graphs by calculating the net inflow and outflow at intermediate middlemen who ideally retain zero balance. AutoAudit~\citep{lee2020autoaudit} utilises adjacency matrix reordering to expose high-density blocks that signify coordinated smurfing. 
	\citet{starnini2021smurf} and SMoTeF~\citep{shadrooh2024smotef} incorporate temporal constraints, pruning data to focus on motifs that emerge within specific time windows. 
	FaSTM$\forall$N~\citep{tariq2023topologyagnostic} tries to find anomalous money flows, or sub-graphs, from a temporal representation of sequential transactions. 
	
	Closely related to smurfing is mixing, which is a money laundering method that is used for cryptocurrencies~\citep{liu2021knowledge}. 
	Coins are transferred to the same entity and are redistributed afterward, resulting in a gather-scatter pattern. 
	
	\citet{prado2018discovering} introduce an anomaly score for mixing detection by identifying mixing services that act as bridges between unrelated user communities. 
	\citet{wu2021detecting} applies positive and unlabelled~(PU)-learning using predefined temporal motifs. These are extracted from two networks, one with just accounts as nodes and the other with accounts and transactions as nodes.
	
	The methodologies and experimental setups of these existing works are summarised and juxtaposed against GARG-AML in Table~\ref{tab:summary literature methodology} and Table~\ref{tab:summary literature experiments}, respectively. Methods for anti-money laundering on fiat currency consider scatter-gather patterns, while research on crypto-currency is mostly concerned with gather-scatter. 
	
	The evaluation in most papers is done on relatively small datasets. 
	The Czech dataset~\citep{czechdata} is used most. 
	It is a small network containing approximately 11,000 nodes and 273,000 edges. 
	This dataset is unlabelled, so the authors of each paper inject synthetic patterns. 
	Evaluation and comparison is often limited to threshold-dependent metrics like precision, recall and F1. 
	GARG-AML addresses these limitations by providing an interpretable, parallelisable scoring framework evaluated on massive, realistic transaction networks using threshold-independent metrics. 
	
	\begin{table*}
		\centering
		\caption{Summary of key aspects of the methodology in the literature.}
		\label{tab:summary literature methodology}
		\small
		\setlength{\tabcolsep}{4pt}
		\begin{tabular}{@{}llccccc@{}}
			\toprule
			Paper & Pattern & \rotatebox{90}{Client Scoring} & \rotatebox{90}{Parallel Calculation} & \rotatebox{90}{All Nodes} & \rotatebox{90}{Code} & \rotatebox{90}{Scalable} \\
			\midrule
			\citet{zhdanova2014no} &  Scatter-gather& \ding{51}& \ding{51}& \ding{51}& & \\
			\citet{li2020flowscope} & Scatter-gather& & & \ding{51}& \ding{51}& \ding{51}\\
			\citet{lee2020autoaudit} & Scatter-gather& \ding{51}& & & \ding{51}& \\
			\citet{starnini2021smurf} & Scatter-gather \& gather-scatter& & \ding{51}& & & \ding{51}\\
			\citet{shadrooh2024smotef} & Scatter-gather& \ding{51}& & & & \ding{51}\\
			\citet{prado2018discovering} & Gather-scatter& \ding{51}& \ding{51}& \ding{51}& & \\
			\citet{wu2021detecting} & Gather-scatter& & & & & \ding{51}\\
			\midrule
			This work & Scatter-gather \& gather-scatter& \ding{51}& \ding{51}& \ding{51}& \ding{51}& \ding{51}\\
			\bottomrule
		\end{tabular}
	\end{table*}
	
	\begin{table*}
		\centering
		\caption{Summary of key aspects of the experimental set-ups in the literature.}
		\label{tab:summary literature experiments}
		\small
		\setlength{\tabcolsep}{4pt}
		\begin{tabularx}{\linewidth}{@{}l>{\raggedright\arraybackslash}X>{\raggedright\arraybackslash}X>{\raggedright\arraybackslash}X@{}}
			\toprule
			Paper & Dataset & Size Dataset (nodes) & Metrics  \\
			\midrule
			\citet{zhdanova2014no} & Simulated& 10,000& Precision, recall\\
			\citet{li2020flowscope} & Proprietary \& Czech Financial& 6,130,000 \& 11,374& F1, Time\\
			\citet{lee2020autoaudit} & Czech Financial \& Enron& 11,374 \& 16,771& Accuracy\\
			\citet{starnini2021smurf} & Proprietary& 18,000,000 (reduced to 21,000)& Time\\
			\citet{shadrooh2024smotef} & Czech Financial \& Enron \& Darpa & 11,374 \& 16,771 \& 371,000& Time, Precision, Recall, F1\\
			\citet{prado2018discovering} & Bitcoin& 942,204& Recall\\
			\citet{wu2021detecting} & WalletExplorer& $\pm2,500,000$& TPR, FPR, geometric~mean\\
			\midrule
			This work & Simulated \& IBM AML& 100 \& 10,000 \& 100,000 \& 515,080 \& 2,054,390& Time, AUC-ROC, AUC-PR, Precision, F1, Precision@K, Rank\\
			\bottomrule
		\end{tabularx}
	\end{table*}
	
	\section{GARG-AML}
	\label{sec: methodology}
	GARG-AML provides each individual node with a score expressing how likely it is that this entity is involved in a smurfing scheme. 
	This is done by analysing the second-order neighbourhood of the nodes. 
	We start by formally defining some crucial concepts. 
	Afterwards, we introduce two versions for GARG-AML, one for undirected and one for directed networks. 
	The full GARG-AML method is given in Algorithm~\ref{alg:gargaml}. 
	
	\subsection{Preliminaries and Notation}
	The process of money laundering is traditionally divided into three stages~\citep{levi2006money,UNODC}.
	During placement, illicit funds enter the financial system, often in jurisdictions where regulatory checks are lighter. 
	During layering, transactions mix illegally and legitimately obtained money, obscuring their origin. 
	During integration, funds are used to make seemingly legitimate purchases, after which the money is effectively laundered. 
	
	During the placement and layering stage, money mules are often used~\citep{10.1108/JFC-02-2020-0028}. 
	These can either be people actively recruited from outside the organisation~\citep{10.1108/JMLC-05-2020-0053,leukfeldt2015cyber} or unaware victims whose bank accounts have been hacked~\citep{10.1108/JFC-02-2020-0028}. 
	Their accounts are used to receive money from criminal activity and to quickly transfer it to the criminal organisation's account. 
	The use of accounts in the name of outsiders circumvents strict KYC rules and puts distance between the criminal activity and the criminals' accounts~\citep{10.1108/JMLC-05-2020-0053}.
	
	One subcategory of money mules is smurfs~\citep{starnini2021smurf, zhdanova2014no}. 
	Smurfs are used to break up large transactions and funnel the money through many accounts in smaller instances. 
	This helps the criminals to avoid business rules that are often only monitoring transactions per counterparty (e.g., between person A and B, and A and C, separately, but not aggregated among A,B and C).
	
	A network $G(V,E)$ is defined by a set of nodes, $V = \{v_1, \ldots, v_k\}$, and edges, $E\subset V\times V$. 
	We represent the network using the adjacency matrix $A$. 
	The elements of $A$ indicate whether pairs of nodes are connected via an edge. 
	Mathematically, it is defined as:
	\[ A_{ij} = \begin{cases}
		1 & \text{if  }(v_i, v_j)\in E \\
		0 & \text{elsewhere}
	\end{cases} \]
	
	$\mathcal{N}_k(v)$ represents the set of nodes that are at distance exactly $k$ from node $v$, with $\mathcal{N}_0(v) = \{v\}$.
	GARG-AML relies on the second-order neighbourhood of node $v$. 
	This is the sub-graph consisting of all nodes at a distance of at most two from $v$, capturing both direct counterparties ($\mathcal{N}_1(v)$) and the `counterparties of counterparties' ($\mathcal{N}_2(v)$).
	
	Suppose that $\mid \mathcal{N}_1(v)\mid = n$ and $\mid \mathcal{N}_2(v)\mid = m-1$; then, the adjacency matrix of the second-order neighbourhood of $v$ is of size $(m+n)\times(m+n)$.
	
	The $k$th-order neighbourhood of node $v$ is the sub-network $G(V', E')$ where $V'=\bigcup\limits_{l\leq k} \mathcal{N}_{l}(v)$ and $E' = \{ (u,w)\mid u,w\in V' \wedge (u,w)\in E \}$. 
	Hence, it is the sub-network that represents all nodes at distance at most $k$ and all the interactions among these nodes. 
	
	\subsection{Undirected Network}
	\label{subsec: gargaml undirected}
\begin{figure}
		\centering
		\includegraphics[width=0.36\linewidth]{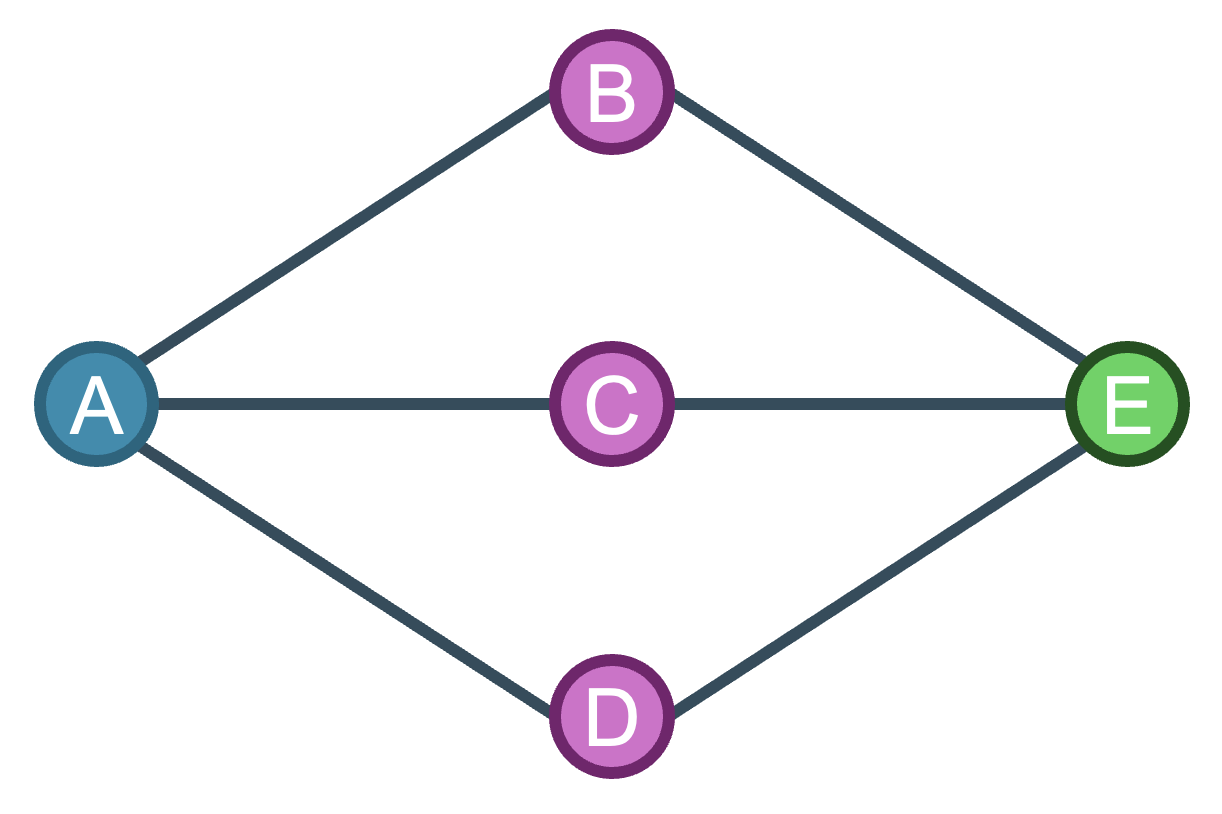}
		\caption{Pure smurfing pattern. $A$ is the node that is scored. The first-order neighbours ($B, C, D$) are shown in purple, the second-order neighbour ($E$) is shown in green.}
		\label{fig:example}
	\end{figure}

	We introduce GARG-AML using an example based on the smurfing pattern illustrated in Figure~\ref{fig:example}. 
	We will consider only the scatter-gather pattern, as this is most widely recognised as smurfing. 
	The methodology is the same for gather-scatter patterns. 
	
	In this example, node $A$ is given a score.
	The adjacency matrix of its second-order neighbourhood, given in Equation~\eqref{eq:mat}, is constructed by arranging the nodes in a specific order: node $A$ is placed first, followed by the second-order neighbour $E$, and then its first-order neighbours $B, C$ and $D$. 
	This ordering results in an adjacency matrix that is divided into four distinct blocks. 
	The diagonal blocks contain only zeros, whereas the off-diagonal blocks contain ones.
	
	\begin{equation}
		\begin{array}{r}
			A \\ E \\ B \\ C \\ D
		\end{array}
		\begin{pmatrix}
			0 & 0 & 1 &1 &1\\
			0 & 0 & 1 &1 &1\\
			1 & 1 & 0 &0 &0 \\
			1 & 1 & 0 &0 &0 \\
			1 & 1 & 0 &0 &0 \\
		\end{pmatrix}
		\label{eq:mat}
	\end{equation}
	
	Given these insights, we define the general GARG-AML method, starting with the adjacency matrix.
	The first row of the adjacency matrix, which corresponds to entries $(A_{1,j})$, represents the connections of node $v$ under consideration. 
	The next $m-1$ rows, i.e., $(A_{i,j})$ for ${2\leq i\leq m}$, represent the connections of nodes at distance two from $v$. 
	The final $n$ rows, i.e., $(A_{i,j})$ for ${m+1\leq i\leq m+n}$, represent the connections of nodes at distance one. 
	
	Enforcing this node ordering---the target node, second-order neighbours, and first-order neighbours---for the adjacency matrix ensures that the diagonal blocks remain sparse (low density), and the off-diagonal blocks have a high density in case of pure smurfing patterns. 
	The exact order of nodes within each group is not important. 
	Any permutation within the first-order neighbours, such as $C,B,D$, is equally valid.
	
	We consider three blocks of the adjacency matrix for score calculation: the upper-left, upper-right, and lower-right blocks. 
	These are defined as follows:
	\begin{eqnarray}
		\text{block}_1 & = & (A_{i,j})_{1\leq i,j \leq m} \label{eq: undirected block 1} \\
		\text{block}_2 & = & (A_{i,j})_{1\leq i\leq m, m+1\leq j \leq m+n} \label{eq: undirected block 2} \\
		\text{block}_3 & = & (A_{i,j})_{m+1\leq i,j \leq m+n} \label{eq: undirected block 3}
	\end{eqnarray}
	
	Our method uses the density, i.e., the relative number of non-zero elements in each block. 
	GARG-AML measures the similarity between an observed pattern and the pure smurfing template by analysing the expected density and actual density (or sparsity) of each block. 
	This results in a single quantitative score expressing similarity to the pure smurfing pattern.
	
	Several structural properties of the adjacency matrix are taken into account for more rigorous density calculations. 
	First, self-loops are excluded; thus, diagonal elements are always zero. Furthermore, in $\text{block}_1$ (upper-left block), the first row and column can never contain ones, as a `1' would indicate a direct connection between a node and its neighbours at distance two, a contradiction. 
	This upper-left block serves as a check that the nodes at distance two do not make transactions to each other. 
	Similarly, for $\text{block}_2$ (upper-right block), all the elements in the first row must be ones since these represent connections between the node under investigation and its first-order neighbours. 
	
	We consider only the \textit{free} entries, i.e., those that are not certain upfront, for density calculations. 
	We define the block size as the number of free elements. 
	This is reflected in the numerator (number of non-zero elements) and denominator (block size).
	\begin{eqnarray}
		\text{score}_1 & = & \frac{\text{sum}(\text{block}_1)}{m^2 - 3m + 2} \label{eq:score1} \\
		\text{score}_2 & = & \frac{\text{sum}(\text{block}_2)-n}{mn-n} \label{eq:score2} \\
		\text{score}_3 & = & \frac{\text{sum}(\text{block}_3)}{n^2-n} \label{eq:score3} 
	\end{eqnarray}
	Smurfing is likely to be present when the density of $\text{block}_2$ (upper-right block) is high, while $\text{block}_1$ (upper-left block) and $\text{block}_3$ (lower-right) exhibit low densities. 
	Therefore, the final GARG-AML score is calculated as:
	\begin{equation}
		\text{score}_\text{total} = \text{score}_2 - \frac{l_1\cdot\text{score}_1 + l_3\cdot\text{score}_3}{l_1+l_3} \label{eq: score undirected}, 
	\end{equation}
	where $l_i$ is a weight given to each term in the score calculation. 
	If $l_i=1$, the score is the average of $\text{score}_1$ and $\text{score}_3$. 
	We take a weighted average using the block sizes:
	\begin{eqnarray}
		l_1 & = & m^2 - 3m + 2 \label{eq:l1} \\
		l_3 & = & n^2-n. \label{eq:l3}
	\end{eqnarray}
	When there is only one destination account, there is only one node at distance two (as in Figure~\ref{fig:example}). 
	This results in $l_1$ being 0. 
	Then, the calculations simplify to $\text{score}_\text{total} = \text{score}_2-\text{score}_3$, and we only analyse whether all money ends up at the destination node (high $\text{score}_2$) and none of the intermediary accounts interact with each other (low $\text{score}_3$).
	
	The final score in Equation~\eqref{eq: score undirected} ranges from $-1$ to $1$. 
	Second-order neighbourhoods resembling a smurfing pattern result in a score close to 1. 
	Additional examples are provided in \ref{app:example undir} as illustration.  
	
	\subsection{Directed Network}
	\label{subsec: gargaml directed}
\begin{figure}
		\centering
		\includegraphics[width=0.6\linewidth]{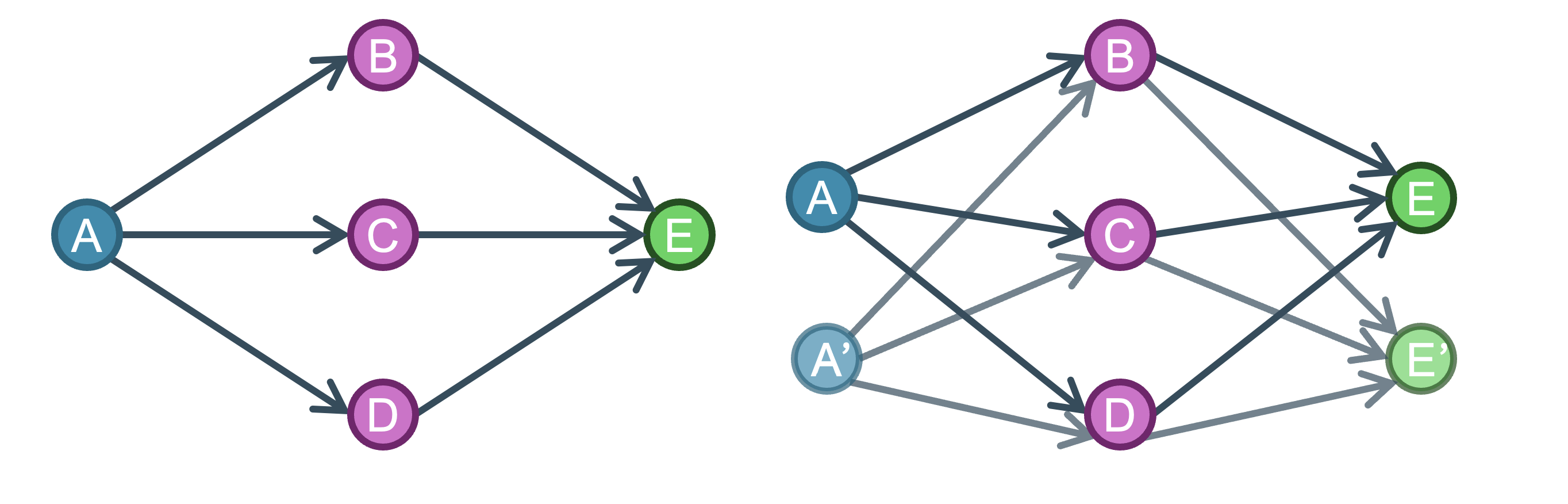}
		\caption{The directed version of the smurfing pattern example, extended with an additional sending node (A') and receiving node (E'). Nodes at level 0 are put in blue, at level 1 in purple and at level 2 in green.}
		\label{fig:exampleDirected}
	\end{figure}

	A defining trait of smurfing is the uni-directional flow of funds. 
	Money moves from one account to others with minimal/no return flow~\citep{starnini2021smurf,lee2020autoaudit,li2020flowscope,shadrooh2024smotef}, forming \textit{pure smurfing} patterns (see Figure~\ref{fig:exampleDirected}). 
	To capture this behaviour, we extend GARG-AML to directed networks.
	
	We start with an example. 
	In the directed network of Figure~\ref{fig:exampleDirected} node $A$ only sends funds to the smurfs, while node $E$ only receives from them. 
	This leads to a non-symmetric adjacency matrix, requiring the analysis of nine matrix blocks instead of three (see Equation~\eqref{eq:mat_dir}). 
	To ensure consistency in block-wise density calculations, especially with multiple senders or receivers, nodes at distance two are classified as senders (level 0) and receivers (level 2).
	
	We define two types of second-order neighbourhoods in network $G(V,E)$. 
	The \textit{strong} second-order neighbourhood, denoted \(\mathcal{N}^s_2(v)\), considers nodes reachable from $v$ via a directed path of length two. 
	The \textit{weak} second-order neighbourhood, \(\mathcal{N}^w_2(v)\), is based on the undirected network. 
	In both, only nodes at exact distance two are included.
	
	A node $w$ is assigned to level 0 if
	\begin{equation}
		w \in \mathcal{N}^w_2(v)\setminus \mathcal{N}^s_2(v).
		\label{eq:layers}
	\end{equation}
	To account for reverse flow patterns, we repeat this selection on the reversed network. 
	This means that a node $w$ at distance two of node $v$ is put at level 0 if there is no directed path of length two from $v$ to $w$ nor from $w$ to $v$. 
	The other second-order neighbours are put at level 2.
	
	For the extended network in Figure~\ref{fig:exampleDirected}, node $A'$ is put at the same level as node $A$ and node $E'$ at the same level as node $E$.
	The resulting adjacency matrix is then
	\begin{equation}
		\begin{array}{l}
			A \\ A' \\ \midrule B \\ C \\ D \\ \midrule E \\ E'
		\end{array}
		\left( \begin{array}{cc|ccc|cc}
			0 & 0 & 1 &1 &1 &0 &0 \\
			0 & 0 & 1 &1 &1 &0 &0 \\
			\midrule
			0 & 0 & 0 &0 &0 &1 &1 \\
			0 & 0 & 0 &0 &0 &1 &1 \\
			0 & 0 & 0 &0 &0 &1 &1 \\
			\midrule
			0 & 0 & 0 &0 &0 &0 &0 \\
			0 & 0 & 0 &0 &0 &0 &0
		\end{array}\right).
		\label{eq:mat_dir}
	\end{equation}
	
	Structural constraints of the adjacency matrix are considered when calculating block-wise densities. 
	For the upper-left block, $\text{score}_{00}$, the first row and column are excluded since all other level-0 nodes are nodes at distance two. 
	Additionally, diagonal elements are excluded for $\text{score}_{00}$, $\text{score}_{11}$ and $\text{score}_{22}$, as self-loops are not present.
	
	Adjustments are also made to $\text{score}_{20}$ and $\text{score}_{02}$, which represent connections between the node and its second-order neighbours. 
	The first row of $\text{block}_{02}$ (from the node to its second-order neighbours) and first column of $\text{block}_{20}$ (from the second-order neighbours to the node) are always zero.
	
	No adjustments are applied to the remaining blocks, as the directionality among level-1 and level-2 nodes does not impose clear structural constraints.
	
	We take the sizes to be $\mid \mathcal{N}_1(v)\mid = n$,  $\mid \mathcal{N}^s_2(v)\mid = m$, and $\mid \mathcal{N}^w_2(v)\setminus \mathcal{N}^s_2(v)\mid = l-1$, resulting in $l$ nodes at level 0, $n$ nodes at level 1, and $m$ nodes at level 2. 
	The adjacency matrix is partitioned into nine blocks, as shown in Equation~\eqref{eq:blocks dir}, with block labels using zero-based indexing to reflect node levels 0, 1, and 2.
	
	\begin{equation}{
			\left( \begin{array}{ccc}
				\text{b}_{00} & \text{b}_{01} & \text{b}_{02} \\
				\text{b}_{10} & \text{b}_{11} & \text{b}_{12} \\
				\text{b}_{20} & \text{b}_{21} & \text{b}_{22} 
			\end{array}\right),}
		\label{eq:blocks dir}
	\end{equation}
	where
	\begin{eqnarray}
		\text{b}_{00} &=& (A_{i,j})_{1\leq i,j \leq l} \nonumber\\ 
		\text{b}_{01} &=& (A_{i,j})_{1\leq i \leq l, l+1 \leq j \leq l+n} \nonumber\\ 
		\text{b}_{02} &=&(A_{i,j})_{1\leq i \leq l, l+n+1 \leq j \leq l+n+m} \nonumber\\
		\text{b}_{10} &=& (A_{i,j})_{l+1\leq i \leq l+n, 1 \leq j \leq l} \nonumber\\
		\text{b}_{11} &=& (A_{i,j})_{l+1\leq i,j\leq l+n} \nonumber\\
		\text{b}_{12} &=& (A_{i,j})_{l+1\leq i \leq l+n, l+n+1 \leq j \leq l+n+m} \nonumber\\
		\text{b}_{20} &=& (A_{i,j})_{l+n+1\leq i \leq l+n+m, 1 \leq j \leq l} \nonumber\\ \text{b}_{21} &=& (A_{i,j})_{l+n+1\leq i \leq l+n+m, l+1 \leq j\leq l+n} \nonumber\\ 
		\text{b}_{22} &=& (A_{i,j})_{l+n+1\leq i, j \leq l+n+m}\nonumber
	\end{eqnarray}
	
	The final score takes the weighted average of the two blocks expected to have high density, according to the smurfing pattern in Equation~\eqref{eq:mat_dir}, and subtracts the weighted average of the remaining blocks, which are expected to be sparse. 
	The final GARG-AML score is between $-1$ and $1$, to have consistent comparison and interpretation across datasets. 
	\begin{eqnarray}
		\text{score} & = & \text{mean}(\text{score}_{01}, \text{score}_{12}) \label{eq: score sender directed}  \\
		& & - \text{mean}(\text{score}_{00}, \text{score}_{02}, \text{score}_{10}, \nonumber\\ && \text{score}_{11}, \text{score}_{20}, \text{score}_{21}, \text{score}_{22}) \nonumber
	\end{eqnarray}
	Additional examples are provided in \ref{app:example dir} as illustration.  
	
	The GARG-AML scores offer three key advantages. 
	First, scores near 1 offer a transparent signal for automated detection and human investigation by reflecting structural smurfing resemblances.
	Second, the resulting scores and densities can be easily incorporated into existing business rules or machine learning workflows. 
	Third, each score is computed independently using a node’s second-order neighbourhood, allowing for efficient parallel computation and low memory usage, even on large transaction networks.
	
	\subsection{Integration in Existing Detection Systems}
	This section illustrates how GARG-AML can easily be integrated in existing detection methods.
	
	Rule-based systems can integrate new scenarios based on the GARG-AML score and its interaction with other features. 
	The rules remain highly interpretable, given that the GARG-AML scores are within a fixed range, with higher scores indicating that the network resembles smurfing. 
	More complex scenarios can be constructed using the individual densities and sizes of the blocks. 
	These are obtained during the intermediary calculations, so these do not increase calculation costs.
	
	Detection models based on machine learning can also be extended. 
	The scores or densities can be interpreted as additional client features. 
	The existing pipelines can be kept to retrain the models.
	
	\section{Experimental Setup}
	\label{sec: experimental setup}
	\subsection{Data}
	\label{subsec:metho-data}
	Due to the high sensitivity of financial transaction and fraud data, which precludes the use of proprietary banking datasets, we evaluate the detection power of GARG-AML using a two-tiered approach. 
	First, we utilise self-generated networks created through various simulation methods to maintain full control over network size, topology, and the specific smurfing patterns present, allowing us to illustrate the model's scalability and its performance against diverse camouflaging tactics. 
	Second, we test the framework on realistic open-source datasets that mimic real-life transaction flows, including millions of transactions with a realistically low ratio of confirmed money laundering cases.
	This establishes a robust evaluation procedure while ensuring all experiments remain fully reproducible.
	
	\subsubsection{Synthetic Data}
	\label{subsec:data synthetic}
	We generate synthetic networks of varying sizes using three generation methods, i.e., Barab\'{a}si-Albert~\citep{barabasi1999emergence}, Erd\H{o}s-R\'{e}nyi~\citep{erdds1959random} and Watts-Strogatz~\citep{wattsStrogatz}. By generating networks with varying properties, we evaluated GARG-AML across a spectrum of topologies.
	
	Each method has specific parameters that we vary to obtain a range of networks to test our method. 
	We summarise these in Table~\ref{tab:params synthetic}. 
	The first parameter is the size of the network, which is subsequently set to 100, 10,000 and 100,000. 
	
	\begin{table*}
		\centering
		\caption{Overview of the parameters used to construct the synthetic data.}
		\label{tab:params synthetic}
		\small
		\setlength{\tabcolsep}{3pt}
		\begin{tabular}{@{}llccc@{}}
			\toprule
			Parameter & Values & \makecell[cb]{Barab\'{a}si-\\Albert} & \makecell[cb]{Erd\H{o}s-\\R\'{e}nyi} & \makecell[cb]{Watts-\\Strogatz} \\ \midrule
			Number of initial nodes & \{100; 10,000; 100,000\} & \ding{51} &  \ding{51} & \ding{51} \\
			Number of edges & \{ 1; 2; 5 \} & \ding{51} & \textcolor{gray!75}{\ding{55}} & \ding{51} \\
			Edge (rewiring) probability & \{ 0.001; 0.01 \} & \textcolor{gray!75}{\ding{55}} & \ding{51} & \ding{51} \\
			Number of smurfing patterns & \{ 3; 5 \} & \ding{51} & \ding{51} & \ding{51} \\
			\midrule
			Total number of combinations & & 18 & 12 & 36 \\ \bottomrule
		\end{tabular}
	\end{table*}
	
	The definitions of the other parameters differ depending on the method. 
	For Barab\'{a}si-Albert, \textit{the number of edges} represents the number of nodes to which each new node connects, resulting in a large difference in degree across the nodes. For Watts-Strogatz, on the other hand, this is half the initial degree that all the nodes have before some of the edges are rewired. 
	
	For the Erd\H{o}s-R\'{e}nyi method, \textit{the edge (rewiring) probability} is the probability of having an edge between any two nodes. 
	The Watts-Strogatz method starts with a fixed number of nodes, and uses this parameter to express the probability that an edge changes its end node (rewiring). 
	
\begin{figure}
		\centering
		\includegraphics[width=0.7\linewidth]{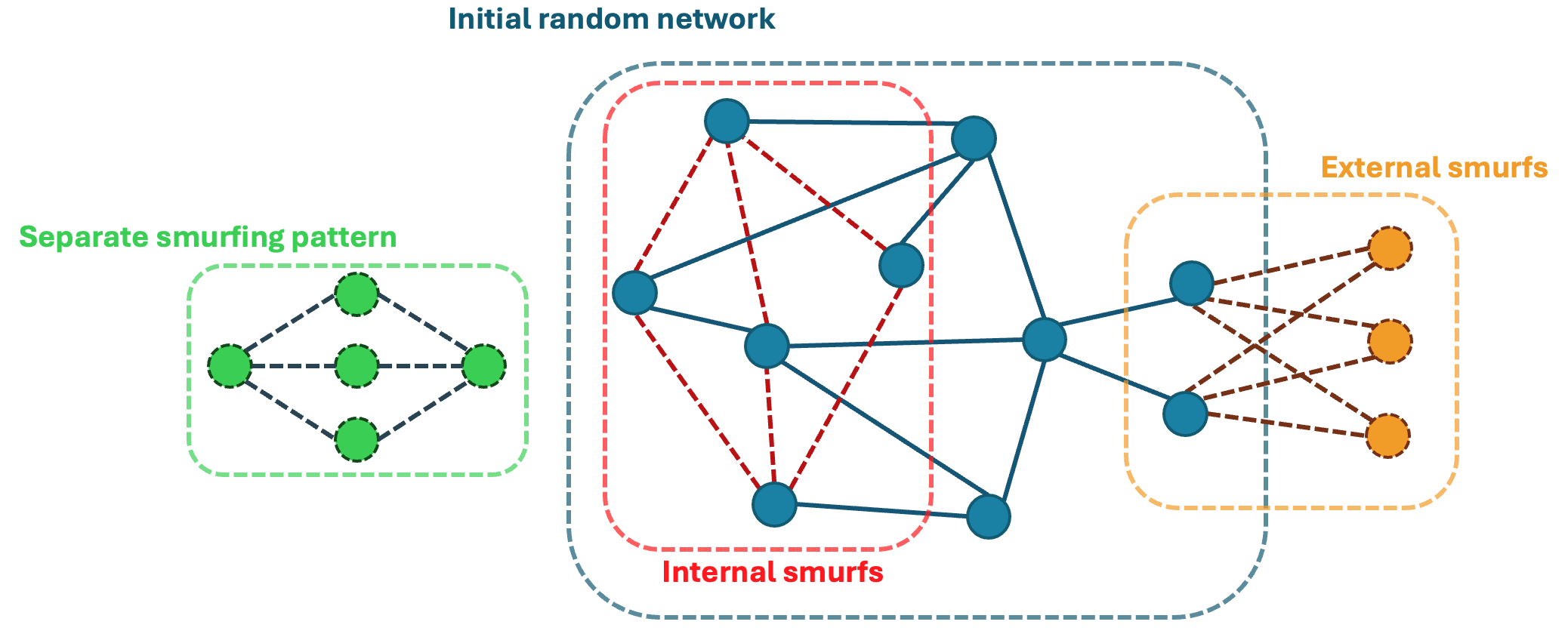}
		\caption{A random network with the three methods of smurfing pattern injection. Solid lines indicate nodes and edges of the original network. Dashed lines indicate injected nodes and edges.}
		\label{fig:injection}
	\end{figure}

	Smurfing patterns are injected after network generation. 
	The three types of smurfing patterns (visualised in Figure~\ref{fig:injection}) considered in the experiment are:
	\begin{enumerate}
		\item \textbf{Separate:} The smurfing patterns are not connected to the rest of the network; 
		\item \textbf{External smurfs:} Mules are added between two existing entities in the network; 
		\item \textbf{Internal smurfs:} Existing nodes in the network are assigned the roles of sender, receiver and mule, and corresponding transactions are added.  
	\end{enumerate}
	
	In our experiments, each of the smurfing patterns is included either three or five times; see Table~\ref{tab:params synthetic}. 
	The number of smurfs in each pattern is randomly chosen between two and ten. 
	This allows for both smaller and larger smurfing patterns. 
	A summary of all synthetic datasets is presented in Table~\ref{tab:summary data synthetic}.
	
	\begin{table*}[p]
		\centering
		\caption{Summary of the numbers of nodes and edges and money laundering percentages for the synthetic datasets according to the network generation parameters. $m$ is the number of edges and $p$ the edge (rewiring) probability (see Table~\ref{tab:params synthetic}).}
		\label{tab:summary data synthetic}
		\footnotesize
		\setlength{\tabcolsep}{4pt}
		\begin{tabular}{@{}lrrrrrrrr@{}}
			\toprule
			& & & \multicolumn{3}{c}{3 smurfing patterns} & \multicolumn{3}{c}{5 smurfing patterns} \\
			\cmidrule(lr){4-6} \cmidrule(l){7-9}
			Generator & $m$ & $p$ & Nodes & Edges & \makecell[cb]{Labels\\(\%)} & Nodes & Edges & \makecell[cb]{Labels\\(\%)} \\
			\midrule
			\multicolumn{9}{@{}l}{\textit{100 initial nodes}} \\
			Barab\'{a}si-Albert & 1 & -- & 145 & 199 & 46.90 & 180 & 291 & 70.00 \\
			Barab\'{a}si-Albert & 2 & -- & 155 & 345 & 60.00 & 179 & 385 & 69.83 \\
			Barab\'{a}si-Albert & 5 & -- & 141 & 582 & 50.35 & 175 & 678 & 74.29 \\
			Erd\H{o}s-R\'{e}nyi & -- & 0.001 & 151 & 122 & 52.32 & 182 & 196 & 70.33 \\
			Erd\H{o}s-R\'{e}nyi & -- & 0.01 & 153 & 155 & 60.78 & 163 & 161 & 65.03 \\
			Watts-Strogatz & 1 & 0.001 & 144 & 204 & 48.61 & 176 & 271 & 65.91 \\
			Watts-Strogatz & 1 & 0.01 & 146 & 216 & 52.05 & 152 & 248 & 69.74 \\
			Watts-Strogatz & 2 & 0.001 & 131 & 268 & 40.46 & 171 & 376 & 70.18 \\
			Watts-Strogatz & 2 & 0.01 & 146 & 316 & 52.74 & 180 & 380 & 67.22 \\
			Watts-Strogatz & 5 & 0.001 & 151 & 618 & 52.32 & 171 & 669 & 69.01 \\
			Watts-Strogatz & 5 & 0.01 & 145 & 616 & 53.10 & 165 & 655 & 66.67 \\
			\midrule
			\multicolumn{9}{@{}l}{\textit{10,000 initial nodes}} \\
			Barab\'{a}si-Albert & 1 & -- & 10,037 & 10,101 & 0.69 & 10,059 & 10,171 & 1.15 \\
			Barab\'{a}si-Albert & 2 & -- & 10,042 & 20,105 & 0.72 & 10,081 & 20,215 & 1.38 \\
			Barab\'{a}si-Albert & 5 & -- & 10,043 & 50,103 & 0.77 & 10,072 & 50,167 & 1.20 \\
			Erd\H{o}s-R\'{e}nyi & -- & 0.001 & 10,044 & 5,101 & 0.71 & 10,079 & 5,263 & 1.40 \\
			Erd\H{o}s-R\'{e}nyi & -- & 0.01 & 10,050 & 50,388 & 0.79 & 10,067 & 49,998 & 1.08 \\
			Watts-Strogatz & 1 & 0.001 & 10,054 & 10,132 & 0.84 & 10,081 & 10,168 & 1.13 \\
			Watts-Strogatz & 1 & 0.01 & 10,052 & 10,134 & 0.85 & 10,070 & 10,182 & 1.20 \\
			Watts-Strogatz & 2 & 0.001 & 10,038 & 20,104 & 0.70 & 10,075 & 20,180 & 1.19 \\
			Watts-Strogatz & 2 & 0.01 & 10,046 & 20,134 & 0.85 & 10,071 & 20,190 & 1.24 \\
			Watts-Strogatz & 5 & 0.001 & 10,039 & 50,090 & 0.63 & 10,081 & 50,196 & 1.27 \\
			Watts-Strogatz & 5 & 0.01 & 10,040 & 50,108 & 0.72 & 10,060 & 50,148 & 1.03 \\
			\midrule
			\multicolumn{9}{@{}l}{\textit{100,000 initial nodes}} \\
			Barab\'{a}si-Albert & 1 & -- & 100,044 & 100,099 & 0.07 & 100,063 & 100,167 & 0.11 \\
			Barab\'{a}si-Albert & 2 & -- & 100,038 & 200,095 & 0.07 & 100,069 & 200,175 & 0.12 \\
			Barab\'{a}si-Albert & 5 & -- & 100,043 & 500,093 & 0.07 & 100,065 & 500,159 & 0.12 \\
			Erd\H{o}s-R\'{e}nyi & -- & 0.001 & 100,033 & 498,969 & 0.06 & 100,087 & 501,039 & 0.14 \\
			Erd\H{o}s-R\'{e}nyi & -- & 0.01 & 100,044 & 4,998,019 & 0.07 & 100,076 & 4,997,365 & 0.13 \\
			Watts-Strogatz & 1 & 0.001 & 100,050 & 100,124 & 0.08 & 100,061 & 100,166 & 0.11 \\
			Watts-Strogatz & 1 & 0.01 & 100,045 & 100,096 & 0.07 & 100,058 & 100,134 & 0.10 \\
			Watts-Strogatz & 2 & 0.001 & 100,044 & 200,114 & 0.07 & 100,065 & 200,154 & 0.11 \\
			Watts-Strogatz & 2 & 0.01 & 100,041 & 200,116 & 0.08 & 100,059 & 200,154 & 0.11 \\
			Watts-Strogatz & 5 & 0.001 & 100,042 & 500,114 & 0.07 & 100,074 & 500,176 & 0.12 \\
			Watts-Strogatz & 5 & 0.01 & 100,045 & 500,116 & 0.08 & 100,068 & 500,166 & 0.11 \\
			\bottomrule
		\end{tabular}
	\end{table*}
	
	Since we have information on which nodes are involved in which pattern, we not only test the methods for money laundering detection in general; we analyse the performance on the different patterns separately, representing increasing levels of obfuscation.
	
	\subsubsection{Open-Source Data}
	As post-hoc injection of specific patterns may lead to detection bias~\citep{deprez2024networkevaluation} and altered topology~\citep{egressy2024provably}, we also include data without this injection. 
	We use the HI-Small and LI-Large datasets, made publicly available by~\citet{altman2024realistic} and \citet{egressy2024provably}, representing a relatively small dataset with high money laundering activity and a large dataset with low money laundering activity, respectively. 
	Table~\ref{tab:summary data} summarises the number of nodes, edges and the percentage of cases labelled as money laundering. Especially LI-Large closely resembles real-world transaction networks. 
	
	\begin{table}
		\centering
		\caption{The summary of the two IBM datasets, representing the number of nodes, edges and the percentage of instances labelled as money laundering (ML).}
		\label{tab:summary data}
		\begin{tabular}{l|rrr}
			\toprule
			Dataset & \# nodes & \# edges & percentage ML  \\
			\midrule
			HI-Small & 515,080 & 5,078,345 & $0.1\%$ \\
			LI-Large & 2,054,390 & 176,066,557 & $0.06\%$ \\
			\bottomrule
		\end{tabular}
	\end{table}
	
	These two datasets are generated using the same simulator~\citep{altman2024realistic, egressy2024provably}. 
	It models financial transactions in a virtual world involving individuals, companies, and banks. 
	A subset of these transactions is labelled as money laundering. 
	The data contains detailed labels, categorising transaction streams into eight distinct types: fan-out, fan-in, gather-scatter, scatter-gather, simple cycle, random, bipartite, and stack.
	The distribution of labelled patterns is shown in Table~\ref{tab:IBM distribution}.
	
	It is worth noting that some transactions are labelled as money laundering but do not match any of the eight defined structural patterns. 
	These are grouped under the category ``not classified'', indicating that they are suspicious but not attributable to a specific laundering archetype. 
	
	\begin{table} 
		\centering
		\caption{The distribution of the money laundering transaction patterns for the two IBM datasets.}
		\label{tab:IBM distribution}
		\begin{tabular}{l|rr}
			\toprule
			Type of pattern & HI-Small  & LI-Large  \\
			\midrule
			Fan-out         & 342       & 1,985       \\
			Fan-in          & 318       & 1,935       \\
			Gather-scatter  & 716       & 3,857     \\
			Scatter-gather  & 626       & 3,694     \\
			Cycle           & 287       & 1,871       \\
			Random          & 191       & 1,458       \\
			Bipartite       & 263       & 1,807       \\
			Stack           & 466       & 2,854     \\
			Not classified  & 1,968     & 81,143    \\
			\midrule 
			Total number of transactions & 5,078,345 & 176,066,557 \\
			\bottomrule
		\end{tabular}
	\end{table}
	
	The original labels in these datasets are provided at the edge level, while GARG-AML produces features at the node level. 
	To bridge this gap, we aggregate labels by computing the ratio of labelled laundering transactions to the total number of transactions per client. 
	This ratio is interpreted as a propensity score for money laundering.
	
	A node is labelled as a money launderer if the proportion of its laundering-labelled transactions exceeds a certain cut-off. 
	Depending on the cut-off ratio we set, the number of labels and with that the available information will change drastically.
	This ultimately has an effect on the learning capabilities of the models. 
	We vary these across the full range, to see the effect on the models and to analyse which models perform best under what circumstances.
	We compare each node’s propensity score against a set of predefined cut-offs: $0\%$, $10\%, 20\%, 30\%, 50\%$ and $90\%$. 
	The resulting label distributions are presented in Table~\ref{tab:labels data HI-Small} and Table~\ref{tab:labels data LI-Large}, for the HI-Small and LI-Large datasets, respectively. 
	
	A smaller cut-off means that only a small portion of a node's transactions should be illicit to label the node as laundering money, with $0\%$ implying that already one transaction is enough.
	A large cut-off means that most of the transactions should be illicit before the node is labelled as laundering money. 
	The lower cut-offs represent ideal scenarios where almost every illicit transaction is caught.
	Higher cut-offs mimic reality, where it is often the case that laundering only becomes apparent when many transactions have been made, and where it is easier to detect if limited obfuscation via other transactions is present.
	
	In many cases, fewer than $0.1\%$ of accounts are assigned a positive label, highlighting the extreme class imbalance and difficulty of the classification task.
	Additionally, accounts with more than $90\%$ of laundering-labelled transactions are extremely rare, which reflects the realistic behaviour of laundering entities mixing illicit flows with legitimate transactions to obfuscate detection.
	
	\begin{table}
		\centering
		\caption{Percentage of instances labelled as 1 in the HI-Small dataset per cut-off. }
		\label{tab:labels data HI-Small}
		\begin{tabular}{lrrrrrr}
			\toprule
			In percentage $(\%)$ &    0 &    10 &    20 &    30 &    50 &    90 \\
			\midrule
			Is Laundering  &  1.234 &  0.445 &  0.175 &  0.103 &  0.020 &  0.013 \\
			Fan-out        &  0.070 &  0.029 &  0.013 &  0.009 &  0.002 &  0.001 \\
			Fan-in         &  0.066 &  0.022 &  0.006 &  0.005 &  0.002 &  0.001 \\
			Gather-scatter &  0.133 &  0.056 &  0.018 &  0.014 &  0.003 &  0.001 \\
			Scatter-gather &  0.072 &  0.036 &  0.023 &  0.010 &  0.003 &  0.000 \\
			Cycle          &  0.053 &  0.019 &  0.013 &  0.005 &  0.000 &  0.000 \\
			Random         &  0.041 &  0.015 &  0.007 &  0.002 &  0.000 &  0.000 \\
			Bipartite      &  0.095 &  0.030 &  0.007 &  0.004 &  0.001 &  0.001 \\
			Stack          &  0.129 &  0.044 &  0.021 &  0.011 &  0.001 &  0.001 \\
			\bottomrule
		\end{tabular}
	\end{table}
	
	\begin{table}
		\centering
		\caption{Percentage of instances labelled as 1 in the LI-Large dataset per cut-off. }
		\label{tab:labels data LI-Large}
		\begin{tabular}{lrrrrrr}
			\toprule
			In percentage $(\%)$ &     0 &    10 &    20 &    30 &    50 &    90 \\
			\midrule
			Is Laundering  & 6.292 & 0.216 & 0.041 & 0.019 & 0.002 & 0.000 \\
			Fan-out        & 0.105 & 0.005 & 0.002 & 0.001 & 0.000 & 0.000 \\
			Fan-in         & 0.101 & 0.005 & 0.002 & 0.002 & 0.000 & 0.000 \\
			Gather-scatter & 0.184 & 0.008 & 0.004 & 0.003 & 0.000 & 0.000 \\
			Scatter-gather & 0.106 & 0.015 & 0.005 & 0.002 & 0.000 & 0.000 \\
			Cycle          & 0.085 & 0.009 & 0.002 & 0.001 & 0.000 & 0.000 \\
			Random         & 0.078 & 0.006 & 0.001 & 0.000 & 0.000 & 0.000 \\
			Bipartite      & 0.172 & 0.004 & 0.001 & 0.000 & 0.000 & 0.000 \\
			Stack          & 0.197 & 0.010 & 0.001 & 0.000 & 0.000 & 0.000 \\
			\bottomrule
		\end{tabular}
	\end{table}
	
	\subsection{Model Specifications}
	\label{subsec:model specification}
\begin{figure*}
		\centering
		\includegraphics[width=0.85\linewidth]{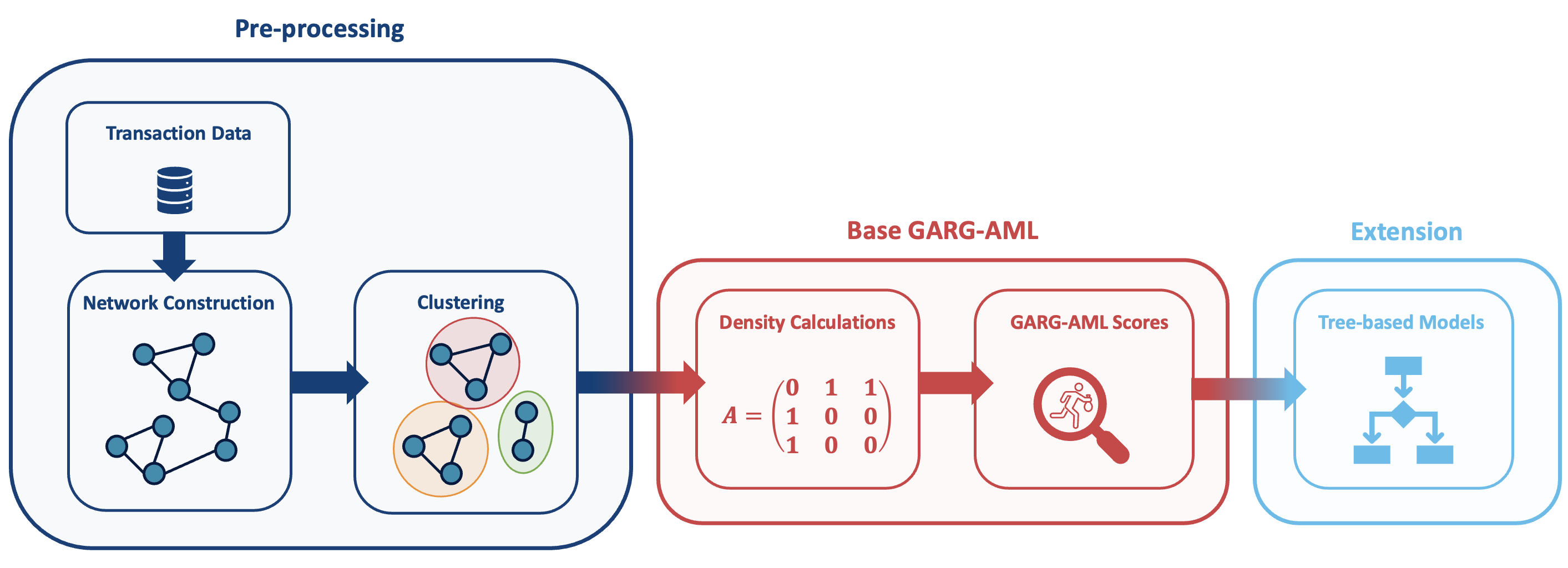}
		\caption{The pipeline of the GARG-AML method.}
		\label{fig:model pipeline}
	\end{figure*}

	The full GARG-AML calculation pipeline consists of multiple stages, beginning with pre-processing steps that prepare the network data for analysis, followed by the computation of the GARG-AML scores. 
	We introduce a possible extension that applies tree-based methods, since these have been shown to have strong performance for AML~\citep{10.1108/JMLC-07-2019-0055,10.1108/JMLC-04-2024-0083,10.1108/JMLC-09-2024-0152,10.1108/JMLC-07-2024-0114}. 
	This extension illustrates the versatility of GARG-AML. 
	An overview of the complete pipeline is shown in Figure~\ref{fig:model pipeline} and summarised in Algorithm~\ref{alg:gargaml}.
	
	\begin{algorithm}
		\caption{ GARG-AML Score Calculation }\label{alg:gargaml}
		\textbf{Input}: Network $G(V,E)$\\
		\textbf{Output}: GARG-AML propensity score(s)
		\begin{algorithmic}[1]
			\State Calculate communities using Louvain method~\citep{blondel2008louvain}
			\State Remove inter-community edges
			\If {Directed score}{
				
				\For {node $v\in V$}
				\State Extract second-order neighbourhood of $v$
				\State Distribute neighbours over level 0, 1 and 2
				\State Define the adjacency matrix with levels 0, 1 and 2. 
				\State Define the nine blocks using Equation~\eqref{eq:blocks dir}
				\State Calculate score via Equation~\eqref{eq: score sender directed}
				\EndFor 
				
			}\Else{
				
				\For {node $v\in V$}
				\State Extract second-order neighbourhood of $v$
				\State Define the adjacency matrix using $v$, $\mathcal{N}_2(v)$ and $\mathcal{N}_1(v)$
				\State Define the three blocks using Equations~\eqref{eq: undirected block 1}--\eqref{eq: undirected block 3}
				\State Calculate score via Equation~\eqref{eq: score undirected}
				\EndFor 
				
			}
			\EndIf
			\State \textbf{return}: scores
		\end{algorithmic}
	\end{algorithm}
	
	Transaction networks often exhibit a scale-free structure~\citep{starnini2021smurf}, which is characterised by a majority of nodes having only a few connections, while a few nodes, called hubs, have a disproportionately high number of connections. 
	As a result, second-order neighbourhoods that contain such hubs encompass an unmanageable number of nodes for score computation.
	
	Therefore, we first partition the network into communities using the Louvain method~\citep{blondel2008louvain}. 
	This method consistently shows good performance for community detection, and scales well to large graphs~\citep{lancichinetti2009community, yang2016comparative}. 
	Communities are identified on the basis of the undirected version of the transaction network.
	Only intra-community edges are retained and edges between communities are discarded. 
	
	The Louvain method can produce overly large communities~\citep{fortunato2007resolution, newman2004finding}. 
	To mitigate this, we use the implementation from NetworkX~\citep{networkx}, which includes a resolution parameter inspired by the improvements proposed by~\citet{traag2019leiden}. 
	In our experiments, we set this hyperparameter to 10 to obtain smaller, more manageable communities.
	The choice of applying the Louvain method and the selection of the resolution parameter can break up the money laundering patterns if they span multiple communities. 
	The sensitivity of GARG-AML to these choices is analysed in Section~\ref{subsec:sensitivity}. 
	
	After pre-processing, we calculate the block densities as described above. 
	We come to the final undirected and directed score via Equation~\eqref{eq: score undirected} and Equation~\eqref{eq: score sender directed}, respectively. 
	
	As a possible extension, we apply a decision tree and gradient boosting classifier that use features based on the GARG-AML scores. 
	A decision tree is a non-parametric model that performs greedy binary partitioning of the features on an impurity criterion. 
	The resulting model can be seen as a collection of IF-ELSE rules. 
	The gradient boosting classifier is an ensemble learning method that constructs an additive model in a forward stage-wise manner. 
	It optimises a differentiable loss function by sequentially training `weak learners' (shallow decision trees), where each subsequent tree is fit to the negative gradient of the loss function (the residuals) of the previous ensemble.
	
	To capture the coordinated nature of smurfing, the base GARG-AML scores are supplemented with neighbourhood summary statistics (min, mean, max, and standard deviation of neighbour degrees and scores). 
	This structural context helps the model differentiate between suspicious multi-mule topologies and benign linear chains by automatically filtering out these specific false positives.
	Applying this on the undirected and directed network results in four additional models. 
	
	The validation is set up in a transductive way. 
	This is done since there is a large distributional shift over time for the IBM datasets. 
	We use a 70-30 stratified train-test split for evaluation. 
	For the GARG-AML scores, the split is performed on the resulting feature table, not the network itself.
	The split for GraphSAGE is done on the network itself. 
	
	For the decision tree, we use the Gini criterion and each leaf should contain at least ten samples. 
	The gradient boosting classifier consists of 100 trees with a maximal depth of three and a learning rate of 0.1. 
	Each leaf node should contain at least 10 observations. 
	
	\subsection{Baselines}
	We benchmark GARG-AML and its extension against established methods in smurfing detection, AutoAudit~\citep{lee2020autoaudit} and FlowScope~\citep{li2020flowscope}, with their default hyperparameters, except that FlowScope's block size is capped at 10 accounts per mode, as in the reference tutorial (Table~\ref{tab:hyperparameters baselines}).
	Neither method outputs a score per account, so both need a small adaptation.
	
	FlowScope returns $k=3$ dense flow blocks in the order in which they are found.
	To reflect the decreasing level of certainty, the accounts in the first block get score $1$, those in the second $\frac{2}{3}$ and those in the third $\frac{1}{3}$; all other accounts get score $0$.
	
	AutoAudit returns a reordering of the indices of the adjacency matrix: first the accounts of each detected smurfing pattern, in the order in which the patterns were found, then all remaining accounts.
	We use this reordering as a ranking: the account at position $j$ ($j = 0, \dots, N-1$) among $N$ accounts gets score $\frac{N-j}{N}$, so every account receives a distinct score.
	Only the head of the ordering comes from the detected patterns; the remaining accounts keep their index order, which depends on the arbitrary numbering of the nodes.
	To ensure robustness, we repeat the experiments three times with a random permutation of the node indices and report the average results.
	
	In addition to AutoAudit and FlowScope, we test GARG-AML against GraphSAGE~\citep{hamilton2017inductive}. 
	GraphSAGE is a graph neural network (GNN) method that we include since GNNs have been shown to be capable deep learning methods for anti-money laundering~\citep{altman2024realistic,egressy2024provably}. 
	A major drawback is that GNNs are black-box methods that do not allow for interpretable results. 
	Compared to the other methods, GraphSAGE is also not specifically designed for smurfing detection. 
	We include GraphSAGE in our comparison to have a better view of the interpretability-vs-performance trade-off. 
	We use GraphSAGE as a supervised classifier, so enough labels and data are needed for training. 
	Therefore, it is only included in the experiments on the IBM data. 
	
	We evaluate GraphSAGE under two feature configurations, both run on the full undirected transaction graph without Louvain pre-processing. 
	In the topology configuration, each account is described only by its degree and log-degree. 
	Since a two-layer GraphSAGE has the same two-hop receptive field as GARG-AML, this gives both methods the same structural information. 
	The attributes configuration adds twelve per-account aggregates from the raw transactions: counts of sent and received transactions; the sum, mean, standard deviation and maximum of outgoing and incoming amounts; and the number of distinct currencies and counterparty banks.
	
	\begingroup
	\small
	\setlength{\LTcapwidth}{\textwidth}
	\begin{longtable}{@{}>{\raggedright\arraybackslash}p{0.27\textwidth}>{\raggedright\arraybackslash}p{0.58\textwidth}l@{}}
		\caption{Hyperparameters of the baseline models. \emph{Default}: the value of the reference implementation (spartan2 for FlowScope; the authors' AA-Smurf code for AutoAudit, where these constants are hard-coded and the method is presented as parameter-free; PyTorch Geometric and PyTorch for GraphSAGE). \emph{Set}: fixed by us before any experiment. No hyperparameter search was run for any model, so no value was selected on, or tuned against, a test fold.}
		\label{tab:hyperparameters baselines}\\
		\toprule
		Hyperparameter & Value & Origin \\
		\midrule
		\endfirsthead
		\multicolumn{3}{@{}l}{\footnotesize\itshape \tablename~\thetable{} (continued)} \\
		\toprule
		Hyperparameter & Value & Origin \\
		\midrule
		\endhead
		\multicolumn{3}{r@{}}{\footnotesize\itshape Continued on next page} \\
		\endfoot
		\bottomrule
		\multicolumn{3}{@{}p{\textwidth}@{}}{\footnotesize\rule{0pt}{2.5ex}$^{a}$ spartan2's default imposes no size limit; 10 per mode is the setting of the spartan2 FlowScope tutorial.\newline $^{b}$ A training fold with fewer than 10 positives gets no validation split and trains for the full 50 epochs (64 of 460 fits on HI-Small, 30 of 220 on LI-Large).} \\
		\endlastfoot
		\multicolumn{3}{@{}l}{\textbf{FlowScope}~\citep{li2020flowscope}} \\*
		Number of blocks $k$ & 3 & default \\
		Penalty weight $\alpha$ ($\lambda$ in the paper) & 4 & default \\
		Search strategy & fast greedy decreasing (\texttt{level} $=0$) & default \\
		Maximum block size & 10 accounts per mode (source, middle, target) & set$^{a}$ \\
		Flow graph & 3-partite, with the directed transaction graph as both hops & set \\
		Edge weight & amount received (IBM); 1 (synthetic) & set \\
		Account score & $1$, $\frac{2}{3}$, $\frac{1}{3}$ for blocks 1--3; $0$ otherwise & set \\
		Alert rule (precision, F1) & score $> 0.5$, i.e.\ blocks 1 and 2 & set \\
		Repetitions & 1 (the algorithm is deterministic) & -- \\
		\midrule
		\multicolumn{3}{@{}l}{\textbf{AutoAudit} (AA-Smurf)~\citep{lee2020autoaudit}} \\*
		Min.\ intermediaries per pair & 3 (candidate source--target pairs need at least 3 two-hop paths) & default \\
		Solution reported & last one whose code length is at least 10\% above the minimum & default \\
		Maximum iterations & none (until no candidate pattern lowers the code length) & default \\
		Adjacency matrix & binary, directed, dense $|V| \times |V|$ & default \\
		Node-index permutations & 3 (seeds 0, 1, 2); metrics averaged over them & set \\
		Account score & $(N-j)/N$ for the account at position $j$ of the returned ordering of $N$ accounts & set \\
		Alert rule (precision, F1) & score $> 0.5$, i.e.\ the top half of the ordering & set \\
		\midrule
		\multicolumn{3}{@{}l}{\textbf{GraphSAGE}~\citep{hamilton2017inductive}} \\*
		Architecture & 2 \texttt{SAGEConv} layers (receptive field: the second-order neighbourhood), 64 hidden units, ReLU, dropout 0.2 after the first layer; one logit per account & set \\
		Aggregator & mean, with root weight, no output normalisation & default \\
		Optimiser & Adam, learning rate $10^{-3}$ & set \\
		Weight decay & 0 & default \\
		Loss & binary cross-entropy, positive-class weight $n_{\mathrm{neg}}/n_{\mathrm{pos}}$ of the training split & set \\
		Mini-batches & neighbour sampling with fan-out (25, 10); 1,024 seed accounts per batch & set \\
		Training length & at most 50 epochs; early stopping on validation AUC-PR with patience 5, best epoch restored & set \\
		Validation split & stratified 10\% of the training fold$^{b}$ & set \\
		Input graph & undirected, unweighted, parallel transactions merged, self-loops removed, no Louvain filtering & set \\
		Node features & \emph{topology}: degree, log-degree (2); \emph{attributes}: also sent and received counts, sum/mean/std/max of outgoing and incoming amounts, distinct currencies and counterparty banks (14); $\log(1+x)$ on counts and amounts, then standardised & set \\
		Inference & full graph (HI-Small); sampled, fan-out (25, 10), batches of 4,096 (LI-Large) & set \\
		Alert rule (precision, F1) & predicted probability $> 0.5$ & set \\
		Random seed & 1997 & set \\
	\end{longtable}
	\endgroup
	
	\subsection{Evaluation}
	Two aspects are critical for business adoption. 
	A model must (1) catch launderers with few false positives (model performance), and (2) process millions of daily transactions quickly (scalability).
	Accordingly, the evaluation of the methods focuses on these two dimensions.
	
	The performance is evaluated using precision, precision at K (P@K), F1-score, the area under the receiver operating characteristic curve (AUC-ROC) and the area under the precision-recall curve (AUC-PR), also called average precision. 
	
	The precision expresses the relative proportion of predicted cases (positive) that are actually positive (true positive); $\frac{\text{TP}}{\text{P}}$. 
	Financial institutions have limited resources to investigate the generated alerts. 
	Only a small portion of alerts can be investigated. 
	The P@K is the relative number of true positives in the top $K$ predictions, and hence expresses the quality of the most suspicious cases. 
	For the synthetic datasets, we set $K=10$.
	For the IBM datasets, we set $K=1000$, given that these are much larger datasets with many more labels. 
	If precision or precision@K is high, investigators have a good indication that the generated alerts are also high quality, containing limited false positives. 
	The F1-score is the harmonic mean of the precision and recall of the model. 
	It strikes a balance between limiting false positives (high precision) and detecting all fraudulent cases (high recall). 
	
	These metrics require binary predictions, so their values depend on a chosen classification threshold. 
	In our evaluation, an account is flagged when its score exceeds $0.5$. 
	For the GARG-AML scores, which range from $-1$ to $1$, this is the raw score.
	For the tree-based extensions and GraphSAGE, it is the predicted probability of laundering.
	For FlowScope, it means the first two blocks, and for AutoAudit the top half of the ordering.
	
	Both AUC metrics, on the other hand, compare models across classification thresholds. 
	The AUC-ROC quantifies the model's ability to discriminate between classes by plotting the true positive rate against the false positive rate.  
	While it is a widely used metric for binary classification, it may not be sufficient for anti-money laundering (AML) datasets, which are often highly imbalanced, typically with fewer than $0.1\%$ observations flagged as laundering money.
	
	The AUC-PR is more indicative of performance under extreme class imbalance, where the positive class (smurfing) is rare~\citep{davis2006relationship,ozenne2015precision}.
	It plots precision against recall.
	The AUC-PR measures the model's ability to identify positive examples correctly while minimising false positives across all possible threshold values. 
	The AUC-PR specifically focuses on the model's precision in identifying the positive class at various recall thresholds.
	
	Scalability is assessed in terms of calculation time. 
	The theoretical time complexity for the baseline models is $O\left(k\mid E\mid\log\left(\mid V\mid\right)\right)$ for FlowScope~\citep{li2020flowscope} and $O\left(\mid V\mid\cdot\mid E\mid +\mid V\mid^2 \right)$ for AutoAudit~\citep{lee2020autoaudit}. 
	
	For GARG-AML, the main calculations are the selection of the second-order neighbourhood and the assignment of the neighbours to the levels. 
	Assuming that we are working in sparse graphs, both steps have complexity $O\left(\mid V\mid\right)$. 
	This brings us to a total worst case time complexity of $O\left(n^2\right)$. 
	Hence, theoretically, GARG-AML scales as AutoAudit, while FlowScope is much faster. 
	However, we can speed up the calculations by running them in parallel. 
	In addition, second-order neighbours are generally much smaller than $n$, meaning that those steps can run very quickly. 
	In addition to runtime, memory usage is a key factor. 
	We report any instances where methods fail due to memory limitations. 
	
	All the experiments were conducted on the Flemish Supercomputer~(VSC) with an Intel Xeon Platinum 8360Y CPU and 200~GB of memory, with GraphSAGE trained on an NVIDIA A100 SXM4 GPU.
	Each experiment was constrained to a time budget of 16 hours, the maximum on the VSC.
	
	\subsection{Statistical Test}
	To test whether the observed differences in performance are not due to random chance, we determine for each model its rank compared to the other models based on the AUC-ROC and the AUC-PR. 
	The best-performing method receives rank 1, the second-best method receives rank 2, and so on.
	
	The statistical significance of the difference in ranks is evaluated using the Friedman test~\citep{Friedman01121937, Friedman_2} and the post hoc Nemenyi test~\citep{nemenyi1963distribution}, as described by~\citet{JMLR:v7:demsar06a}. 
	It compares the performance of $k$ methods across $N$ datasets. 
	If we take $r_i^j$ to be the rank of method $j$ on dataset $i$, the average performance across all datasets is equal to \(R_j = \frac{1}{N}\sum_i r_i^j\). 
	
	The null hypothesis of the Friedman test is that the average ranks $R_j$ are equal, and the Friedman statistic is equal to
	\[ Q = \frac{12N}{k(k+1)} \left[ \sum_j R_j^2 -\frac{k(k+1)^2}{4} \right]. \]
	The Friedman statistic follows a $\chi^2_{k-1}$ distribution under this null hypothesis.  
	
	Rejecting the null hypothesis indicates that significant differences are present among the methods. 
	In that case, we proceed with the post hoc Nemenyi test.
	It indicates that the performances of two methods are significantly different at level $\alpha$ if their respective average ranks differ by at least the critical difference (CD):
	\[ \text{CD} = q_\alpha \sqrt{\frac{k(k+1)}{6N}}, \]
	where critical levels $q_\alpha$ are provided by~\citet{JMLR:v7:demsar06a}. 
	
	Our experiments compare the ranks of the $k=8$ methods across the $N=66$ synthetic datasets at a significance level of $\alpha=0.05$.
	We visualise the results using CD diagrams~\citep{JMLR:v7:demsar06a} for the AUC-ROC and AUC-PR. 
	
	\section{Results and Discussion}
	\label{sec: results and discussion}
	\subsection{Scalability}
\begin{figure}
		\centering
		\includegraphics[width=0.6\linewidth]{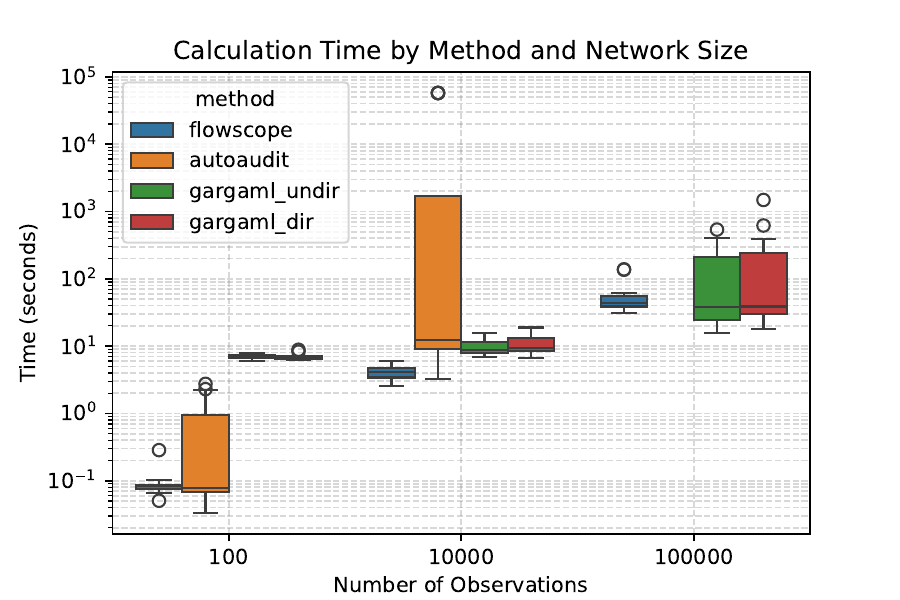}
		\caption{Box plot of the run time on a logarithmic scale across network sizes. From left to right: FlowScope, AutoAudit, undirected GARG-AML and directed GARG-AML scores. The out-of-time results of AutoAudit on the dataset of size 10,000 are capped at 16 hours (57,600 seconds). AutoAudit is not included for the size-100,000 datasets as it runs out of memory on these datasets.}
		\label{fig:boxplot_time_log}
	\end{figure}

\begin{figure}
		\centering
		\includegraphics[width=0.6\linewidth]{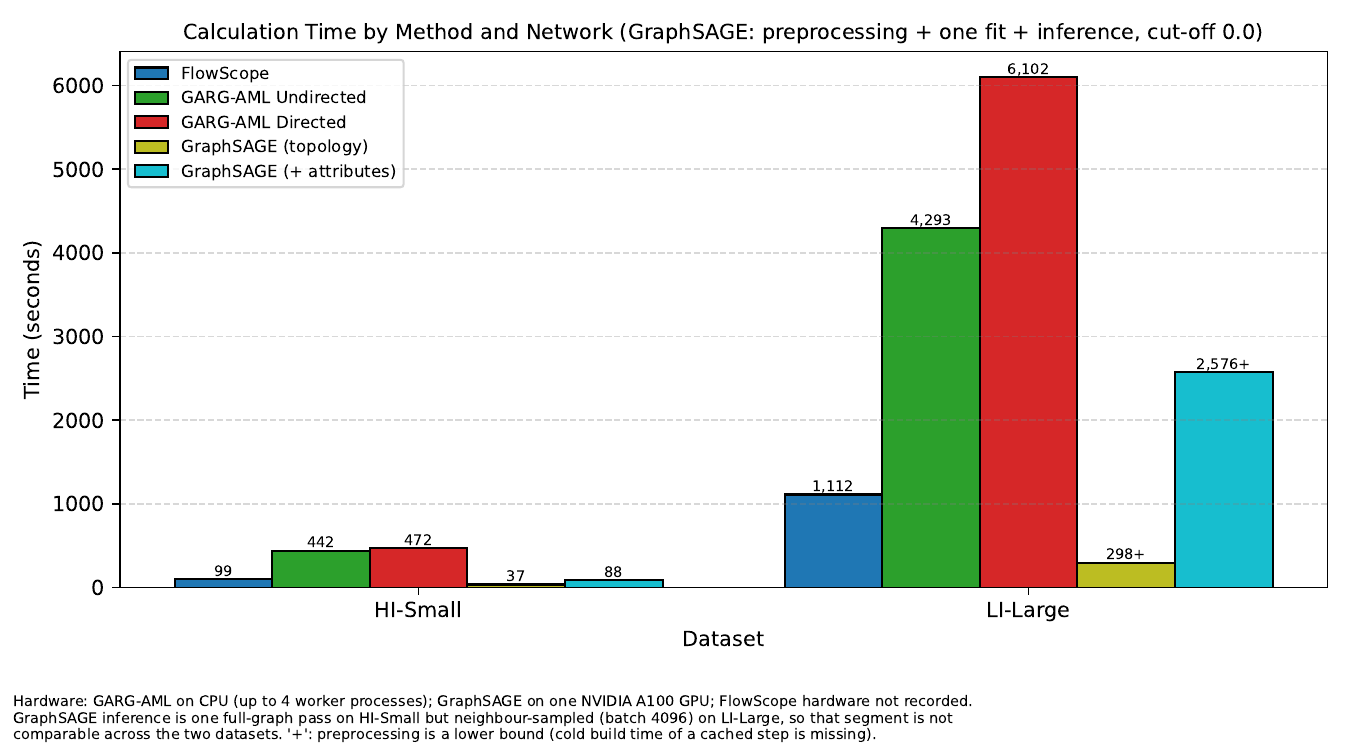}
		\caption{Bar plot of the run time for the two IBM datasets. From left to right: FlowScope, undirected GARG-AML and directed GARG-AML scores. AutoAudit is not included as it runs out of memory on these datasets.}
		\label{fig:boxplot_time_ibm}
	\end{figure}

	Figure~\ref{fig:boxplot_time_log} and Figure~\ref{fig:boxplot_time_ibm} show the calculation times across the various network sizes of the synthetic and IBM datasets, respectively. 
	The time in Figure~\ref{fig:boxplot_time_log} is shown on a logarithmic scale, to provide a better distinction across dataset sizes. 
	These results indicate that FlowScope is the fastest method, although GARG-AML seems to scale well to larger datasets, contrary to what we expected from the theoretical worst-case time complexity. 
	In contrast, AutoAudit, although quite fast in the beginning, has difficulties with the largest networks. 
	The distribution of the calculation time of AutoAudit is much wider with more outliers, indicating a high sensitivity to the network topology.  
	
	
	Additionally, AutoAudit consistently runs out of time, exceeding 16 hours, on Watts-Strogatz networks of size 10,000. 
	We set the out-of-time results equal to 16 hours (57,600 seconds) in Figure~\ref{fig:boxplot_time_log}, resulting in the observed outliers. 
	For the larger networks, i.e., the synthetic network of size 100,000 and the IBM datasets, AutoAudit exceeds the available memory. 
	Hence, AutoAudit is omitted from Figure~\ref{fig:boxplot_time_ibm}. 
	
	Figure~\ref{fig:boxplot_time_ibm} also shows that the GraphSAGE models have the smallest calculation time on the HI-Small dataset, and give great scalability on the LI-Large dataset as well.
	Training with additional node features takes substantially longer.
	This calculation speed is, however, only possible due to the deployment of GPU resources.
	
	\subsection{Performance}
	\label{sub:perf}
	\subsubsection{Synthetic Data}
	\label{subsub:perf synth}
	The results are presented according to the injected patterns in Table~\ref{tab:perf_synth}, with more detailed results given in \ref{app:synthetic}.
	The results indicate that, on average, the undirected GARG-AML scores show competitive performance. 
	This performance is boosted when combining the GARG-AML scores with the tree-based learners. 
	In comparison, the performance of the baselines, FlowScope and AutoAudit, is less stable.
	
	A major concern for transaction monitoring is the high false positive rate. 
	The precision shows that more than half of the alerts from FlowScope and AutoAudit were false positives. 
	GARG-AML scores combined with tree-based learners result in around $70\%$ of alerts being actual laundering accounts, except for the `existing mules' pattern.
	The results for P@10 show that half of the top ranked nodes by the base undirected GARG-AML score are laundering money. 
	This goes up to more than 9 out of 10 for the tree-based models.
	This points to a strong efficiency gain for financial institutions.
	No P@10 results are reported for FlowScope and AutoAudit. 
	The definition of the scores for these two methods results in many ties, giving a `top 10' that consists of more than ten observations. 
	This would skew the comparison to the other methods. 
	
	\begin{table*}[p]
		\centering
		\caption{Performance comparison across patterns. For the two GARG-AML score rows, Precision and F1 use the rule score $>0.5$; AUC-ROC, AUC-PR and P@10 do not depend on a cut. Entries are the mean$_{\pm\text{standard deviation}}$ over the synthetic datasets.}
		\label{tab:perf_synth}
		\footnotesize
		\setlength{\tabcolsep}{3.2pt}
		\begin{tabular}{@{}lccccc@{}}
			\toprule
			Method & Precision & F1-score & AUC-ROC & AUC-PR & P@10 \\
			\midrule
			\multicolumn{6}{@{}l}{\textit{All laundering}} \\
			\midrule
			FlowScope & $0.417_{\pm 0.348}$ & $0.192_{\pm 0.182}$ & $0.547_{\pm 0.097}$ & $0.272_{\pm 0.290}$ & -- \\
			AutoAudit & $0.219_{\pm 0.312}$ & $0.194_{\pm 0.268}$ & $0.374_{\pm 0.314}$ & $0.336_{\pm 0.334}$ & -- \\
			GARG-AML Undirected & $0.536_{\pm 0.456}$ & $0.464_{\pm 0.369}$ & $0.721_{\pm 0.140}$ & $0.435_{\pm 0.375}$ & $0.529_{\pm 0.470}$ \\
			GARG-AML Directed & $0.316_{\pm 0.363}$ & $0.148_{\pm 0.162}$ & $0.598_{\pm 0.147}$ & $0.280_{\pm 0.334}$ & $0.295_{\pm 0.316}$ \\
			GARG-AML Undir. Tree & $0.935_{\pm 0.079}$ & $0.797_{\pm 0.129}$ & $0.892_{\pm 0.074}$ & $0.806_{\pm 0.129}$ & $0.977_{\pm 0.060}$ \\
			GARG-AML Undir. Boost & $0.901_{\pm 0.184}$ & $0.805_{\pm 0.176}$ & $0.887_{\pm 0.108}$ & $0.789_{\pm 0.220}$ & $0.942_{\pm 0.184}$ \\
			GARG-AML Dir. Tree & $0.894_{\pm 0.105}$ & $0.721_{\pm 0.189}$ & $0.877_{\pm 0.075}$ & $0.732_{\pm 0.198}$ & $0.926_{\pm 0.129}$ \\
			GARG-AML Dir. Boost & $0.866_{\pm 0.176}$ & $0.734_{\pm 0.222}$ & $0.863_{\pm 0.115}$ & $0.718_{\pm 0.272}$ & $0.902_{\pm 0.198}$ \\
			\midrule
			\multicolumn{6}{@{}l}{\textit{Separate}} \\
			\midrule
			FlowScope & $0.174_{\pm 0.313}$ & $0.165_{\pm 0.286}$ & $0.584_{\pm 0.200}$ & $0.196_{\pm 0.274}$ & -- \\
			AutoAudit & $0.113_{\pm 0.171}$ & $0.158_{\pm 0.236}$ & $0.509_{\pm 0.442}$ & $0.453_{\pm 0.447}$ & -- \\
			GARG-AML Undirected & $0.293_{\pm 0.276}$ & $0.383_{\pm 0.334}$ & $0.875_{\pm 0.115}$ & $0.372_{\pm 0.377}$ & $0.327_{\pm 0.351}$ \\
			GARG-AML Directed & $0.107_{\pm 0.187}$ & $0.077_{\pm 0.089}$ & $0.768_{\pm 0.117}$ & $0.161_{\pm 0.146}$ & $0.100_{\pm 0.101}$ \\
			GARG-AML Undir. Tree & $0.865_{\pm 0.207}$ & $0.852_{\pm 0.213}$ & $0.975_{\pm 0.046}$ & $0.872_{\pm 0.177}$ & $0.782_{\pm 0.188}$ \\
			GARG-AML Undir. Boost & $0.867_{\pm 0.222}$ & $0.863_{\pm 0.219}$ & $0.938_{\pm 0.193}$ & $0.874_{\pm 0.216}$ & $0.776_{\pm 0.228}$ \\
			GARG-AML Dir. Tree & $0.848_{\pm 0.242}$ & $0.705_{\pm 0.218}$ & $0.950_{\pm 0.066}$ & $0.747_{\pm 0.199}$ & $0.723_{\pm 0.206}$ \\
			GARG-AML Dir. Boost & $0.770_{\pm 0.284}$ & $0.747_{\pm 0.261}$ & $0.883_{\pm 0.226}$ & $0.712_{\pm 0.298}$ & $0.683_{\pm 0.261}$ \\
			\midrule
			\multicolumn{6}{@{}l}{\textit{Existing mules}} \\
			\midrule
			FlowScope & $0.223_{\pm 0.203}$ & $0.206_{\pm 0.196}$ & $0.596_{\pm 0.114}$ & $0.152_{\pm 0.152}$ & -- \\
			AutoAudit & $0.062_{\pm 0.094}$ & $0.087_{\pm 0.129}$ & $0.318_{\pm 0.271}$ & $0.071_{\pm 0.090}$ & -- \\
			GARG-AML Undirected & $0.036_{\pm 0.084}$ & $0.056_{\pm 0.126}$ & $0.405_{\pm 0.188}$ & $0.064_{\pm 0.096}$ & $0.032_{\pm 0.101}$ \\
			GARG-AML Directed & $0.077_{\pm 0.137}$ & $0.083_{\pm 0.142}$ & $0.554_{\pm 0.146}$ & $0.087_{\pm 0.129}$ & $0.076_{\pm 0.127}$ \\
			GARG-AML Undir. Tree & $0.330_{\pm 0.373}$ & $0.276_{\pm 0.315}$ & $0.738_{\pm 0.185}$ & $0.332_{\pm 0.288}$ & $0.339_{\pm 0.282}$ \\
			GARG-AML Undir. Boost & $0.362_{\pm 0.378}$ & $0.293_{\pm 0.316}$ & $0.705_{\pm 0.215}$ & $0.326_{\pm 0.284}$ & $0.323_{\pm 0.276}$ \\
			GARG-AML Dir. Tree & $0.289_{\pm 0.368}$ & $0.204_{\pm 0.260}$ & $0.692_{\pm 0.180}$ & $0.260_{\pm 0.236}$ & $0.262_{\pm 0.237}$ \\
			GARG-AML Dir. Boost & $0.371_{\pm 0.375}$ & $0.274_{\pm 0.270}$ & $0.670_{\pm 0.199}$ & $0.284_{\pm 0.277}$ & $0.286_{\pm 0.273}$ \\
			\midrule
			\multicolumn{6}{@{}l}{\textit{New mules}} \\
			\midrule
			FlowScope & $0.020_{\pm 0.051}$ & $0.014_{\pm 0.030}$ & $0.458_{\pm 0.066}$ & $0.069_{\pm 0.098}$ & -- \\
			AutoAudit & $0.043_{\pm 0.072}$ & $0.059_{\pm 0.094}$ & $0.241_{\pm 0.221}$ & $0.056_{\pm 0.082}$ & -- \\
			GARG-AML Undirected & $0.242_{\pm 0.231}$ & $0.317_{\pm 0.286}$ & $0.755_{\pm 0.182}$ & $0.216_{\pm 0.210}$ & $0.197_{\pm 0.247}$ \\
			GARG-AML Directed & $0.118_{\pm 0.164}$ & $0.115_{\pm 0.141}$ & $0.436_{\pm 0.170}$ & $0.129_{\pm 0.169}$ & $0.124_{\pm 0.156}$ \\
			GARG-AML Undir. Tree & $0.781_{\pm 0.292}$ & $0.613_{\pm 0.263}$ & $0.936_{\pm 0.064}$ & $0.690_{\pm 0.197}$ & $0.653_{\pm 0.199}$ \\
			GARG-AML Undir. Boost & $0.730_{\pm 0.273}$ & $0.685_{\pm 0.275}$ & $0.911_{\pm 0.128}$ & $0.706_{\pm 0.264}$ & $0.673_{\pm 0.240}$ \\
			GARG-AML Dir. Tree & $0.683_{\pm 0.323}$ & $0.603_{\pm 0.294}$ & $0.944_{\pm 0.069}$ & $0.643_{\pm 0.245}$ & $0.632_{\pm 0.239}$ \\
			GARG-AML Dir. Boost & $0.756_{\pm 0.285}$ & $0.682_{\pm 0.295}$ & $0.873_{\pm 0.214}$ & $0.650_{\pm 0.317}$ & $0.617_{\pm 0.285}$ \\
			\bottomrule
		\end{tabular}
	\end{table*}
	
	These results are an average over widely different label distributions (see Table~\ref{tab:summary data synthetic}). 
	The precision, F1-scores and the AUC-PR are highly sensitive to the label imbalance. 
	Their baselines depend on the ratio of positive labels to the total number of labels, which changes with each dataset.
	The AUC-ROC baseline on the other hand is always equal to 0.5. 
	This makes all but the AUC-ROC values in Table~\ref{tab:perf_synth} hard to compare.
	
\begin{figure}
		\centering
		\includegraphics[width=0.8\linewidth]{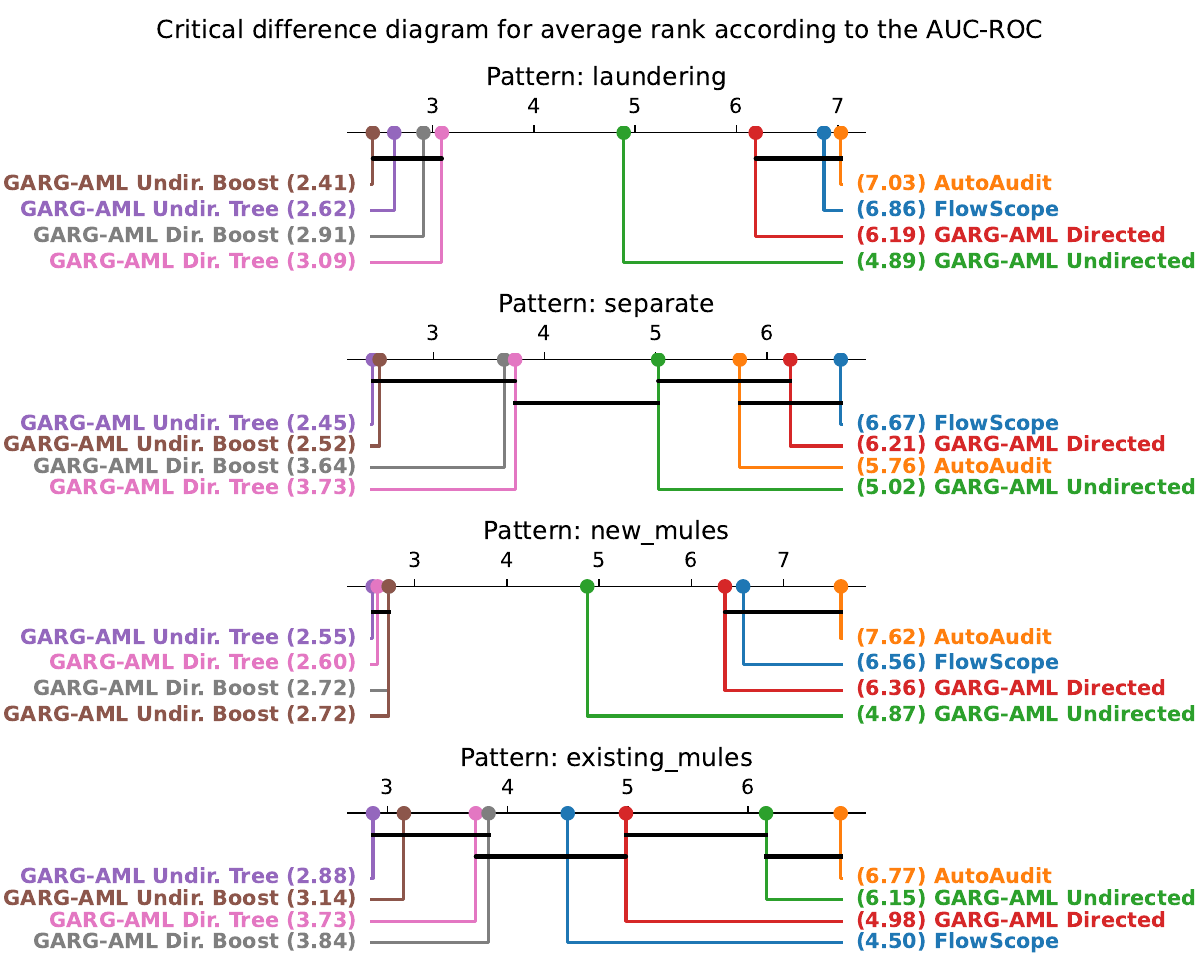}
		\caption{Critical difference plots of the average ranks according to the AUC-ROC across patterns.}
		\label{fig:CD_ROC}
	\end{figure}

\begin{figure}
		\centering
		\includegraphics[width=0.8\linewidth]{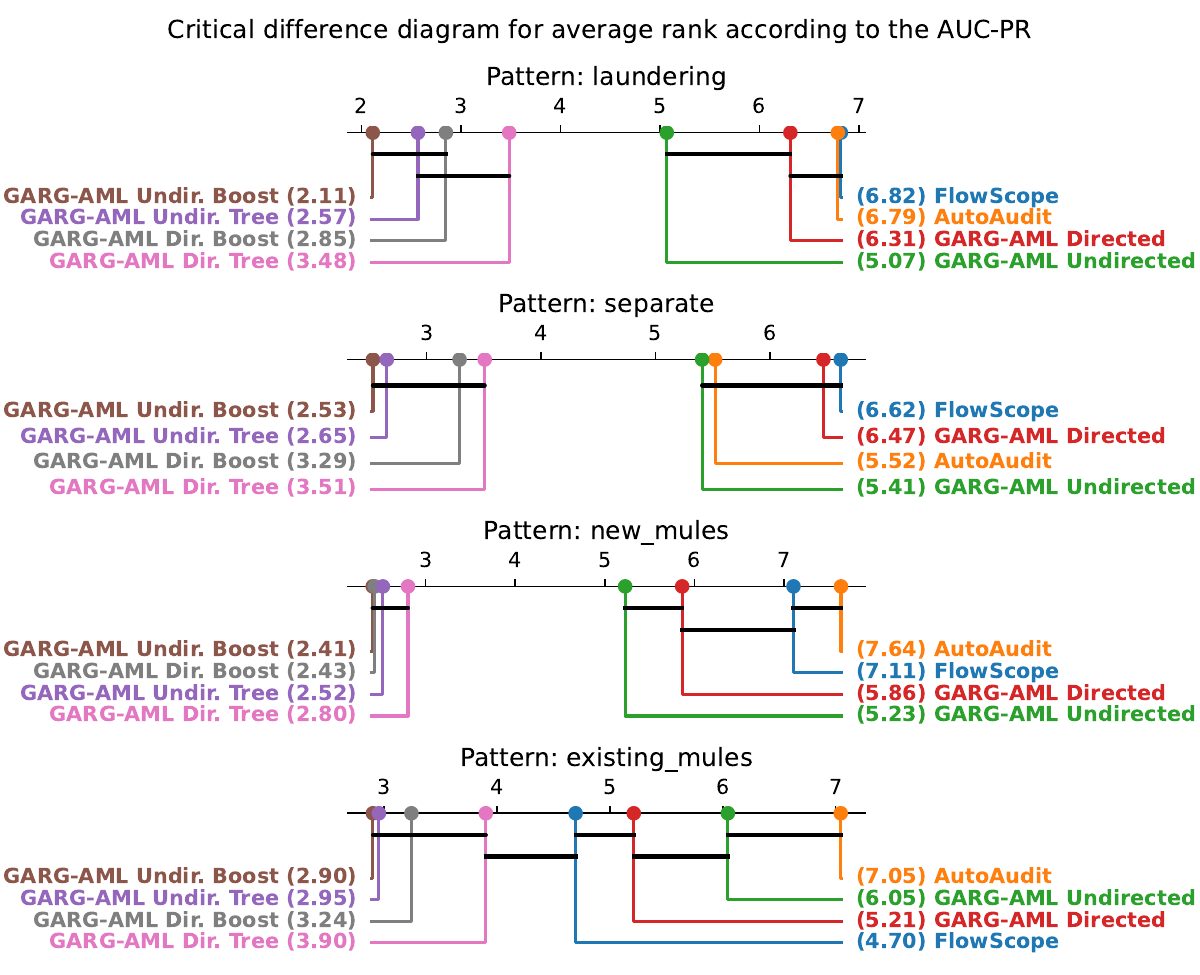}
		\caption{Critical difference plots of the average ranks according to the AUC-PR across patterns.}
		\label{fig:CD_PR}
	\end{figure}

	To mitigate this, we analyse the average rank of each model's AUC-ROC and AUC-PR for a more robust evaluation. 
	The results are shown in Figure~\ref{fig:CD_ROC} and Figure~\ref{fig:CD_PR} for the AUC-ROC and AUC-PR, respectively. 
	We find that the tree-based learners score better than the other methods, even when applying a simple decision tree.
	
	Table~\ref{tab:Friedmann_synth} illustrates that the Friedman test rejects the null hypothesis for each case, which indicates that significant differences between the ranks are present. 
	The post hoc Nemenyi test indicates that the tree-based methods differ significantly from most other methods. 
	The undirected scores on their own already obtain strong performance. For the existing mules patterns, however, it seems that the best performing models do not differ significantly from FlowScope. 
	
	One caveat to be noted is that the underlying hypotheses of the Friedman and Nemenyi statistical tests assume independence amongst the datasets. 
	This is not strictly the case for the synthetic datasets, as they share generators and parameters. 
	However, given the magnitude of the $p$-values and the heterogeneity in the graph topology, the main conclusions remain valid. 
	
	\begin{table}
		\centering
		\caption{Results for the Friedman test for the eight different configurations ($\alpha=0.05$).}
		\label{tab:Friedmann_synth}
		\begin{tabular}{llrrc}
			\toprule
			Metric &        Pattern &  $\chi^2_7$ &       $p$ &   Reject $H_0$  \\
			\midrule
			AUC-ROC &     laundering &   303.99 &    8.5e-62 &    True \\
			AUC-ROC &       separate &   212.54 &    2.5e-42 &    True \\
			AUC-ROC &      new mules &   345.30 &    1.2e-70 &    True \\
			AUC-ROC & existing mules &   150.18 &    3.7e-29 &    True \\
			AUC-PR  &     laundering &   303.86 &    9.1e-62 &    True \\
			AUC-PR  &       separate &   224.27 &    8.2e-45 &    True \\
			AUC-PR  &      new mules &   379.42 &    6.2e-78 &    True \\
			AUC-PR  & existing mules &   179.33 &    2.7e-35 &    True \\
			\bottomrule
		\end{tabular}
	\end{table}
	
	\subsubsection{IBM Data}
	Table~\ref{tab:results ibm small} and Table~\ref{tab:results ibm large} present the results for the smurfing patterns, i.e., gather-scatter and scatter-gather, and the full `Is Laundering' label. 
	The cut-offs for binary label definition presented here are 0.1, 0.5 and 0.9. 
	The results for all patterns and cut-offs are provided in \ref{app:full ibm}. 
	Missing results indicate that there were too few labels for the model to learn from. 
	The tree-based and GraphSAGE models could not be trained on the scatter-gather label of the HI-Small dataset with cut-off 0.9 (Table~\ref{tab:results ibm small}). 
	No positive gather-scatter and scatter-gather labels were present in the LI-Large dataset for cut-off 0.9 (Table~\ref{tab:results ibm large}). Likewise, the tree-based and GraphSAGE models could not be trained on the scatter-gather label of the LI-Large dataset with cut-off 0.5 (Table~\ref{tab:results ibm large}).
	
	\begin{sidewaystable}
		\centering
		\caption{Performance comparison across graph patterns and cut-offs for the HI-Small dataset. For the two GARG-AML score rows, Precision and F1 use the rule score $>0.5$; AUC-ROC, AUC-PR and P@1000 do not depend on a cut. The best result per column is in bold; the best result excluding GraphSAGE is underlined.}
		\label{tab:results ibm small}
		\scriptsize
		\setlength{\tabcolsep}{3.4pt}
		\begin{tabular}{@{}l ccccc ccccc ccccc@{}}
			\toprule
			& \multicolumn{5}{c}{\textbf{Cut-off 0.1}} & \multicolumn{5}{c}{\textbf{Cut-off 0.5}} & \multicolumn{5}{c}{\textbf{Cut-off 0.9}} \\
			\cmidrule(lr){2-6} \cmidrule(lr){7-11} \cmidrule(l){12-16}
			\textbf{Model} & \makecell[cb]{Preci-\\sion} & F1 & \makecell[cb]{AUC-\\ROC} & \makecell[cb]{AUC-\\PR} & \makecell[cb]{P@\\1000} & \makecell[cb]{Preci-\\sion} & F1 & \makecell[cb]{AUC-\\ROC} & \makecell[cb]{AUC-\\PR} & \makecell[cb]{P@\\1000} & \makecell[cb]{Preci-\\sion} & F1 & \makecell[cb]{AUC-\\ROC} & \makecell[cb]{AUC-\\PR} & \makecell[cb]{P@\\1000} \\
			\midrule
			\multicolumn{16}{@{}l}{\textit{Is Laundering}} \\
			\midrule
			FlowScope             & 0.000 & 0.000 & 50.0 & 0.445 & 0.004 & 0.000 & 0.000 & 50.0 & 0.020 & 0.001 & 0.000 & 0.000 & 50.0 & 0.013 & 0.000 \\
			GARG-AML Undirected   & 0.006 & 0.012 & 61.2 & 0.559 & 0.016 & 0.000 & 0.001 & 66.2 & 0.037 & 0.001 & 0.000 & 0.001 & \underline{81.7} & 0.041 & 0.001 \\
			GARG-AML Directed     & 0.004 & 0.009 & 45.5 & 0.398 & 0.006 & 0.000 & 0.001 & 53.0 & 0.028 & 0.000 & 0.000 & 0.001 & 63.3 & 0.030 & 0.000 \\
			GARG-AML Undir. Tree  & 0.515 & 0.043 & 63.6 & 3.989 & 0.163 & 0.000 & 0.000 & 61.3 & 0.151 & 0.005 & 0.000 & 0.000 & 69.5 & 0.047 & 0.000 \\
			GARG-AML Undir. Boost & 0.418 & 0.024 & 77.2 & 3.840 & 0.135 & \textbf{\underline{0.022}} & \textbf{\underline{0.013}} & \underline{74.6} & 0.168 & 0.005 & 0.000 & 0.000 & 80.0 & 0.047 & 0.000 \\
			GARG-AML Dir. Tree    & 0.484 & \underline{0.049} & 58.7 & 4.664 & \underline{0.175} & 0.000 & 0.000 & 55.1 & \underline{0.648} & \underline{0.007} & 0.000 & 0.000 & 54.1 & \underline{0.049} & \underline{0.002} \\
			GARG-AML Dir. Boost   & \textbf{\underline{0.570}} & 0.041 & \underline{80.0} & \underline{5.050} & 0.144 & 0.000 & 0.000 & 70.0 & 0.090 & 0.002 & \underline{0.002} & \underline{0.003} & 66.3 & 0.039 & 0.001 \\
			GraphSAGE (topology)  & 0.008 & 0.016 & 77.5 & 1.440 & 0.040 & 0.000 & 0.001 & 81.5 & 0.116 & 0.004 & 0.001 & 0.001 & 85.1 & 0.060 & 0.001 \\
			GraphSAGE (+ attributes) & 0.040 & \textbf{0.078} & \textbf{96.8} & \textbf{41.698} & \textbf{0.717} & 0.003 & 0.006 & \textbf{96.1} & \textbf{3.759} & \textbf{0.038} & \textbf{0.003} & \textbf{0.005} & \textbf{97.0} & \textbf{7.005} & \textbf{0.025} \\
			\midrule
			\multicolumn{16}{@{}l}{\textit{Gather-scatter}} \\
			\midrule
			FlowScope             & 0.000 & 0.000 & 50.0 & 0.056 & 0.000 & 0.000 & 0.000 & 50.0 & 0.003 & 0.000 & 0.000 & 0.000 & 50.0 & 0.001 & 0.000 \\
			GARG-AML Undirected   & 0.001 & 0.002 & 65.7 & 0.082 & 0.004 & 0.000 & 0.000 & 53.5 & 0.003 & 0.000 & \underline{0.000} & \underline{0.000} & \underline{83.4} & 0.004 & 0.000 \\
			GARG-AML Directed     & 0.001 & 0.001 & 49.6 & 0.057 & 0.001 & 0.000 & 0.000 & 48.8 & 0.003 & 0.000 & \underline{0.000} & 0.000 & 65.7 & 0.003 & 0.000 \\
			GARG-AML Undir. Tree  & 0.000 & 0.000 & 71.4 & \underline{5.006} & 0.056 & 0.000 & 0.000 & 49.6 & 0.003 & 0.000 & 0.000 & 0.000 & 49.6 & 0.001 & 0.000 \\
			GARG-AML Undir. Boost & 0.197 & 0.083 & \underline{84.1} & 4.983 & 0.054 & 0.000 & 0.000 & \underline{63.9} & 0.090 & 0.002 & 0.000 & 0.000 & 57.6 & \underline{0.059} & \underline{0.002} \\
			GARG-AML Dir. Tree    & \textbf{\underline{0.263}} & 0.033 & 67.7 & 4.936 & \underline{0.078} & 0.000 & 0.000 & 57.7 & 0.248 & 0.002 & 0.000 & 0.000 & 50.0 & 0.001 & 0.000 \\
			GARG-AML Dir. Boost   & 0.173 & \underline{0.089} & 83.9 & 4.986 & 0.076 & \textbf{\underline{0.006}} & \textbf{\underline{0.010}} & 57.1 & \underline{0.290} & \underline{0.004} & 0.000 & 0.000 & 58.3 & 0.019 & 0.000 \\
			GraphSAGE (topology)  & 0.004 & 0.007 & 94.5 & 2.688 & 0.027 & 0.000 & 0.000 & 89.3 & 0.033 & 0.000 & 0.000 & 0.000 & 99.4 & 0.102 & 0.000 \\
			GraphSAGE (+ attributes) & 0.073 & \textbf{0.136} & \textbf{99.2} & \textbf{49.264} & \textbf{0.247} & 0.003 & 0.005 & \textbf{95.3} & \textbf{1.079} & \textbf{0.008} & \textbf{0.003} & \textbf{0.006} & \textbf{99.9} & \textbf{0.636} & \textbf{0.006} \\
			\midrule
			\multicolumn{16}{@{}l}{\textit{Scatter-gather}} \\
			\midrule
			FlowScope             & 0.000 & 0.000 & 50.0 & 0.036 & 0.001 & 0.000 & 0.000 & 50.0 & 0.003 & 0.001 & \textbf{\underline{0.000}} & 0.000 & 50.0 & 0.000 & \textbf{\underline{0.000}} \\
			GARG-AML Undirected   & 0.001 & 0.001 & 57.3 & 0.040 & 0.001 & 0.000 & 0.000 & 41.3 & 0.002 & 0.000 & \textbf{\underline{0.000}} & \textbf{\underline{0.000}} & \textbf{\underline{86.0}} & \textbf{\underline{0.001}} & \textbf{\underline{0.000}} \\
			GARG-AML Directed     & 0.000 & 0.000 & 41.9 & 0.030 & 0.000 & 0.000 & 0.000 & 31.5 & 0.002 & 0.000 & \textbf{\underline{0.000}} & 0.000 & 47.2 & 0.000 & \textbf{\underline{0.000}} \\
			GARG-AML Undir. Tree  & \textbf{\underline{0.538}} & 0.319 & 82.1 & \textbf{\underline{32.409}} & \underline{0.122} & 0.000 & 0.000 & 57.1 & 1.129 & 0.002 & -- & -- & -- & -- & -- \\
			GARG-AML Undir. Boost & 0.436 & 0.294 & \underline{87.7} & 18.704 & 0.098 & 0.006 & 0.012 & 72.0 & 0.198 & 0.004 & -- & -- & -- & -- & -- \\
			GARG-AML Dir. Tree    & 0.500 & \textbf{\underline{0.327}} & 77.3 & 24.886 & 0.101 & 0.000 & 0.000 & 64.3 & \textbf{\underline{11.416}} & 0.004 & -- & -- & -- & -- & -- \\
			GARG-AML Dir. Boost   & 0.041 & 0.071 & 82.5 & 2.668 & 0.025 & \textbf{\underline{0.010}} & \textbf{\underline{0.020}} & \underline{73.3} & 0.410 & \underline{0.005} & -- & -- & -- & -- & -- \\
			GraphSAGE (topology)  & 0.002 & 0.004 & 94.2 & 0.813 & 0.021 & 0.000 & 0.000 & 86.4 & 0.066 & 0.001 & -- & -- & -- & -- & -- \\
			GraphSAGE (+ attributes) & 0.025 & 0.050 & \textbf{98.7} & 26.451 & \textbf{0.153} & 0.000 & 0.000 & \textbf{88.7} & 2.904 & \textbf{0.007} & -- & -- & -- & -- & -- \\
			\bottomrule
		\end{tabular}
	\end{sidewaystable}
	
	\begin{sidewaystable}
		\centering
		\caption{Performance comparison across graph patterns and cut-offs for the LI-Large dataset. For the two GARG-AML score rows, Precision and F1 use the rule score $>0.5$; AUC-ROC, AUC-PR and P@1000 do not depend on a cut. The best result per column is in bold; the best result excluding GraphSAGE is underlined.}
		\label{tab:results ibm large}
		\scriptsize
		\setlength{\tabcolsep}{3.4pt}
		\begin{tabular}{@{}l ccccc ccccc ccccc@{}}
			\toprule
			& \multicolumn{5}{c}{\textbf{Cut-off 0.1}} & \multicolumn{5}{c}{\textbf{Cut-off 0.5}} & \multicolumn{5}{c}{\textbf{Cut-off 0.9}} \\
			\cmidrule(lr){2-6} \cmidrule(lr){7-11} \cmidrule(l){12-16}
			\textbf{Model} & \makecell[cb]{Preci-\\sion} & F1 & \makecell[cb]{AUC-\\ROC} & \makecell[cb]{AUC-\\PR} & \makecell[cb]{P@\\1000} & \makecell[cb]{Preci-\\sion} & F1 & \makecell[cb]{AUC-\\ROC} & \makecell[cb]{AUC-\\PR} & \makecell[cb]{P@\\1000} & \makecell[cb]{Preci-\\sion} & F1 & \makecell[cb]{AUC-\\ROC} & \makecell[cb]{AUC-\\PR} & \makecell[cb]{P@\\1000} \\
			\midrule
			\multicolumn{16}{@{}l}{\textit{Is Laundering}} \\
			\midrule
			FlowScope             & 0.000 & 0.000 & 50.0 & 0.216 & 0.003 & 0.000 & 0.000 & 50.0 & 0.002 & 0.000 & 0.000 & 0.000 & 50.0 & 0.000 & 0.000 \\
			GARG-AML Undirected   & 0.003 & \underline{0.006} & 55.1 & 0.255 & 0.009 & 0.000 & 0.000 & 68.0 & 0.003 & 0.000 & 0.000 & 0.000 & \underline{91.3} & 0.003 & 0.000 \\
			GARG-AML Directed     & 0.003 & 0.005 & 47.2 & 0.208 & 0.007 & 0.000 & 0.000 & 68.4 & 0.004 & 0.000 & 0.000 & 0.000 & 83.8 & 0.005 & 0.000 \\
			GARG-AML Undir. Tree  & 0.000 & 0.000 & 55.4 & 0.361 & 0.025 & 0.000 & 0.000 & 51.3 & \underline{0.012} & \underline{0.002} & 0.000 & 0.000 & 59.9 & \textbf{\underline{0.068}} & \textbf{\underline{0.002}} \\
			GARG-AML Undir. Boost & 0.000 & 0.000 & 80.7 & 0.860 & 0.024 & 0.000 & 0.000 & 59.3 & 0.003 & 0.000 & 0.000 & 0.000 & 64.8 & 0.004 & 0.000 \\
			GARG-AML Dir. Tree    & \textbf{\underline{0.100}} & 0.001 & 55.3 & 0.424 & \underline{0.038} & 0.000 & 0.000 & 50.0 & 0.002 & 0.000 & 0.000 & 0.000 & 50.0 & 0.000 & 0.000 \\
			GARG-AML Dir. Boost   & 0.000 & 0.000 & \underline{81.8} & \underline{0.989} & 0.026 & \underline{0.000} & \underline{0.000} & \underline{69.1} & 0.005 & 0.001 & \underline{0.000} & \underline{0.000} & 74.1 & 0.003 & 0.000 \\
			GraphSAGE (topology)  & 0.004 & 0.008 & 76.9 & 0.630 & 0.027 & 0.000 & 0.000 & 85.7 & 0.021 & 0.000 & 0.000 & 0.000 & \textbf{92.5} & 0.005 & 0.000 \\
			GraphSAGE (+ attributes) & 0.018 & \textbf{0.036} & \textbf{96.9} & \textbf{18.328} & \textbf{0.480} & \textbf{0.000} & \textbf{0.001} & \textbf{96.0} & \textbf{1.583} & \textbf{0.009} & \textbf{0.000} & \textbf{0.000} & 79.5 & 0.017 & 0.000 \\
			\midrule
			\multicolumn{16}{@{}l}{\textit{Gather-scatter}} \\
			\midrule
			FlowScope             & 0.000 & 0.000 & 50.0 & 0.008 & 0.001 & \underline{0.000} & \underline{0.000} & 50.0 & 0.000 & \underline{0.000} & -- & -- & -- & -- & -- \\
			GARG-AML Undirected   & 0.000 & 0.000 & 59.5 & 0.009 & 0.000 & \underline{0.000} & \underline{0.000} & 59.0 & 0.001 & \underline{0.000} & -- & -- & -- & -- & -- \\
			GARG-AML Directed     & \underline{0.000} & \underline{0.000} & 60.9 & 0.010 & 0.000 & \underline{0.000} & \underline{0.000} & 60.8 & 0.001 & \underline{0.000} & -- & -- & -- & -- & -- \\
			GARG-AML Undir. Tree  & 0.000 & 0.000 & 49.9 & 0.008 & 0.000 & \underline{0.000} & \underline{0.000} & 50.0 & 0.000 & \underline{0.000} & -- & -- & -- & -- & -- \\
			GARG-AML Undir. Boost & 0.000 & 0.000 & 60.2 & 0.013 & \underline{0.002} & \underline{0.000} & \underline{0.000} & 39.5 & \underline{0.001} & \underline{0.000} & -- & -- & -- & -- & -- \\
			GARG-AML Dir. Tree    & 0.000 & 0.000 & 50.6 & 0.011 & \underline{0.002} & \underline{0.000} & \underline{0.000} & 50.0 & 0.000 & \underline{0.000} & -- & -- & -- & -- & -- \\
			GARG-AML Dir. Boost   & \underline{0.000} & 0.000 & \underline{65.1} & \underline{0.021} & 0.000 & \underline{0.000} & \underline{0.000} & \underline{67.1} & 0.001 & \underline{0.000} & -- & -- & -- & -- & -- \\
			GraphSAGE (topology)  & 0.000 & 0.001 & 87.7 & 0.109 & 0.005 & 0.000 & 0.000 & 96.8 & 0.005 & 0.000 & -- & -- & -- & -- & -- \\
			GraphSAGE (+ attributes) & \textbf{0.006} & \textbf{0.012} & \textbf{98.6} & \textbf{16.385} & \textbf{0.087} & \textbf{0.001} & \textbf{0.002} & \textbf{99.8} & \textbf{0.098} & \textbf{0.001} & -- & -- & -- & -- & -- \\
			\midrule
			\multicolumn{16}{@{}l}{\textit{Scatter-gather}} \\
			\midrule
			FlowScope             & 0.000 & 0.000 & 50.0 & 0.015 & 0.000 & \textbf{\underline{0.000}} & \textbf{\underline{0.000}} & \textbf{\underline{50.0}} & 0.000 & \textbf{\underline{0.000}} & -- & -- & -- & -- & -- \\
			GARG-AML Undirected   & 0.000 & 0.000 & 58.8 & 0.018 & 0.000 & \textbf{\underline{0.000}} & \textbf{\underline{0.000}} & 41.6 & \textbf{\underline{0.000}} & \textbf{\underline{0.000}} & -- & -- & -- & -- & -- \\
			GARG-AML Directed     & 0.000 & 0.000 & 59.6 & 0.018 & 0.000 & \textbf{\underline{0.000}} & \textbf{\underline{0.000}} & 40.8 & 0.000 & \textbf{\underline{0.000}} & -- & -- & -- & -- & -- \\
			GARG-AML Undir. Tree  & 0.000 & 0.000 & 49.8 & 0.022 & \underline{0.005} & -- & -- & -- & -- & -- & -- & -- & -- & -- & -- \\
			GARG-AML Undir. Boost & \underline{0.002} & \underline{0.003} & 65.3 & 0.045 & 0.001 & -- & -- & -- & -- & -- & -- & -- & -- & -- & -- \\
			GARG-AML Dir. Tree    & 0.000 & 0.000 & 51.0 & 0.022 & 0.003 & -- & -- & -- & -- & -- & -- & -- & -- & -- & -- \\
			GARG-AML Dir. Boost   & 0.000 & 0.001 & \underline{71.3} & \underline{0.046} & 0.002 & -- & -- & -- & -- & -- & -- & -- & -- & -- & -- \\
			GraphSAGE (topology)  & 0.000 & 0.001 & 89.7 & 0.122 & 0.000 & -- & -- & -- & -- & -- & -- & -- & -- & -- & -- \\
			GraphSAGE (+ attributes) & \textbf{0.015} & \textbf{0.030} & \textbf{99.6} & \textbf{32.092} & \textbf{0.190} & -- & -- & -- & -- & -- & -- & -- & -- & -- & -- \\
			\bottomrule
		\end{tabular}
	\end{sidewaystable}
	
	The precision and F1-scores are often (close) to zero.
	This is a result of the choice of the widely used threshold value of 0.5.
	Almost no observations get a score of 0.5 or higher, meaning that almost all observations are classified as normal. 
	This highlights that careful tuning is needed for these thresholds, but also that comparing models on just a few thresholds can be problematic.
	Therefore, we decide to rely mostly on the threshold-independent metrics (AUC-ROC and AUC-PR) to form our conclusions.
	
	In most cases, GraphSAGE performs best. 
	The AUC-PR and P@1000 are substantially higher when the GraphSAGE model incorporates the node features. 
	This might indicate that the node features on their own already carry a large laundering signal.
	We note that the comparison is not purely like-for-like. 
	The GraphSAGE results on a given pattern are of the model trained on that specific pattern, while the GARG-AML scores are constructed purely on gather-scatter and scatter-gather patterns.
	GARG-AML can also not distinguish between the two, so any scatter-gather pattern picked up during the gather-scatter evaluation results in additional false positives. 
	This puts GraphSAGE at a small advantage from the start. 
	
	Among the GARG-AML variants, the boosted extensions give the best AUC-ROC at cut-offs 0.1 and 0.5, whereas at cut-off 0.9 the base undirected score is best. 
	GraphSAGE has the highest AUC-ROC overall in most cases.
	The undirected GARG-AML scores reach a moderate AUC-ROC for the general laundering cases (55--68\% at cut-offs 0.1 and 0.5), which include more patterns than just smurfing, but remain well below GraphSAGE with attributes (96--97\%). 
	
	For the lower cut-off values, we see a significant improvement of AUC-PR across the tree-based models. 
	This results from the models being able to better capture the interaction among a node's GARG-AML score and those of its neighbours. 
	This reduces the false positives compared to the base score, resulting in a much higher AUC-PR. 
	Therefore, on lower cut-offs, we conclude that the tree-based methods are better suited for smurfing detection. 
	
	For the highest cut-offs, the base undirected GARG-AML score reaches an AUC-ROC above $80\%$ for the full `Is Laundering' label and for several patterns. 
	In these cases, only clients are labelled that have $90\%$ or more of their transactions involved in money laundering. 
	Therefore, these patterns closely resemble pure smurfing, making them easily recognisable by GARG-AML. 
	
	The tree-based models, on the other hand, seem to perform poorly at these highest cut-offs. 
	The class imbalance (see also Table~\ref{tab:labels data HI-Small} and Table~\ref{tab:labels data LI-Large}) is too extreme in these cases, resulting in the model's inability to train properly. 
	This is most apparent on the LI-Large data at cut-off 0.9, where only the `Is Laundering' label has any positives (very few; see Table~\ref{tab:labels data LI-Large}) and the tree-based models reach an AUC-ROC of only $50$--$74\%$. 
	In this extreme case, the base undirected GARG-AML score (AUC-ROC $91.3\%$) is on par with GraphSAGE using topology only ($92.5\%$) and above GraphSAGE with attributes ($79.5\%$). 
	The undirected tree-based model has the highest AUC-PR and P@1000 ($0.068\%$ and $0.002$, versus $0.017\%$ and $0.000$ for GraphSAGE with attributes), even though its AUC-ROC is low. 
	Since these values rest on very few positive accounts, they should be read as indicative only. 
	We therefore conclude that GARG-AML remains competitive with GraphSAGE when labels are extremely scarce, not that it outperforms it.
	
	A general trend observed across the experiments is the counter-intuitive underperformance of the directed version of GARG-AML. 
	It seems that the definition of the score in Equation~\eqref{eq: score sender directed} misses an important part of smurfs. 
	The final score works well from the point of view of the sender, but is less suited when looking at the intermediate smurfing accounts. 
	We give more details on the inherent shortcoming of the directed GARG-AML score in \ref{app: limitations intermediate accounts}.
	Solving this, however, would result in the definition of a collection of rules trying to cover each case individually. 
	Future work should build upon the idea of the nine blocks, and combine these with more advanced methods to come to meaningful results. 
	
	\subsection{Sensitivity to the Pre-processing Step}
	\label{subsec:sensitivity}
	The main pre-processing step of GARG-AML involves applying clustering via the Louvain method and removing inter-community edges. 
	This might remove certain money laundering transactions if these are between nodes in different communities. 
	We analyse different resolution parameters and include an evaluation where hubs are removed instead. 
	Removal of hubs can be interpreted as a bank removing the largest clients (often big utility/insurance/retail companies), assuming that money laundering will most likely not go through these firms. 
	Our analysis covers the impact on community size (for the Louvain method only), edges removed, calculation time and predictive performance when the resolution parameter is varied, and when the pre-processing step is skipped altogether. 
	
	Table~\ref{tab:severance-patterns} shows that each method removes some money laundering transactions. 
	The hub removal removes around $10\%$ of all money laundering transactions, but these are almost exclusively among the `Not Classified' laundering transactions. 
	The Louvain method, on the other hand, affects around one in three laundering transactions and this across most patterns. 
	Only for the scatter-gather pattern, our main pattern of interest, do most transactions stay in the network. 
	
	\begin{table*}
		\centering
		\small
		\caption{Percentage of each pattern's edges discarded by the pre-processing on HI-Small; $n$ is the pattern's edge count. Hub removal deletes the $k$ highest-degree accounts and runs no community detection; $r$ is the Louvain resolution, $\dagger$ the setting used in the main experiments. Pattern rows over-count 87 edges belonging to attempts of two types.}
		\label{tab:severance-patterns}
		\setlength{\tabcolsep}{4pt}
		\begin{tabular}{@{}lrrrrrrrrr@{}}
			\toprule
			&   & \multicolumn{3}{c}{\makecell[cb]{Hub removal\\(top $k$)}} & \multicolumn{5}{c}{Louvain resolution ($r$)} \\
			\cmidrule(lr){3-5}\cmidrule(l){6-10}
			Pattern & $n$ & 5 & 10 & 100 & 1 & 5 & 10$^{\dagger}$ & 20 & 50 \\
			\midrule
			\multicolumn{10}{@{}l}{\emph{GARG-AML targets}} \\
			Gather-scatter & 661 & 0.0 & 0.0 & 0.0 & 49.2 & 58.7 & 57.0 & 58.4 & 60.7 \\
			Scatter-gather & 601 & 0.0 & 0.0 & 0.0 & 22.5 & 9.7 & 4.3 & 1.7 & 0.7 \\
			\midrule
			\multicolumn{10}{@{}l}{\emph{Other injected patterns}} \\
			Fan-out & 331 & 0.0 & 0.0 & 0.0 & 45.3 & 48.9 & 48.3 & 50.8 & 48.3 \\
			Fan-in & 309 & 0.0 & 0.0 & 0.0 & 47.9 & 55.0 & 55.0 & 57.9 & 59.2 \\
			Cycle & 265 & 0.0 & 0.0 & 0.8 & 36.2 & 39.6 & 35.1 & 27.9 & 29.4 \\
			Bipartite & 252 & 0.0 & 0.0 & 0.0 & 29.0 & 32.9 & 32.5 & 30.6 & 29.0 \\
			Stack & 446 & 0.0 & 0.0 & 0.0 & 35.2 & 42.2 & 37.2 & 37.0 & 37.0 \\
			Random & 181 & 0.0 & 0.0 & 0.0 & 38.7 & 44.2 & 33.7 & 36.5 & 37.0 \\
			\midrule
			Not Classified & 1,930 & 23.6 & 28.9 & 32.5 & 4.6 & 19.9 & 24.4 & 29.4 & 34.9 \\
			\midrule
			\textbf{All flagged edges} & 4,889 & 9.3 & 11.4 & 12.9 & 25.4 & 33.1 & 32.8 & 34.6 & 36.9 \\
			All graph edges & 644,246 & 4.5 & 5.5 & 7.1 & 6.0 & 19.6 & 22.8 & 28.6 & 31.6 \\
			\bottomrule
		\end{tabular}
	\end{table*}
	
	We analyse the difference in AUC-ROC and AUC-PR across these configurations. 
	Given that the value of the AUC-PR depends on the prevalence of the labels (a random model has AUC-PR equal to the relative number of positive labels), we divide the AUC-PR by the label prevalence to have a fair comparison. 
	Otherwise, removing nodes would change the AUC-PR values even if the model performance stays stable. 
	
\begin{figure*}
		\centering
		\includegraphics[width=0.9\linewidth]{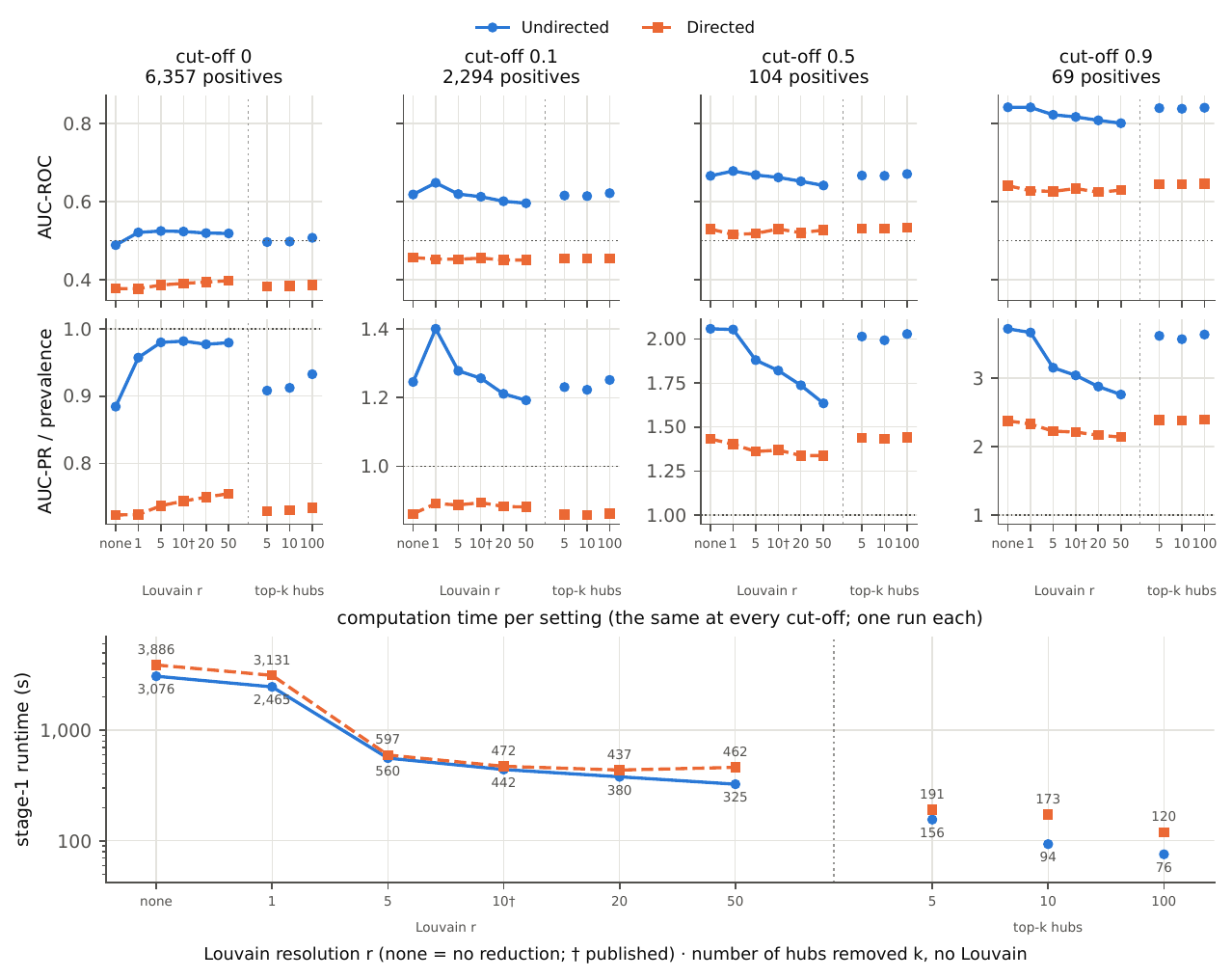}
		\caption{Sensitivity of the GARG-AML score on HI-Small to the pre-processing: AUC-ROC (top) and AUC-PR relative to prevalence (middle) at four label cut-offs, and stage-1 runtime in seconds (bottom; one run per setting, independent of the cut-off). Undirected in blue, directed in orange. ``none'' is no Louvain reduction, numbers are Louvain resolutions († published, $r=10$); right of the dashed line, the top-$k$ hubs are removed instead. No model is fitted. Dotted lines mark a random ranking.}
		\label{fig:sweep-score}
	\end{figure*}

	The effect on performance is given in Figure~\ref{fig:sweep-score} for the HI-Small dataset. 
	The patterns are different for low cut-off and high cut-off levels. 
	Therefore, the results show that there is no single strategy best for all situations. 
	
	At lower cut-off levels, i.e., a small portion of a node's transactions involved in money laundering leads to a money laundering label, the performance is in general higher in networks where clustering is applied. 
	At cut-off 0, higher resolution (smaller communities) is correlated with higher performance. 
	In this case, even a single transaction results in the node being labelled. 
	The other transactions seem to be adding noise to the data. 
	So, removing these and only keeping those within a small, close-knit community seems to result in much clearer laundering patterns. 
	
	At higher cut-off levels, the best models are those based on the full network or the networks where only a few hubs have been deleted. 
	Labels are only assigned if most of the transactions are involved in money laundering. 
	Removing transactions across communities gets rid of important information in these cases. 
	
	The main trade-off to be made is that lower resolution limits, i.e., larger communities, or fewer deleted hubs come at the cost of longer calculation times. 
	We see that removing hubs is the most effective in reducing calculation time, even though the number of removed edges is limited. 
	Increasing the resolution limits, on the other hand, reduces the calculation time dramatically at first, but these improvements flatten out past a resolution limit of 10. 
	
	\section{Limitations}
	\label{sec: limitations}
	\subsection{Limitations in the Data}
\begin{figure}
		\centering
		\includegraphics[width=0.6\linewidth]{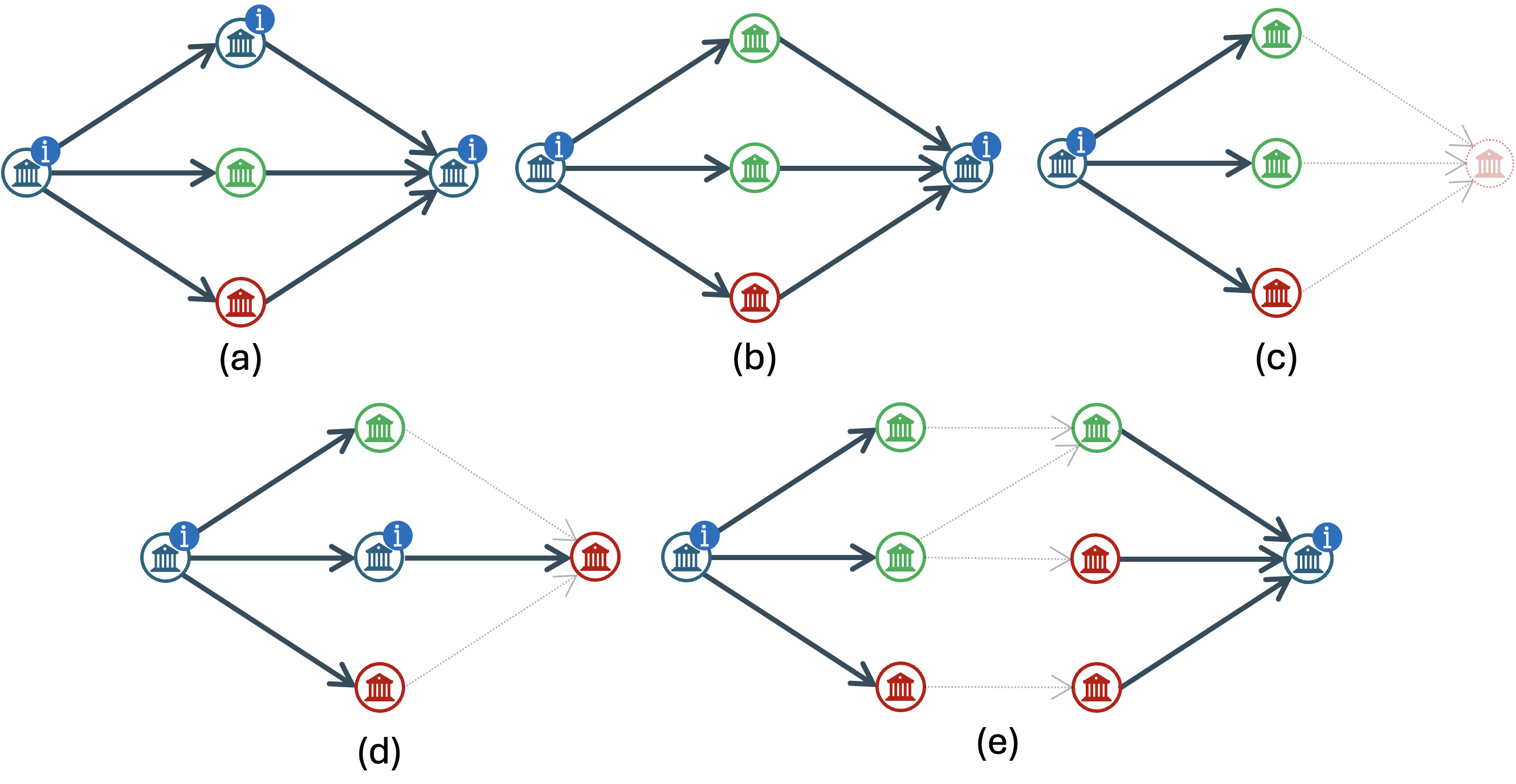}
		\caption{Smurfing over multiple banks. The (i)-icon indicates that the main (blue) bank has all client information about the corresponding account. The dotted lines indicate that the bank is unaware of these transactions/accounts.}
		\label{fig:limitationsData}
	\end{figure}

	GARG-AML assumes full visibility of all transactions, which is rarely the case in practice~\citep{altman2024realistic,tariq2023topologyagnostic,dumitrescu2022anomaly}. 
	Banks only have access to transactions that involve their own clients, and the sender and receiver account numbers. 
	Transactions between two external accounts remain hidden, which makes second-order neighbourhoods incomplete (see Figure~\ref{fig:limitationsData}). 
	This can obscure smurfing structures or lead to incorrect classification. 
	\ref{app:limited view} reports a sensitivity analysis that mimics this limited view on transactions: a single bank sees on average only 7\% of its clients' second-order neighbourhoods and shows no reliable detection performance, whereas pooling the 50 largest simulated banks (42\% visible on average) retains the (modest) performance obtained on the full network.
	
	Initiatives are needed that allow linking transactions on an inter-bank and international level. 
	However, sharing all transaction data is often not possible because of privacy and/or commercial reasons. 
	Federated learning~\citep{pmlr-v54-mcmahan17a} could provide a solution, albeit adapted to network data where links are shared among parties.
	
	\subsection{Limitations in the Methodology}
	There are three main limitations to the presented methodology. 
	The first concerns the use of only one layer of smurfs. 
	It is likely that criminal organisations utilise multiple layers of smurfs to launder their money. 
	While multi-layered schemes exist, they impose a deterrent risk-reward trade-off: bypassing the core organisation increases the risk of funds disappearing and requires greater trust in intermediaries. 
	Furthermore, incremental smurf fees reduce the total illicit reward. 
	Consequently, single-layer analysis already captures a meaningful portion of standard smurfing operations. 
	
	The second limitation is that relying on a single score is restrictive; it may miss complex, real-world smurfing operations~\citep{doi:10.1177/14773708251323349,10.1108/JFC-02-2020-0028,leukfeldt2015cyber}. 
	An improvement would already be to use business rules on the values of the densities and sizes of the blocks. 
	However, this requires extensive expert knowledge on diverse smurfing topologies.
	This is outside the scope of this paper. 
	
	The third methodological limitation is that GARG-AML uses unweighted graph structures, ignoring transaction volume and timing. 
	While these are simplifications, the resulting smurfing score remains interpretable and useful in downstream AML systems. 
	Timing issues can be addressed by restricting analysis to shorter time windows, applying sliding windows, or comparing scores across time.
	
	\section{Conclusion}
	\label{sec: conclusion}
	This paper introduces Graph-Aided Risk Grading for Anti-Money Laundering (GARG-AML), which is a novel approach for detecting smurfing by quantifying the structural similarity of nodes’ neighbourhoods to known money laundering patterns via density-based features in the adjacency matrix.
	
	We validated GARG-AML on several synthetic datasets and compared it to state-of-the-art detection methods. 
	The results of the experiments show that GARG-AML achieves strong detection power, with competitive performance, and excellent scalability for large networks. 
	Unlike approaches that are based on complex optimisation techniques or machine learning models, GARG-AML relies on domain expertise and straightforward density-based calculations. 
	We also showed that in cases where labels are very scarce, GARG-AML is competitive with GNNs.
	
	While GARG-AML is not a complete solution for smurfing detection, it offers two key advantages over existing methods: (1) it produces interpretable features, either as separate densities or a single composite score, that can be seamlessly integrated into AML systems, and (2) it is computationally efficient, with parallel and memory-light calculations, thus enabling practical deployment in real-world settings.
	
	These key properties allow financial institutions to adopt our method easily and strengthen their role as capable guardians. 
	Blocking simple smurfing tactics forces criminals into riskier, less profitable schemes. 
	Ultimately, if the cost and danger of laundering money outweigh the rewards, regulators can effectively put criminal syndicates out of business.
	
	Future work will explore (i) treating individual densities as features in classification models to identify distinct smurfing and money mule topologies, (ii) extending the methodology to incorporate transaction timing and volume possibly by integrating it as a baseline in the SMoTeF pipeline, as well as generalising it to other anomalous transaction patterns, and (iii) constructing a federated learning framework to alleviate the limited visibility of network data across banks. 
	
	Detecting smurfs is also an important research area for network security and intrusion detection~\citep{YANG2009115, shadrooh2024smotef}. 
	Further research could analyse whether similar featurisation of the patterns is possible for denial of service~(DoS) or similar attacks. 
	
	\section*{Acknowledgements}
	This work was supported by the Research Foundation – Flanders (FWO) [grant number 1SHEN24N and G015020N] and by the BNP Paribas Fortis Chair in Fraud Analytics. The resources and services used in this work were provided by the VSC (Flemish Supercomputer Center), funded by the Research Foundation - Flanders (FWO) and the Flemish Government.
	
	\section*{Author Contributions}
	\noindent\textbf{Bruno Deprez:} Conceptualization, Data Curation, Methodology, Formal Analysis, Software, Visualization, Writing---Original Draft.\\
	\textbf{Bart Baesens:} Conceptualization, Supervision, Writing---Review \& Editing.\\
	\textbf{Tim Verdonck:} Conceptualization, Supervision, Writing---Review \& Editing.\\
	\textbf{Wouter Verbeke:} Conceptualization, Project administration, Supervision, Writing---Review \& Editing.
	
	\section*{Declaration of Competing Interest}
	The authors declare that they have no known competing financial interests or personal relationships that could have appeared to influence the work reported in this paper.
	
	\bibliographystyle{plainnat} 
	\bibliography{cas-refs}
	
	\clearpage
	
	\appendix
	\setcounter{table}{0}
	\setcounter{figure}{0}
	\counterwithin{table}{section}
	\counterwithin{figure}{section}
	\renewcommand{\thefigure}{\Alph{section}.\arabic{figure}}
	\renewcommand{\thetable}{\Alph{section}.\arabic{table}}
	\section{Illustration GARG-AML}
	\label{app:example}
\begin{figure}[h]
		\centering
		\includegraphics[width=0.6\linewidth]{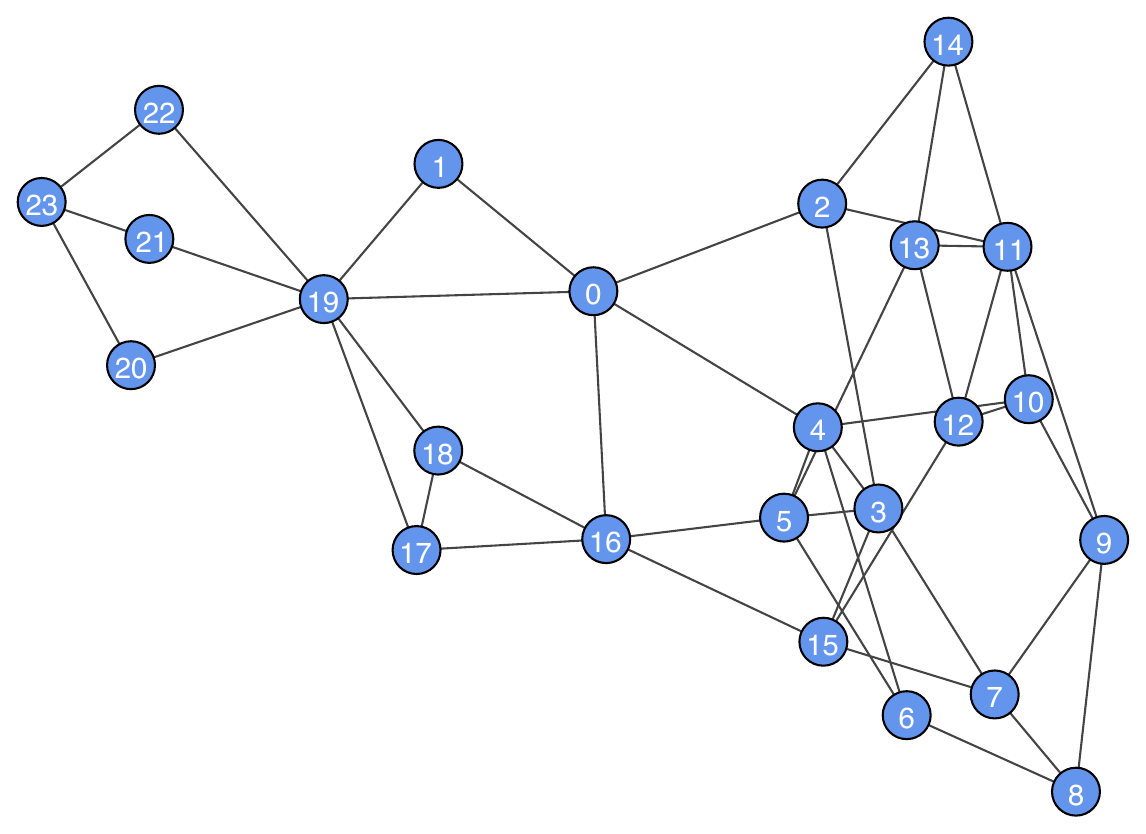}
		\caption{The undirected network. Nodes 20 to 23 are part of the injected smurfing pattern. Node 19 is the destination node.}
		\label{fig:toy full}
	\end{figure}

	We demonstrate GARG-AML as described in Section~\ref{sec: methodology}. 
	We start with a Watts-Strogatz network containing 20 nodes, each connecting to the next two nodes (initial degree equal to four), with rewiring probability $20\%$. 
	We inject a smurfing pattern, where money originates at a new node (23) and gets transferred to an existing node (19) in the network using three mule accounts (20, 21, 22). 
	The full network is visualised in Figure~\ref{fig:toy full}. 
	
	We calculate the GARG-AML scores for two nodes as an illustration.
	We select node 23 because its second-order neighbourhood constitutes a pure smurfing pattern. We also select node 8 to show the impact of the neighbours' interactions on the final score. 
	
	\subsection{Undirected Calculations}
	\label{app:example undir}
\begin{figure}
		\centering
		\includegraphics[width=0.6\linewidth]{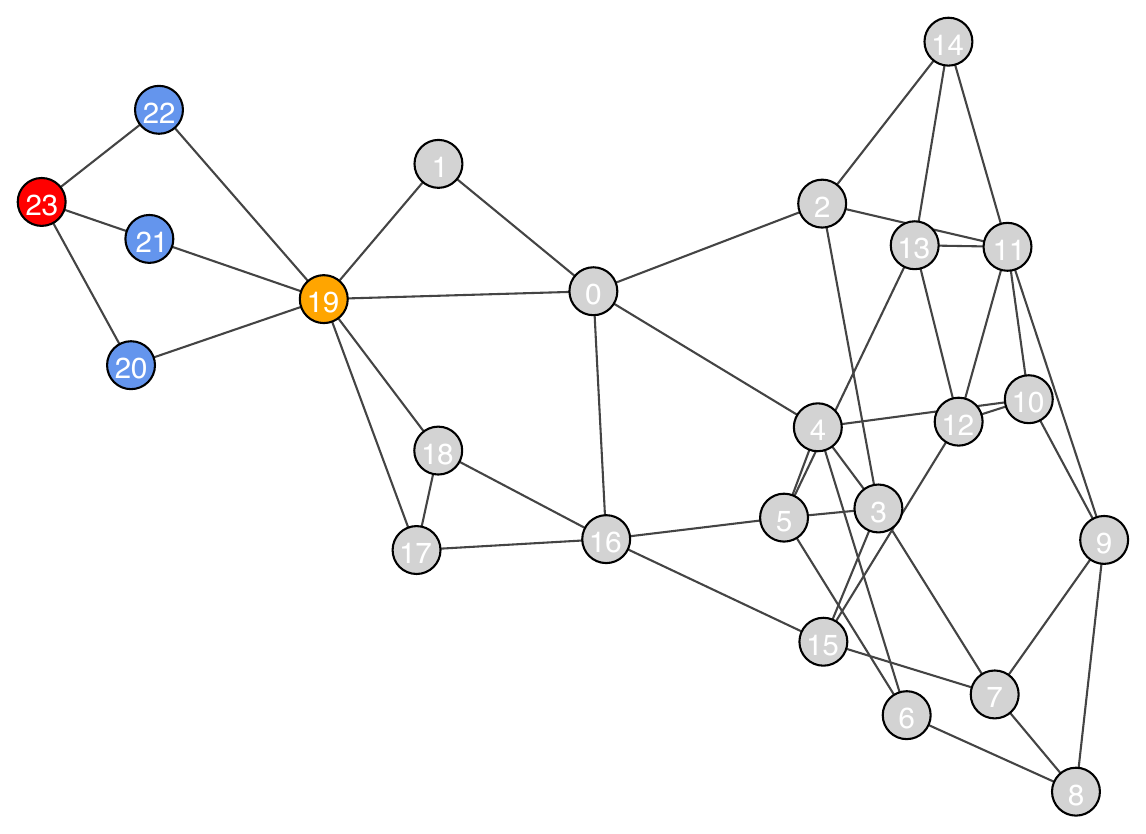}
		\caption{Visualisation of the second-order neighbourhood of node 23 (presented in red). Nodes at distance one are presented in blue, nodes at distance two are presented in orange. All other nodes are given in grey.}
		\label{fig:toy smurfing}
	\end{figure}

	We start with the smurfing pattern using node 23. 
	Its second-order neighbourhood is visualised in Figure~\ref{fig:toy smurfing}.
	The adjacency matrix, with the blocks, is as follows:
	\begin{equation}
		\begin{array}{r}
			23 \\ 19 \\ 20 \\ 21 \\ 22
		\end{array}
		\begin{pmatrix}
			\textcolor{red}{0} & \textcolor{red}{0} & \textcolor{green}{1} &\textcolor{green}{1} &\textcolor{green}{1}\\
			\textcolor{red}{0} & \textcolor{red}{0} & 1 &1 &1\\
			\textcolor{green}{1} & 1 & \textcolor{red}{0} &0 &0 \\
			\textcolor{green}{1} & 1 & 0 &\textcolor{red}{0} &0 \\
			\textcolor{green}{1} & 1 & 0 &0 &\textcolor{red}{0} \\
		\end{pmatrix}
		\label{app-eq:ex_mat_smurfing}
	\end{equation}
	The elements that should be zero are indicated in red. 
	Those elements that are certainly equal to one are indicated in green. 
	
	We take the upper-left block as block 1, the upper-right as block 2 (can also be the lower-left, since the matrix is symmetric), and the lower-right as block 3. Using Equations~\eqref{eq:score1}--\eqref{eq:l3}, this results in the following:
	\begin{eqnarray}
		\text{score}_1 &=& 0 \nonumber \\
		\text{score}_2 &=& 1 \nonumber \\
		\text{score}_3 &=& 0 \nonumber \\
		l_1 &=& 0 \nonumber \\
		l_2 &=& 3 \nonumber \\
		l_3 &=& 6 \nonumber \\
		\text{score}_\text{GARG-AML} & = & \text{score}_2 - \frac{l_1\cdot\text{score}_1 + l_3\cdot\text{score}_3}{l_1+l_3} \nonumber \\
		& =& 1
	\end{eqnarray}
	
\begin{figure}
		\centering
		\includegraphics[width=0.6\linewidth]{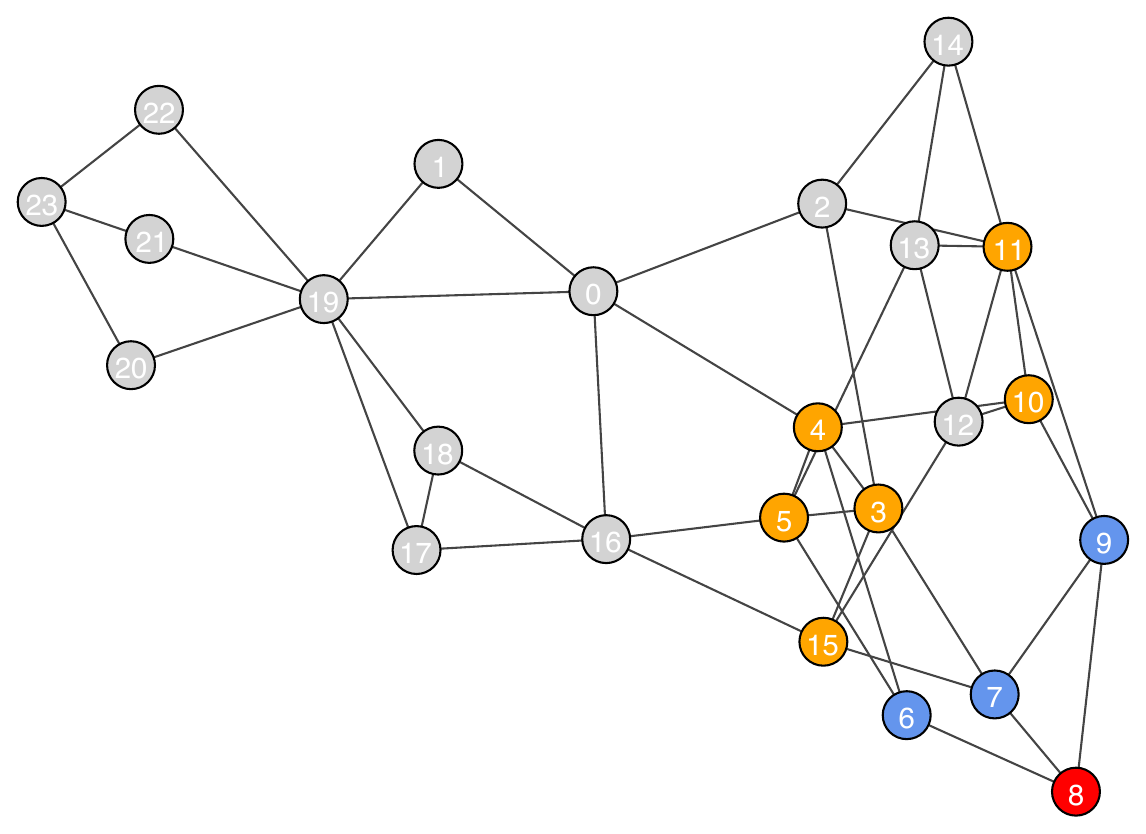}
		\caption{Visualisation of the second-order neighbourhood of node 8 (presented in red). Nodes at distance one are presented in blue, and nodes at distance two are presented in orange. All other nodes are given in grey.}
		\label{fig:toy other}
	\end{figure}

	The second-order neighbourhood of node 8 is shown in Figure~\ref{fig:toy other}.
	We observe an interaction between neighbours at distance one and two, something we do not have for a pure smurfing pattern.
	These are reflected in the adjacency matrix in Equation~\eqref{app-eq:ex_mat_no_smurfing}. 
	The first row contains the node to be scored, node 8. 
	The following rows represent the nodes at distance two, i.e., 3, 4, 5, 10, 11 and 15. 
	The final rows represent the nodes at distance one, i.e., 6, 7 and 9. 
	
	\begin{equation}
		\begin{array}{r}
			8 \\ 3 \\ 4 \\ 5 \\ 10 \\ 11 \\ 15 \\ 6 \\ 7 \\ 9
		\end{array}
		\begin{pmatrix}
			\textcolor{red}{0} & \textcolor{red}{0} & \textcolor{red}{0} & \textcolor{red}{0} & \textcolor{red}{0} & \textcolor{red}{0} & \textcolor{red}{0} & \textcolor{green}{1} & \textcolor{green}{1} & \textcolor{green}{1}\\
			\textcolor{red}{0} & \textcolor{red}{0} & 1 & 1 & 0 & 0 & 1 & 0 & 0 & 1\\
			\textcolor{red}{0} & 1 & \textcolor{red}{0} & 1 & 1 & 0 & 0 & 0 & 1 & 0\\
			\textcolor{red}{0} & 1 & 1 & \textcolor{red}{0} & 0 & 0 & 0 & 0 & 1 & 0\\
			\textcolor{red}{0} & 0 & 1 & 0 & \textcolor{red}{0} & 1 & 0 & 1 & 0 & 0\\
			\textcolor{red}{0} & 0 & 0 & 0 & 1 & \textcolor{red}{0} & 0 & 1 & 0 & 0\\
			\textcolor{red}{0} & 1 & 0 & 0 & 0 & 0 & \textcolor{red}{0} & 0 & 0 & 1\\
			\textcolor{green}{1} & 0 & 0 & 0 & 1 & 1 & 0 & \textcolor{red}{0} & 0 & 1\\
			\textcolor{green}{1} & 0 & 1 & 1 & 0 & 0 & 0 & 0 & \textcolor{red}{0} & 0\\
			\textcolor{green}{1} & 1 & 0 & 0 & 0 & 0 & 1 & 1 & 0 & \textcolor{red}{0}
		\end{pmatrix}
		\label{app-eq:ex_mat_no_smurfing}
	\end{equation}
	
	This results in the following scores: 
	\begin{eqnarray}
		\text{score}_1 &=& 0.4 \nonumber \\
		\text{score}_2 &=& 0.33 \nonumber \\
		\text{score}_3 &=& 0.33 \nonumber \\
		l_1 &=& 30 \nonumber \\
		l_2 &=& 18 \nonumber \\
		l_3 &=& 6 \nonumber \\
		\text{score}_\text{GARG-AML} & = & \text{score}_2 - \frac{l_1\cdot\text{score}_1 + l_3\cdot\text{score}_3}{l_1+l_3} \nonumber \\
		& =& -0.0556
	\end{eqnarray}
	
	\subsection{Directed Calculations}
	\label{app:example dir}
\begin{figure}
		\centering
		\includegraphics[width=0.6\linewidth]{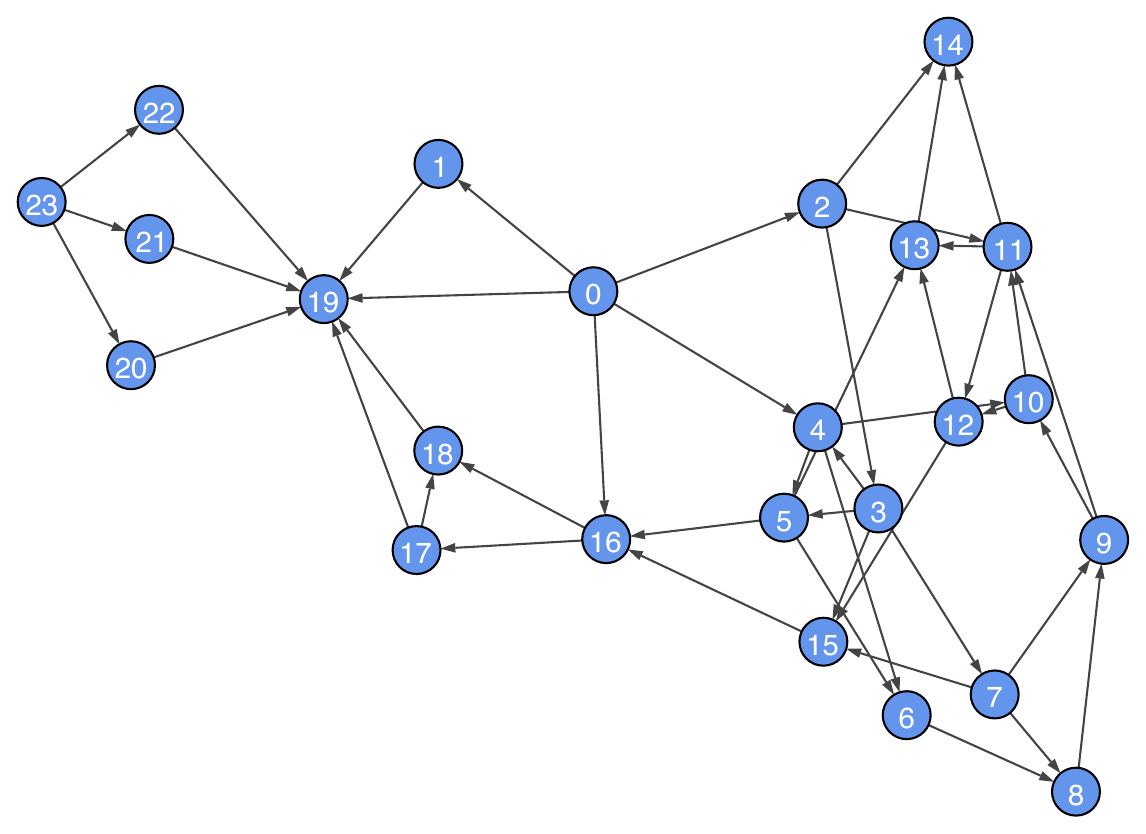}
		\caption{The directed network. Nodes 20 to 23 are part of the injected smurfing pattern, and with node 19 the destination node.}
		\label{fig:toy full dir}
	\end{figure}

	The directed network is illustrated in Figure~\ref{fig:toy full dir}. 
	We calculate the scores for node 23 and 8. 
	
\begin{figure}
		\centering
		\includegraphics[width=0.6\linewidth]{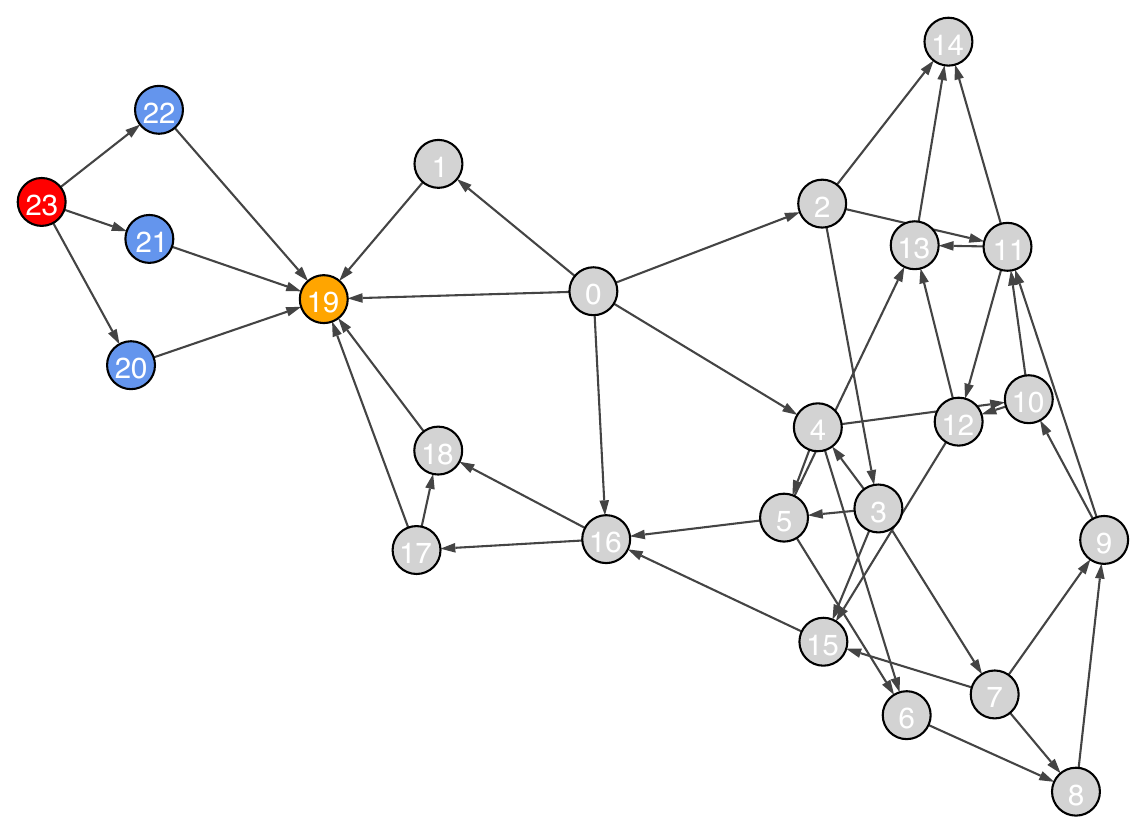}
		\caption{Visualisation of the second-order neighbourhood of node 23 (presented in red). Nodes at distance one are presented in blue, and nodes at distance two are presented in orange. All other nodes are given in grey.}
		\label{fig:toy smurfing dir}
	\end{figure}

	The directed second-order neighbourhood for node 23 is shown in Figure~\ref{fig:toy smurfing dir}, and the corresponding adjacency matrix is provided in Equation~\eqref{app-eq:ex_mat_smurfing_dir}.
	Note that we now have a different ordering of the nodes, using level 0 (node 23), level 1 (nodes 20, 21, 22) and level 2 (node 19), resulting in nine blocks. 
	
	\begin{equation}
		\begin{array}{r}
			23 \\ 20 \\ 21 \\ 22 \\ 19
		\end{array}
		\begin{pmatrix}
			\textcolor{red}{0} & 1 & 1 & 1 & \textcolor{red}{0} \\
			0 & \textcolor{red}{0} & 0 & 0 & 1 \\
			0 & 0 & \textcolor{red}{0} & 0 & 1 \\
			0 & 0 & 0 & \textcolor{red}{0} & 1 \\
			\textcolor{red}{0} & 0 & 0 & 0 & \textcolor{red}{0}
		\end{pmatrix}
		\label{app-eq:ex_mat_smurfing_dir}
	\end{equation}
	
	The density of the nine blocks is calculated for the final score. 
	We use the matrix representation of Equation~\eqref{eq:blocks dir} to represent the densities.
	This results in:
	\begin{equation}{
			\left( \begin{array}{ccc}
				0 & 1 & 0 \\
				0 & 0 & 1 \\
				0 & 0 & 0
			\end{array}\right)}
	\end{equation}
	
	The score for node 23 is calculated as
	\begin{eqnarray}
		\text{score}_\text{GARG-AML} & = & \text{mean}(\text{score}_{01}, \text{score}_{12})   \\
		& &- \text{mean}(\text{score}_{00}, \text{score}_{02}, \text{score}_{10}, \nonumber \\ && \text{score}_{11}, \text{score}_{20}, \text{score}_{21}, \text{score}_{22}) \nonumber \\
		& =& \text{mean}(1,1) - \text{mean}(0,0,0,0,0,0,0) \nonumber\\
		&=& 1.
	\end{eqnarray}
	
\begin{figure}
		\centering
		\includegraphics[width=0.6\linewidth]{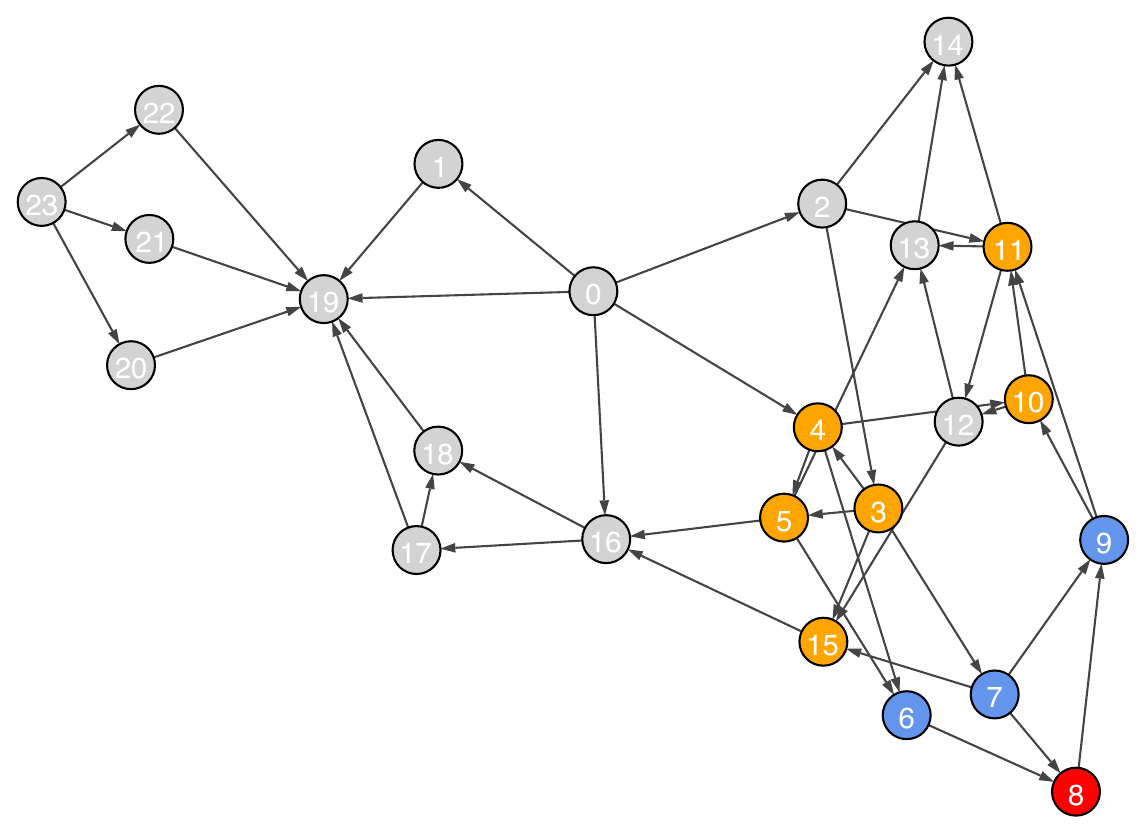}
		\caption{Visualisation of the second-order neighbourhood of node 8 (presented in red). Nodes at distance one are presented in blue, and nodes at distance two are presented in orange. All other nodes are given in grey.}
		\label{fig:toy other dir}
	\end{figure}

	We do the same for node 8. 
	Here, level 0 consists of nodes 8 and 15, level 1 consists of nodes 6, 7 and 9, and level 2 consists of nodes 3, 4, 5, 10 and 11. 
	Contrary to the second-order neighbourhood of node 23, we see in Figure~\ref{fig:toy other dir} that money flows in multiple directions, with nodes sending and receiving money to and from nodes across levels. 
	This indicates that node 8 is not part of a smurfing scheme. 
	The corresponding adjacency matrix is given in Equation~\eqref{app-eq:ex_mat_no_smurfing_dir}.
	
	\begin{equation}
		\begin{array}{r}
			8\\ 15\\ 9\\ 6\\ 7\\ 3\\ 4\\ 5\\ 10\\ 11
		\end{array}
		\begin{pmatrix}
			\textcolor{red}{0}& \textcolor{red}{0}& 1& 0& 0& \textcolor{red}{0}& \textcolor{red}{0}& \textcolor{red}{0}& \textcolor{red}{0}& \textcolor{red}{0}\\
			\textcolor{red}{0}& \textcolor{red}{0}& 0& 0& 0& 0& 0& 0& 0& 0\\
			0& 0& \textcolor{red}{0}& 0& 0& 0& 0& 0& 1& 1\\
			1& 0& 0& \textcolor{red}{0}& 0& 0& 0& 0& 0& 0\\
			1& 1& 1& 0& \textcolor{red}{0}& 0& 0& 0& 0& 0\\
			\textcolor{red}{0}& 1& 0& 0& 1& \textcolor{red}{0}& 1& 1& 0& 0\\
			\textcolor{red}{0}& 0& 0& 1& 0& 0& \textcolor{red}{0}& 1& 1& 0\\
			\textcolor{red}{0}& 0& 0& 1& 0& 0& 0& \textcolor{red}{0}& 0& 0\\
			\textcolor{red}{0}& 0& 0& 0& 0& 0& 0& 0& \textcolor{red}{0}& 1\\
			\textcolor{red}{0}& 0& 0& 0& 0& 0& 0& 0& 0& \textcolor{red}{0}
		\end{pmatrix}
		\label{app-eq:ex_mat_no_smurfing_dir}
	\end{equation}
	
	\begin{equation}{
			\left( \begin{array}{ccc}
				0 & 0.1667 & 0 \\
				0.5 & 0.1667 & 0.133 \\
				0.2 & 0.2 & 0.25
			\end{array}\right)}
	\end{equation}
	
	\begin{eqnarray}
		\text{score}_\text{GARG-AML} & = & \text{mean}(\text{score}_{01}, \text{score}_{12}) \nonumber \\
		& &- \text{mean}(\text{score}_{00}, \text{score}_{02}, \text{score}_{10}, \text{score}_{11}, \nonumber \\ &&\text{score}_{20}, \text{score}_{21}, \text{score}_{22}) \nonumber \\
		& =& \text{mean}(0.1667,0.133) \nonumber \\&&- \text{mean}(0,0,0.5,0.1667,0.2,0.2,0.25) \nonumber\\
		&=& 0.15 - 0.188 \nonumber \\
		&=& -0.038. \nonumber
	\end{eqnarray}
	We can, indeed, see that the score for node 8 is below 0. 
	Hence, we have no suspicion that this node is part of a smurfing scheme. 
	
	\subsection{Limitations for Intermediate Accounts}
	\label{app: limitations intermediate accounts}
\begin{figure}
		\centering
		\includegraphics[width=0.3\linewidth]{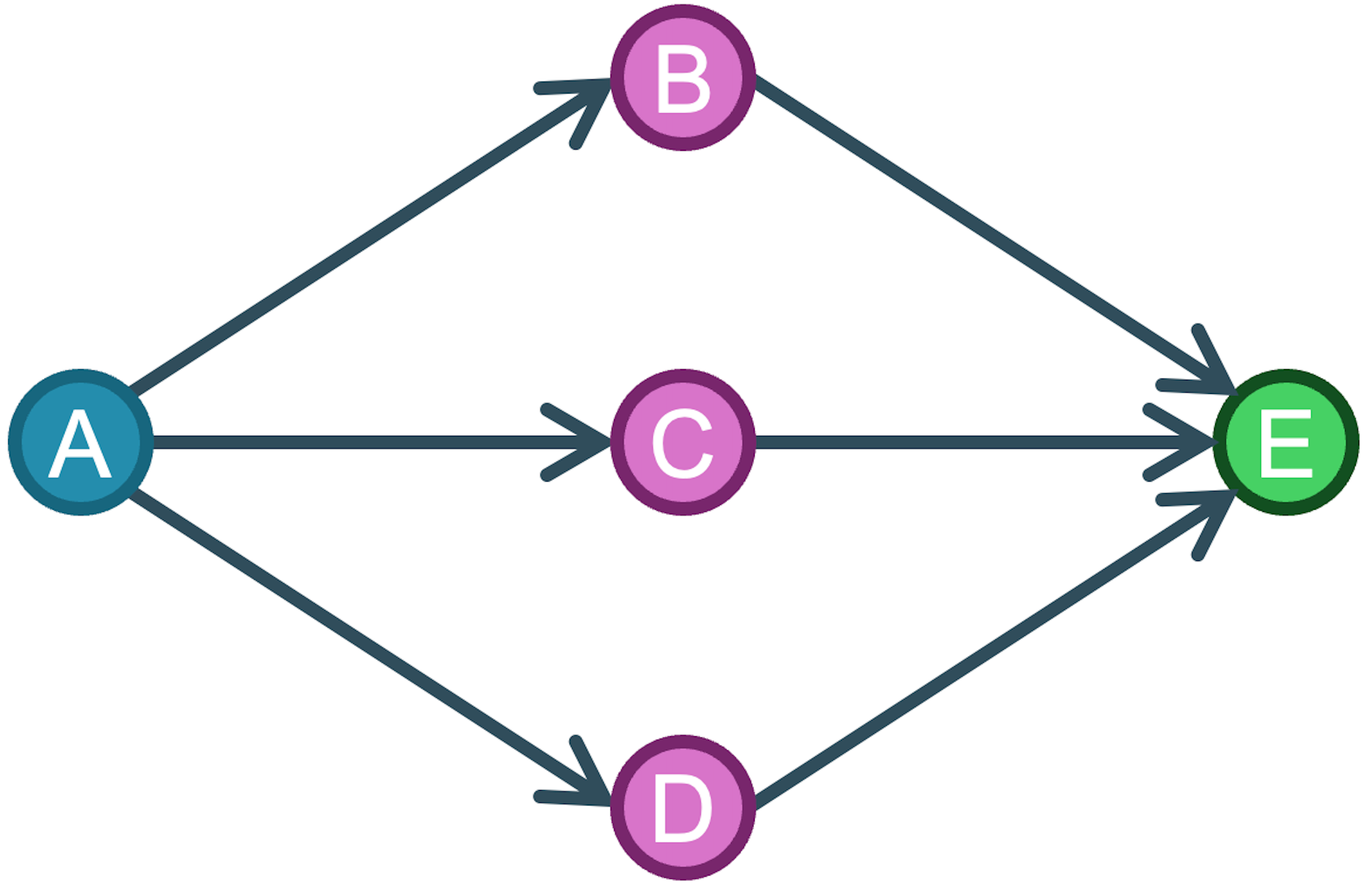}
		\caption{A pure directed smurfing pattern.}
		\label{fig-app:limitations intermediate}
	\end{figure}

	Section~\ref{sub:perf} illustrated the consistent underperformance of the directed scores. 
	We return to the original example from Section~\ref{subsec: gargaml directed}, and include it again in Figure~\ref{fig-app:limitations intermediate}.
	
	We illustrate the calculations from the perspective of node $B$. 
	Equation~\eqref{eq:layers} tells us that both nodes $C$ and $D$ are put at level 0. 
	Nodes $A$ and $E$ are first-order neighbours, so they are put at level 1. 
	Without looking at the adjacency matrix, the main problem is already apparent. 
	There are no nodes at level 2, meaning that their scores are defaulted to 0.
	
	We obtain the following adjacency and density matrix, which ultimately results in a score equal to 0 (weighted mean) or 0.18 (regular mean):
	\begin{equation}
		\begin{array}{r}
			B \\ C \\ D \\ A \\ E
		\end{array}
		\begin{pmatrix}
			\textcolor{red}{0} & 0 & 0 & 0 & 1\\
			0 & \textcolor{red}{0} & 0 & 0 & 1 \\
			0 & 0 & \textcolor{red}{0} & 0 & 1 \\
			1 & 1 & 1 & \textcolor{red}{0} & 0 \\
			0 & 0 & 0 & 0 & \textcolor{red}{0}
		\end{pmatrix}
	\end{equation}
	\begin{equation}{
			\left( \begin{array}{ccc}
				0 & 1/2 & 0 \\
				1/2 & 0 & 0 \\
				0 & 0 & 0
			\end{array}\right)}
	\end{equation}
	Hence, even with a perfect smurfing pattern, the intermediate accounts receive a low score. 
	As these form the majority of nodes involved in smurfing, this skews the results. 
	More research is needed to obtain more meaningful results from this particular representation. 
	
	\newpage
	\section{Results Synthetic Data}
	\label{app:synthetic}
\begin{figure*}
		\centering
		\includegraphics[width=0.9\linewidth]{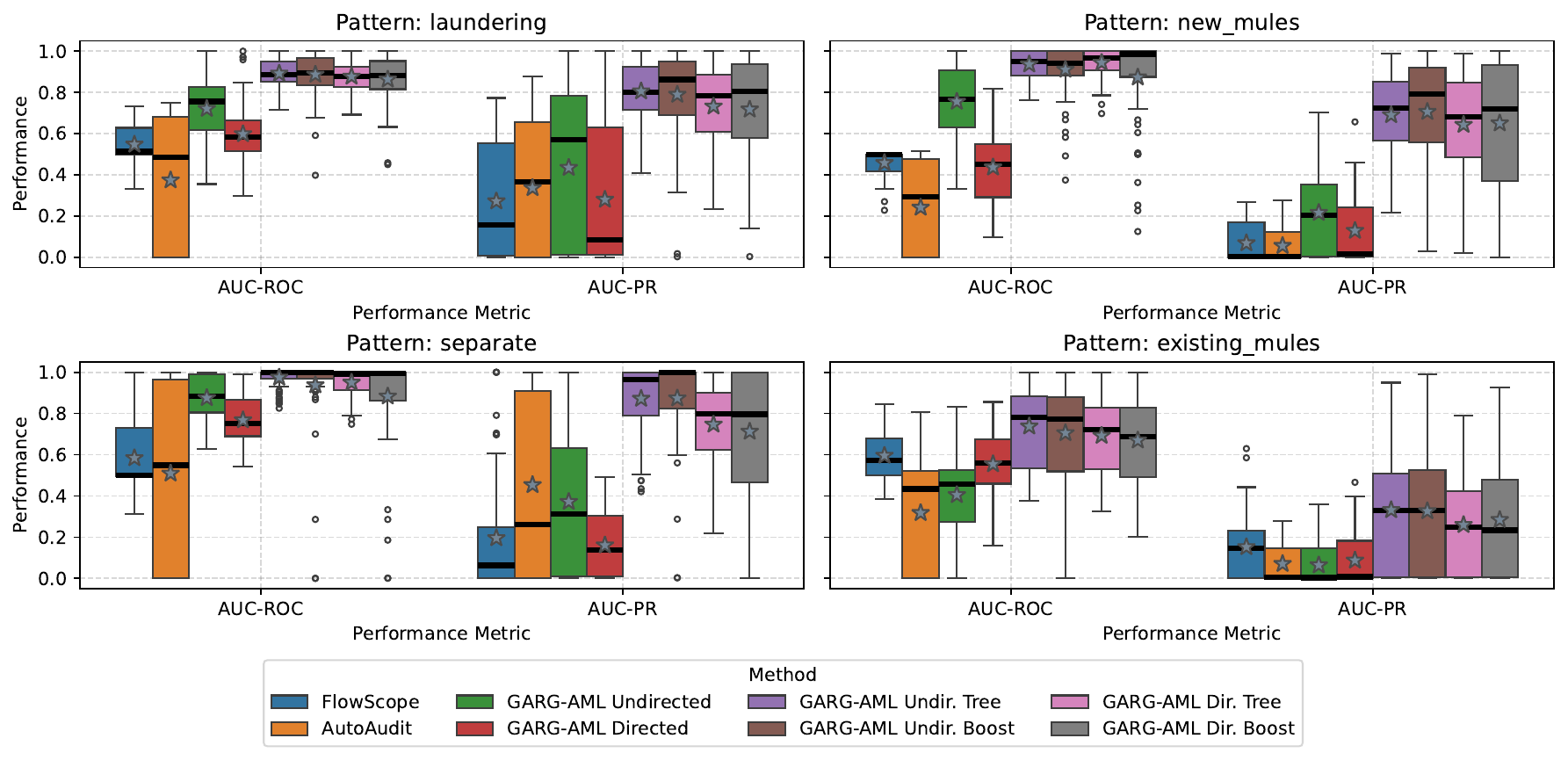}
		\caption{AUC-ROC and AUC-PR values of the compared methods across all synthetic datasets, split by money laundering pattern.}
		\label{fig-app:boxplot_perf_synth}
	\end{figure*}

\begin{figure*}
		\centering
		\includegraphics[width=0.9\linewidth]{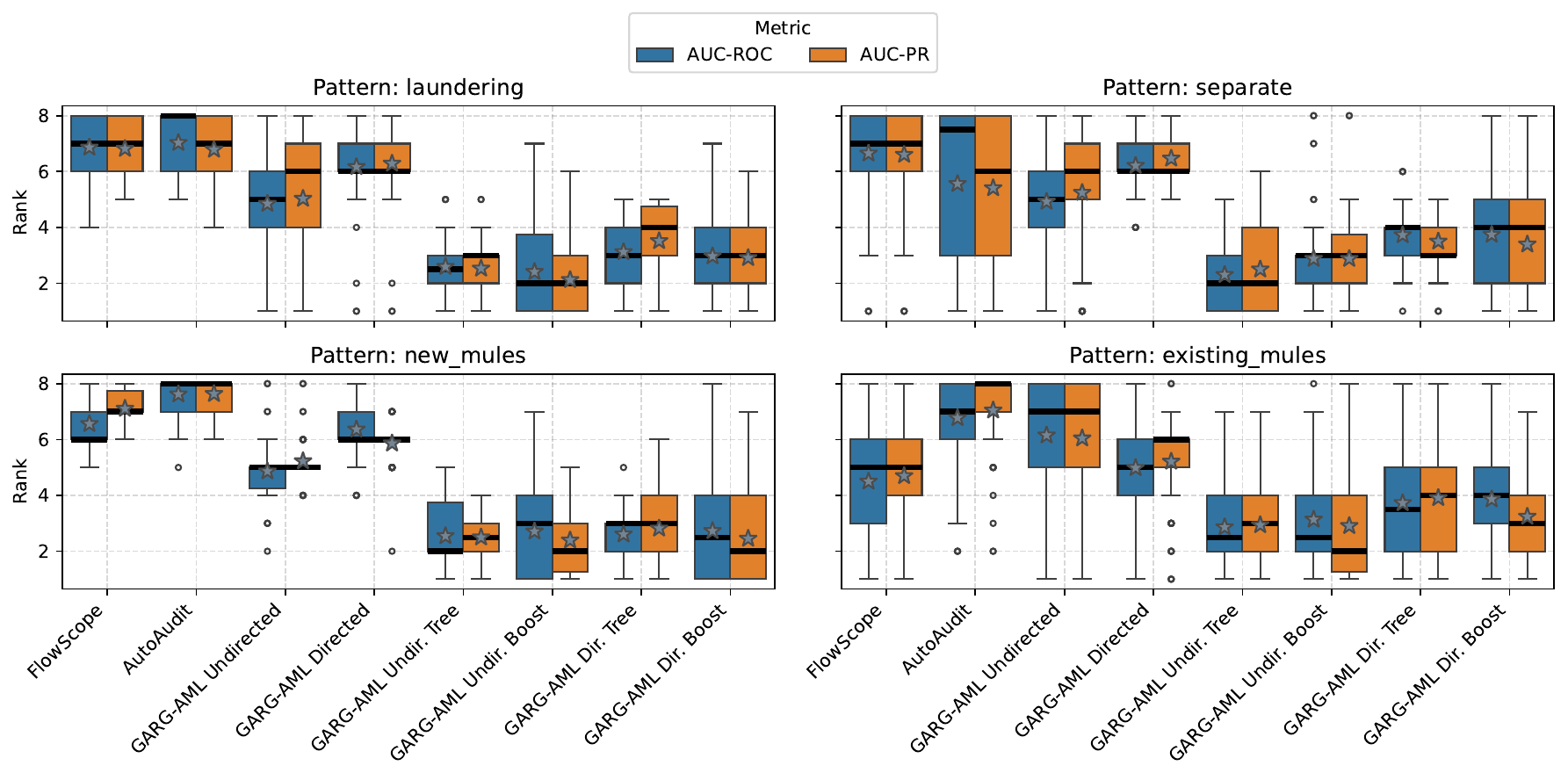}
		\caption{Ranks of the AUC-ROC and AUC-PR values of the compared methods across all synthetic datasets, split by money laundering pattern, where rank 1 indicates the best performance and rank 8 indicates the worst performance.}
		\label{fig-app:boxplot_rank_synth}
	\end{figure*}

	The results are presented according to the injected patterns. 
	These are provided in Figure~\ref{fig-app:boxplot_perf_synth}, with their corresponding ranks presented in Figure~\ref{fig-app:boxplot_rank_synth}.
	
	Figure~\ref{fig-app:boxplot_perf_synth} shows the distribution of AUC-ROC and AUC-PR values across the datasets. 
	The AUC-PR values are more spread out than the AUC-ROC values. 
	For a model with random label predictions, the AUC-ROC is always 0.5, regardless of the label distribution, whereas the AUC-PR equals the proportion of positive labels, which leads to the high variance.
	
	Figure~\ref{fig-app:boxplot_perf_synth} indicates that, on average, the undirected GARG-AML scores show competitive performance. 
	This performance is boosted when combining the GARG-AML scores with the tree-based learners.  
	In comparison, the performance of AutoAudit is less stable. Here, we set all the AUC-ROC and AUC-PR values to 0 when the method runs out of time or out of memory to allow for a full comparison of the ranks.
	
	Figure~\ref{fig-app:boxplot_rank_synth} shows the boxplots of the ranks of the methods (rank 1 means best scoring model, and rank 8 worst). 
	We can see more clearly that models combining the tree-based learners with the scores and summary statistics of the node's neighbours score better than the other methods, even when just applying a simple decision tree.
	
	\newpage
	\section{Results IBM Data}
	\label{app:full ibm}
\begin{figure*}
		\centering
		\includegraphics[width=\linewidth]{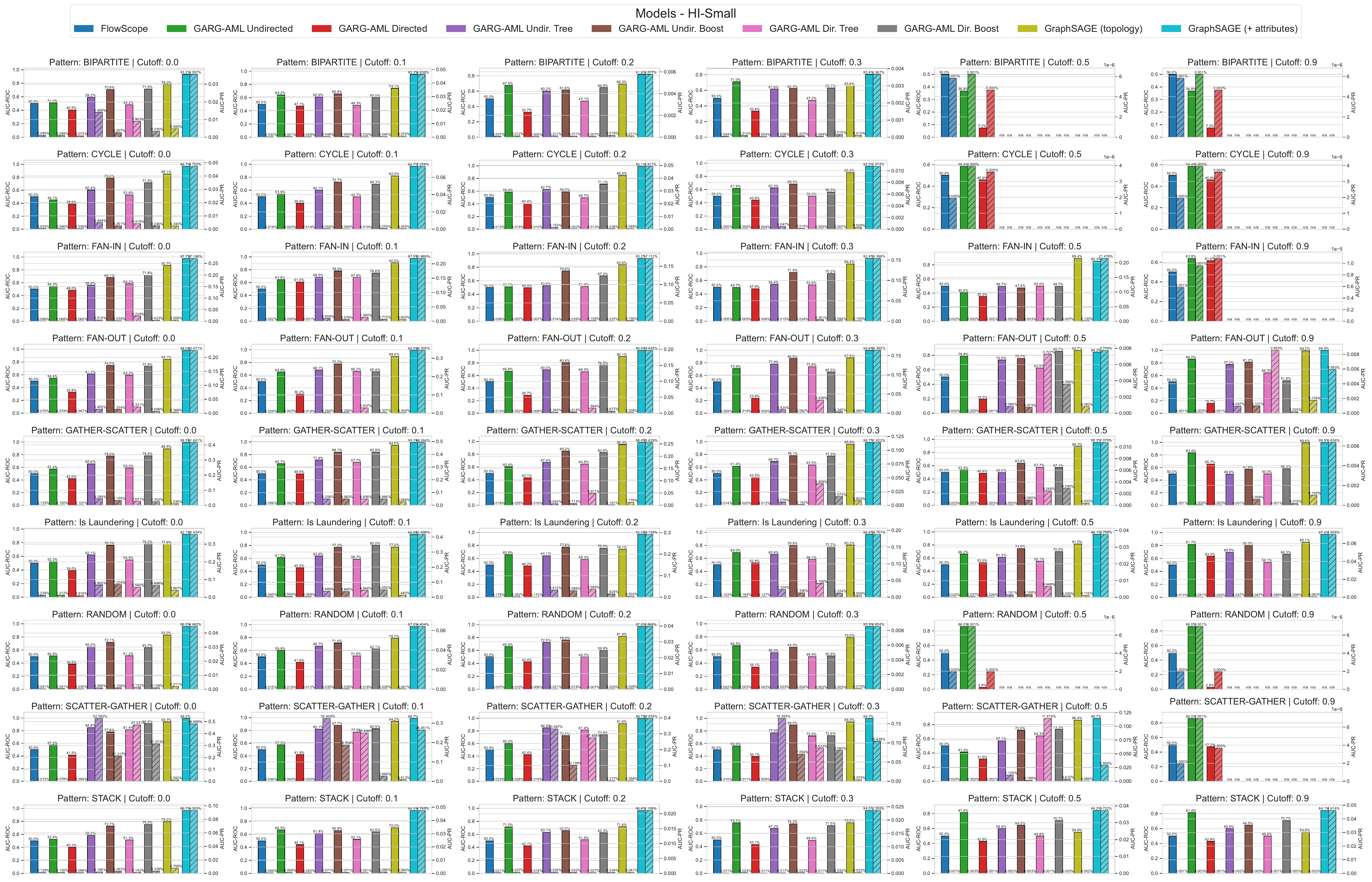}
		\caption{The AUC-ROC (solid bars) and AUC-PR (hatched bars) results on the HI-Small IBM dataset.}
		\label{fig-app:HISmall}
	\end{figure*}

\begin{figure*}
		\centering
		\includegraphics[width=\linewidth]{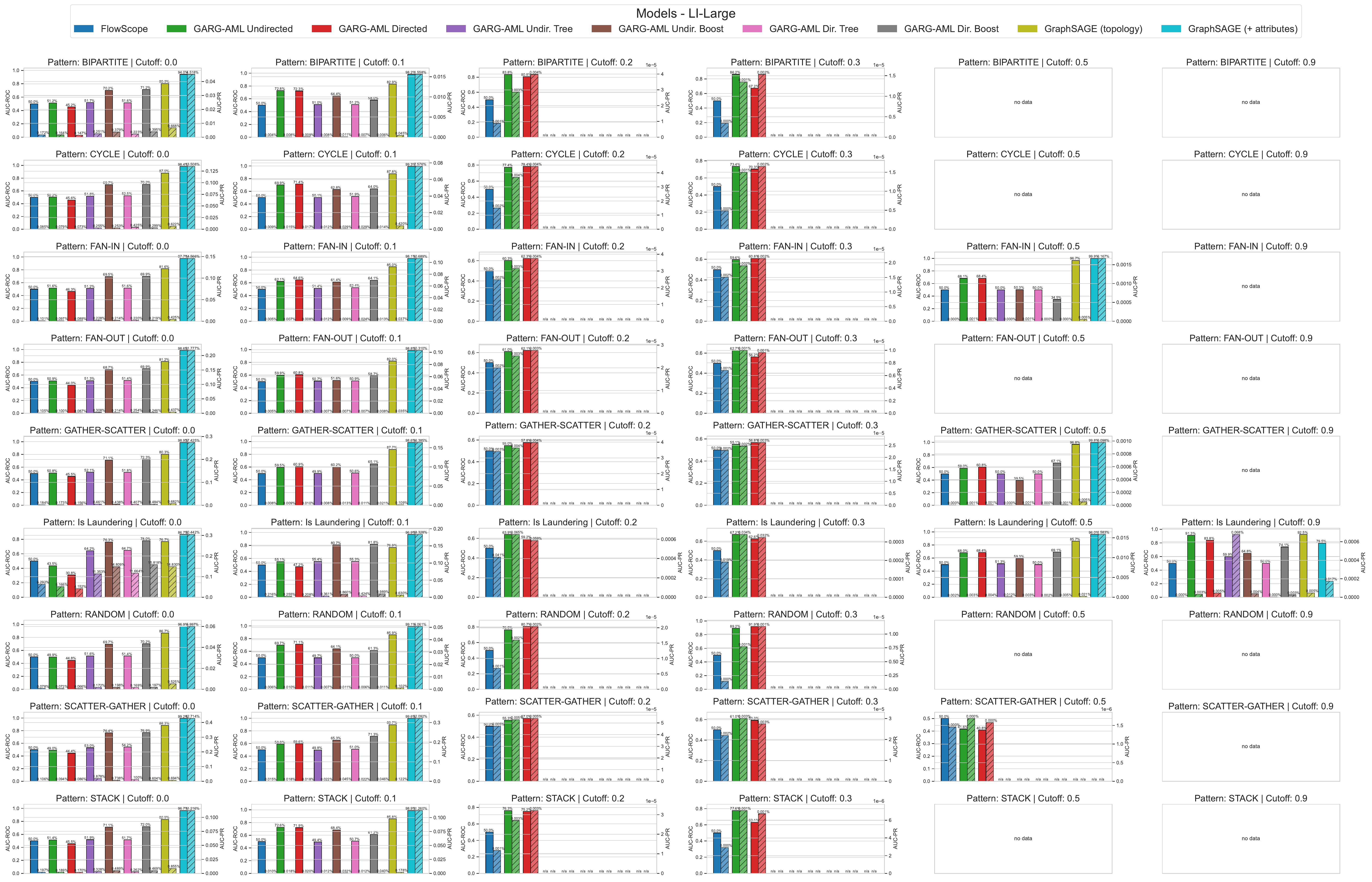}
		\caption{The AUC-ROC (solid bars) and AUC-PR (hatched bars) results on the LI-Large IBM dataset.}
		\label{fig-app:LILarge}
	\end{figure*}

	In this part of the appendix, we provide the full results on the IBM datasets. 
	These include all patterns defined by~\citet{altman2024realistic}. 
	The binary labels are defined using the cut-off values 0, 0.1, 0.2, 0.3, 0.5 and 0.9. 
	The results on the HI-Small and LI-Large datasets are provided in Figure~\ref{fig-app:HISmall} and Figure~\ref{fig-app:LILarge}, respectively. 
	Overall, GraphSAGE trained with node attributes performs best across both datasets, with the exceptions discussed below. 
	Cells marked n/a indicate that no results are available for that model.
	
	Figure~\ref{fig-app:HISmall} shows that, on HI-Small, GraphSAGE with attributes has the highest AUC-ROC for almost all patterns and cut-offs for which the supervised models could be trained. 
	Among the GARG-AML variants, the boosted extensions improve clearly on the base scores at the lower cut-offs (e.g., AUC-ROC of $80.0\%$ and $77.2\%$ versus $61.2\%$ for the full `Is Laundering' label at cut-off 0.1), and the tree-based extensions give a large gain in AUC-PR. 
	For scatter-gather, the pattern GARG-AML is designed for, the tree-based extensions reach an AUC-PR that is on par with or higher than that of GraphSAGE with attributes at most cut-offs (e.g., $32.4\%$ versus $26.5\%$ at cut-off 0.1 and $11.4\%$ versus $2.9\%$ at cut-off 0.5), although their AUC-ROC remains lower. 
	At the highest cut-offs, few positive labels remain, the supervised models often cannot be trained, and the base undirected score is the best GARG-AML variant (e.g., an AUC-ROC of $86.0\%$ for scatter-gather at cut-off 0.9). 
	For the non-smurfing patterns, GraphSAGE clearly outperforms the GARG-AML variants in most cases (e.g., fan-in at cut-off 0.1: AUC-ROC of $97.5\%$ versus $64.6\%$ for the base undirected score).
	
	On LI-Large (Figure~\ref{fig-app:LILarge}), results for the supervised models are not available at cut-offs 0.2 and 0.3, nor for patterns with too few positives at cut-offs 0.5 and 0.9. 
	Where they can be compared, GraphSAGE with attributes has by far the highest AUC-ROC ($87$--$99\%$ at cut-offs 0 and 0.1) and AUC-PR. 
	The boosted GARG-AML extensions improve the AUC-ROC over the base scores (e.g., $80.7\%$ versus $55.1\%$ for `Is Laundering' at cut-off 0.1), but their AUC-PR at cut-off 0.1 remains below $1\%$. 
	The only setting in which the unsupervised scores are competitive with GraphSAGE is `Is Laundering' at cut-off 0.9: the base undirected score has an AUC-ROC of $91.3\%$, similar to GraphSAGE using topology only ($92.5\%$) and higher than GraphSAGE with attributes ($79.5\%$). 
	This label has very few positives (Table~\ref{tab:labels data LI-Large}), so the corresponding AUC-PR and P@1000 values should be interpreted with care.
	
	\newpage
	
	\section{Limited Transaction View}
	\label{app:limited view}
	We include a sensitivity analysis where only transactions involving the bank's own clients are visible.
	This obfuscates the smurfing pattern if money launderers transfer their funds through multiple banks (see Section~\ref{sec: limitations}). 
	
	The IBM datasets include the sender's and receiver's bank information. 
	The experiments presented in this appendix are based on the HI-Small dataset, in which the 515,080 accounts and 5,078,345 transactions are spread over 30,528 banks.
	The main limitation of this dataset is that there are 6,193 banks with a single client, and the median number of clients per bank is 4, which is not a realistic distribution of clients. 
	
	We analyse two set-ups. 
	The first is to consider the largest bank in the dataset, called bank 012. 
	This shows the performance when a bank only has access to transactions involving their own clients. 
	The second is to consider the collection of transactions at the 50 largest banks. 
	This shows the performance when banks share information, e.g., through a specialised collaboration or because they are part of the same banking group. 
	Table~\ref{tab:appendix-coverage} summarises the resulting datasets.
	
	\begin{table*}[ht]
		\centering
		\caption{How much of the HI-Small network one institution observes. Percentages are
			shares of the full network.}
		\label{tab:appendix-coverage}
		\begin{tabular}{lrrrrrr}
			\toprule
			Bank  & \multicolumn{2}{c}{Clients} & \multicolumn{2}{c}{Transactions} & \multicolumn{2}{c}{Laundering tx.} \\
			\cmidrule(lr){2-3} \cmidrule(lr){4-5} \cmidrule(lr){6-7}
			& $n$ & \% & $n$ & \% & $n$ & \% \\
			\midrule
			012 &  2,639 & 0.51 & 99,452 & 1.96 & 146 & 2.82 \\
			top-50 & 55,423 & 10.76 & 1,592,556 & 31.36 & 1,722 & 33.26 \\
			\bottomrule
		\end{tabular}
	\end{table*}
	
\begin{figure*}
		\centering
		\includegraphics[width=0.8\linewidth]{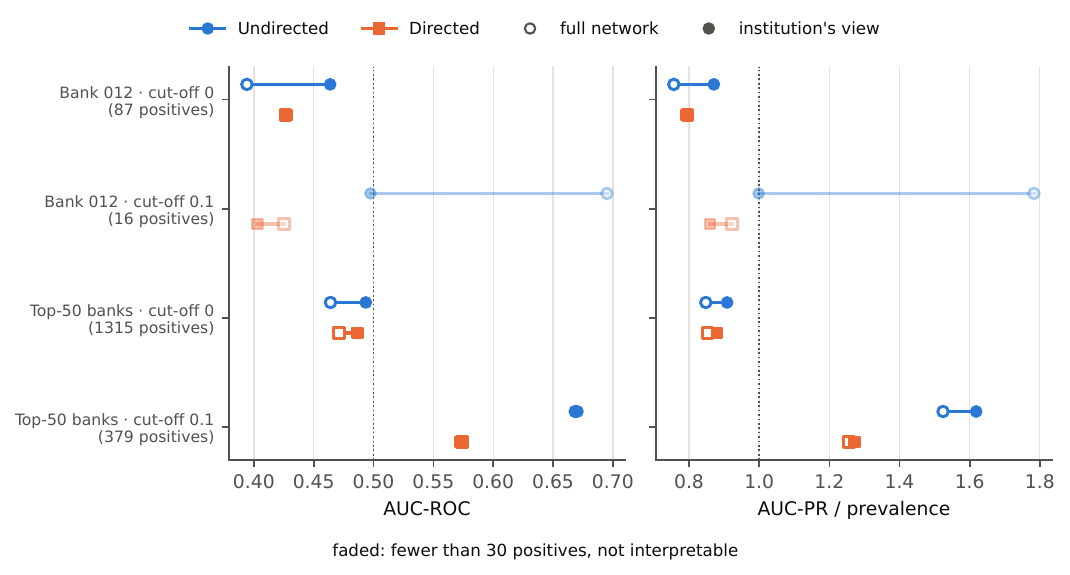}
		\caption{AUC-ROC (left) and AUC-PR relative to prevalence (right) of the GARG-AML score for the clients of an institution, on the full network (open markers) and on the institution's view (filled markers), for the undirected and directed score at label cut-offs 0 and 0.1. The number of positive clients is given per row. Louvain pre-processing is switched off in both views, so the full-network values differ from those in the main text. Faded rows have fewer than 30 positives and are not interpretable. Dotted lines mark a random ranking (AUC-ROC of 0.5; AUC-PR equal to the prevalence).}
		\label{fig:obs-detection}
	\end{figure*}

	For each case, we compare the original GARG-AML scores, where full visibility of the network is assumed, with the scores computed on the limited view of the institution. 
	The change in performance is given in Figure~\ref{fig:obs-detection} for the label cut-offs 0 and 0.1, with 87 and 16 positive clients for bank 012 and 1,315 and 379 for the top-50 banks. 
	Louvain pre-processing is switched off in both views, so the full-network values differ from those in the main text.
	
	The performance of the GARG-AML scores is low in most settings. 
	At cut-off 0, the AUC-ROC is below 0.5 for both institutions, for the undirected and the directed score, on the full network (0.39--0.47) as well as on the institution's view. 
	Only the top-50 banks at cut-off 0.1 reach a clearly better-than-random AUC-ROC (0.67 for the undirected and 0.57 for the directed score), with an AUC-PR of 1.5--1.6 times (undirected) and 1.3 times (directed) the prevalence. 
	Limiting the view does not reduce the performance in the settings with at least 30 positives. 
	For bank 012 at cut-off 0, the AUC-ROC of the undirected score changes from 0.39 to 0.46 and that of the directed score is unchanged (0.43), and for the top-50 banks all changes in AUC-ROC are at most 0.03 (e.g., 0.669 to 0.670 for the undirected score at cut-off 0.1). 
	The only large change is the undirected score of bank 012 at cut-off 0.1 (AUC-ROC from 0.70 to 0.50), but this rests on only 16 positives and is not interpretable. 
	Hence, we find no evidence that the limited view deteriorates performance, but for the single bank there is also no reliable signal to begin with. 
	A possible explanation for the slight improvements is that many of the laundering patterns are one-hop patterns, which remain visible to the institution, while the limited view removes the noise of subsequent transactions.
	
\begin{figure}
		\centering
		\includegraphics[width=0.8\linewidth]{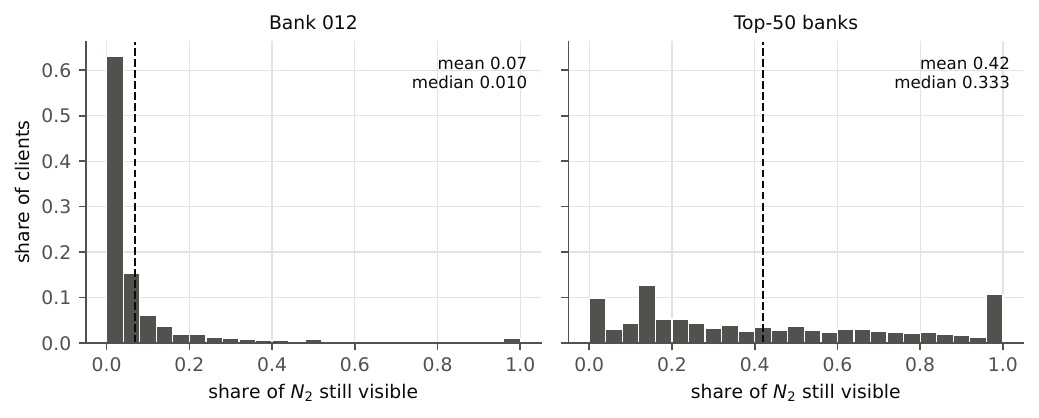}
		\caption{Share of each client's second-order neighbourhood that is still visible in the institution's view of HI-Small: bank 012 (left) and the 50 largest banks pooled (right). The dashed line is the mean. Clients with no second-order neighbourhood on the full network are not shown.}
		\label{fig:obs-n2-survival}
	\end{figure}

	Figure~\ref{fig:obs-n2-survival} shows that in the single bank view, almost all second-order neighbours are invisible to the bank (on average 7\% visible, median 1\%). 
	For the larger banking group consisting of the top-50 banks, the shares are more spread out, with on average 42\% (median 33\%) of the second-order neighbourhood visible. 
	A banking group with a substantial market share (here 31\% of all transactions) therefore observes a considerable part of its clients' second-order neighbourhoods, even when fraudsters spread their transactions over several banks. 
	For most clients, however, a large part of the neighbourhood is still missing, so these results are encouraging but not conclusive: they rely on a single simulated dataset, in which the distribution of clients over banks is not realistic.
	
\end{document}